\documentclass[12pt]{article}

\usepackage[active]{srcltx}
\usepackage{amsmath,graphicx}
\usepackage{epstopdf}
\usepackage{multirow}
\usepackage{amsfonts}
\usepackage{amssymb}
\usepackage{amscd}
\usepackage{epsf}
\usepackage[square, comma, sort&compress]{natbib}

\def\hybrid{\topmargin 10pt    \oddsidemargin 0pt
        \headheight 0pt \headsep 0pt
        \textwidth 6.25in       
        \textheight 9in
        \marginparwidth .875in
        \parskip 5pt plus 1pt   \jot = 1.5ex}

\hybrid

\numberwithin{equation}{section}
\numberwithin{table}{section}
\renewenvironment{thebibliography}[1]
 { \small
   \begin{list}{[\arabic{enumi}]}
    {\usecounter{enumi} \setlength{\parsep}{0pt}
     \setlength{\itemsep}{5pt} \settowidth{\labelwidth}{#1.}
     \sloppy
    }}{\end{list}}

\newcommand{\beq}{\begin{equation}}
\newcommand{\eeq}{\end{equation}}
\newcommand{\bea}{\begin{eqnarray}}
\newcommand{\eea}{\end{eqnarray}}
\newcommand{\ba}{\begin{array}}
\newcommand{\ea}{\end{array}}
\newcommand{\bt}{\begin{tabular}}
\newcommand{\et}{\end{tabular}}
\newcommand{\bc}{\begin{center}}
\newcommand{\ec}{\end{center}}

\def\nn{\nonumber}

\def\0 {\nonumber}

\newcommand{\ep}{\epsilon}

\newcommand{\sym}[2]
   {\omega\!\left(#1,#2\right)}
\newcommand{\symr}[2]
   {\omega_\rho\!\left(#1,#2\right)}
\newcommand{\symJ}[2]
   {\omega_J\!\left(#1,#2\right)}

\newcommand{\cref}{{\bf [check ref]}}

\newcommand{\be}{\begin{eqnarray}}
\newcommand{\ee}{\end{eqnarray}}
\newcommand{\Tr}{{\rm Tr}}
\newcommand{\Li}{{\rm Li}}

\def\IR{\mathbb{R}}
\def\IC{\mathbb{C}}
\def\IZ{\mathbb{Z}}
\def\id{{\bf 1}}
\def\Div{{{\rm Div}}}
\def\Fin{{{\rm Fin}}}

\def\pic #1#2{\hbox{\lower#1pt\hbox{~\mbox{\epsfxsize=20truemm \epsffile{#2}}}}}
\def\pic #1#2#3{\hbox{\lower#1pt\hbox{~\mbox{\includegraphics[scale=#3]{#2}}}}}
\def\picw #1#2#3{\hbox{\raise#1pt\hbox{~\mbox{\includegraphics[scale=#3]{#2}}}}}

\def\Remainder{{R}}
\def\WLa{{\rm a}}

\begin{document}

%
%

\begin{titlepage}

\rightline{\small SPIN-08/033} \rightline{\small ITP-UU-08/41}
\begin{center}

\vskip 2cm
{\Large \bf Scattering Amplitudes, Wilson Loops }\\
\vskip 0.4cm

{\Large \bf and the String/Gauge Theory Correspondence}

\vskip 1.5cm

{\bf Luis F. Alday$^a$ and Radu Roiban$^b$}

\vskip 0.8cm
{\em $^a$ Institute for Theoretical Physics and Spinoza Institute\\
Utrecht University, 3508 TD Utrecht, The Netherlands}\\

{\em $^b$ Department of Physics, Pennsylvania State University\\
University Park, PA 16802, USA}

\vskip 0.3cm {\tt $^a$ l.f.alday@uu.nl,~$^b$radu@phys.psu.edu}
\vskip2cm

\begin{abstract}

\vskip0.6cm

\noindent
We review, in a self-contained and pedagogical manner, recent
developments and techniques for the evaluation of the scattering
amplitudes of planar ${\cal N}=4$ SYM theory at both weak and strong
coupling. Special emphasis is placed on the newly discovered
connection between a special class of amplitudes and the expectation
values of particular cusped light-like Wilson loops.

\end{abstract}

\bigskip

(To be published in Physics Reports)

\end{center}

\end{titlepage}

\pagebreak

{\small
\tableofcontents }

\newpage

\section{Introduction} \label{sec:intro}

There are two approaches to understanding and solving ${\cal N}=4$
Yang-Mills (SYM): on the one hand, being a conformal field theory,
it is uniquely specified by the spectrum of (anomalous) dimensions
of gauge-invariant operators and their three-point correlation
functions, while, on the other hand, like any other quantum field
theory, it is completely specified by its scattering
matrix.\footnote{In the presence of a regulator, the definition of
the scattering matrix in a conformal field theory is no different
than in any massive quantum field theory.} The remarkable
properties of ${\cal N}=4$ SYM theory
  in the planar limit,
in particular its high degree of symmetry, allowed important progress
on both fronts: on the one hand, the integrability of the generator of
scale transformations allows the evaluation of the anomalous
dimensions of infinitely long operators through a Bethe ansatz
\cite{Beisert:2005fw, Beisert:2006ib, BES} while on the other hand the
theory is sufficiently symmetric and with sufficiently good high
energy behavior to allow high order perturbative calculations of its
scattering matrix (see e.g.\cite{Bern:2007ct}).

The strong coupling regime of the theory is directly accessible
through the AdS/CFT duality \cite{Maldacena:1997re,Gubser:1998bc,
Witten:1998qj} (see \cite{Aharony:1999ti} for a review),
which provides a description of ${\cal N}=4$ SYM theory solely in
terms of colorless, gauge invariant quantities.  It casts the analysis
of the strongly coupled planar theory in terms of the weakly-coupled
worldsheet theory for superstrings in $AdS_5\times S^5$. Being in one
to one correspondence with closed string states, local gauge invariant
operators have a natural place in the AdS/CFT duality.  This fact
played a major role in our understanding of the spectrum of operators
of the ${\cal N}=4$ SYM theory (as well as in many other contexts).

Scattering amplitudes describe the scattering of on-shell states
of the theory. As such, they carry color charge and thus it is not
immediately clear whether they can be described directly by the
closed string theory dual.
It is however possible to extend the closed string theory in
$AdS_5\times S^5$ by an open string sector. Depending on the
precise physical problem, they are described either by
semiclassical worldsheet configurations (e.g. when they describe
the expectation value of Wilson loops) or by vertex operators
(e.g. when they capture the scattering amplitudes of open string
states). Appropriately integrated, the correlation functions of
open string vertex operators are what one might define as the
gauge theory scattering amplitudes.
Vertex operators carry Chan-Paton factors and the correlation
functions of vertex operators decompose, in a natural way, into sum of
terms, each of which exhibits a clean separation of the color degrees
of freedom and the dependence on particle momenta.  The factors
carrying the kinematic dependence are known as partial
amplitudes. This decomposition mirrors closely the color decomposition
of gauge theory scattering amplitudes which we will discuss in section
\ref{weak_coupling}.
While non-local quantities, partial amplitudes carry no color charge
and thus could in principle be described by the closed string theory
dual to ${\cal N}=4$ SYM theory.

Strong coupling information extracted along these lines combined with
weak coupling higher-loop calculations lead us to hope that, at least
in some sectors, the scattering matrix of planar ${\cal N}=4$ SYM
theory can be found exactly. The four- and five-gluon amplitudes,
which are currently known to all orders in perturbation theory (up to
a set of undetermined constants), provide a proof of principle in this
direction.

Here we review, in a self-contained and pedagogical manner, some of
the recent developments and techniques for the evaluation of the
scattering amplitudes of planar ${\cal N}=4$ SYM theory at both weak
and strong coupling.
Related reviews of these topics may be found in references \cite{Dixon:2008tu,Alday:2008cg}.
The techniques developed for perturbative
calculations in this theory have been extended to other less symmetric
theories, as well as to QCD.
For a detailed account we refer the reader to the original literature.
We will however outline the attempts of generalizing
the strong coupling arguments to other theories.

Section \ref{weak_coupling} is devoted to weak coupling calculations
of scattering amplitudes. After setting up the notation and describing
some of their general properties, we proceed to outline
techniques for higher-loop high-multiplicity calculations.  While the
discussion is kept general at times, the main focus is planar ${\cal
N}=4$ SYM theory.  The generalized unitarity-based method is the
technique of choice for such calculations, as it combines in a natural
way, order by order in perturbation theory, the consequences of global
symmetries and of gauge invariance.

A common feature of all on-shell scattering amplitudes in massless
gauge theories in four dimensions is the presence of infrared
divergences, originating from low energy virtual particles as well
as from virtual momenta almost parallel to external ones. We will
discuss their structure captured by the soft/collinear
factorization theorem. A surprising feature of certain planar
amplitudes of ${\cal N}=4$ SYM theory, noticed in explicit
calculations, is that the exponential structure of the infrared
divergences extends also to the finite part of the amplitude. We
will describe the conjectured iteration relations based on this
observations, which suggest that any maximally helicity violating
loop amplitude may be written in terms of the corresponding one
loop amplitude.
We end section \ref{weak_coupling} with an outline of potential
departures from these relations and the current state of the art
in testing them.

For a variety of reasons, the identification and evaluation of the
strong coupling counterpart of the partial amplitudes described in
section \ref{weak_coupling} is not entirely straightforward. In
section \ref{strong_coupling} we describe how the AdS/CFT
duality can be used for this purpose. The main result is that, at
strong coupling, partial amplitudes are closely related to a
special class of polygonal, light-like Wilson loops. Thus, they may
be evaluated as the area of certain minimal surfaces with boundary
conditions fixed by the momenta of the massless particles
participating in the scattering process.\footnote{Certain features
of partial amplitudes -- such as the polarization of the scattered
particles -- are however not captured directly by their Wilson loop
interpretation. This information is best captured in the vertex
operator picture for the scattering process.}
The strong coupling calculations exhibit features analogous to their
weak coupling counterparts, such as the presence of long distance/low
energy divergences. Thus, in analogy with the weak coupling situation,
the very definition of scattering amplitudes requires the presence of
a regulator.
Finding gauge-invariant regulators is not completely obvious in
weakly-coupled gauge theories; by contrast, any regulator which
may be realized on the string theory side of the AdS/CFT
correspondence without direct reference to the color degrees of
freedom of the open string sector is manifestly gauge-invariant.
To set-up the computation we begin by introducing a D-brane as an
infrared regulator. Actual computations are, however, carried out
using a string theory analog of dimensional regularization,
obtained by taking the near horizon limit of D$(3-2\epsilon)$
branes. While not yet clear how to extend the calculations beyond
leading order, this regularization scheme has the advantage of
being analogous to dimensional regularization as used in gauge
theory calculations and thus of allowing a direct comparison of
results.

We carry out the calculation of the four-gluon scattering amplitude
both in the strong coupling version of dimensional regularization as
well as using an infrared cut-off which removes, in a gauge-invariant
way, all dangerous low energy modes. This cut-off scheme is
particularly appropriate for understanding the conformal properties of
the amplitudes at strong coupling.

The arguments used to construct the strong coupling interpretation of
gluon partial amplitudes in ${\cal N}=4$ SYM theory may be generalized
to other, less symmetric theories and with a richer field content. We
describe processes involving not only gluons but also local operators,
mesonic operators and quarks. We end section \ref{strong_coupling}
with an overview of other interesting discussions concerning the
strong coupling limit of scattering amplitudes to leading and
subleading orders.

While the arguments leading to it apply directly only in the strong
coupling regime, the result -- that the calculation of certain partial
amplitudes is mathematically equivalent to the calculation of the
expectation value of certain null polygonal Wilson loops -- can be
stated independently of the value of the coupling constant. This
observation led to the conjecture that the same null polygonal Wilson
loops reproduce maximally helicity violating (partial) amplitudes,
order by order in weakly coupled perturbation theory.  Section
\ref{Amp_vs_WL} reviews the evidence in favor of this conjecture,
beginning with the explicit identification at one loop of the
expectation value of the $n$-sided null Wilson loop and the $n$-gluon
maximally helicity violating amplitude. This observation allows, as we
will describe, for a direct strong coupling test of the BDS iteration
relation described in section \ref{weak_coupling}; the result suggests
that the iteration relation needs to be modified in the strong
coupling regime.

Unlike their generic counterparts, light-like Wilson loops are
invariant under conformal transformations on the space they are
defined on (in this case a space closely related to momentum
space\footnote{In the case of ${\cal N}=4$ SYM this space may itself
be identified with the position space. In this formulation light-like
Wilson loops are invariant under conventional conformal
symmetry. Their expectation values may then be mapped back to momentum
space and related to scattering amplitudes.}). While the presence of
divergences requires regularization, it can be argued that any
regularization breaks this symmetry. The anomaly introduced by this
breaking is an important tool for extracting higher-loop information
on the expectation value of Wilson loops. Its key property is that it
can be identified to all orders in perturbation theory due to its
close relation to the structure of cusp singularities. We will review
it in some detail in section \ref{Amp_vs_WL}.
The resulting anomalous Ward identity mirrors the one discussed in
section \ref{strong_coupling} at strong coupling.
The restrictions imposed by conformal symmetry are particularly
strong for Wilson loops corresponding to the scattering of a small
number of particles, fixing uniquely the kinematic dependence of
the expectation value of the four- and five-sided loop.
We end section \ref{Amp_vs_WL} by outlining the current state of the
art in the calculation of expectation values of null polygonal Wilson
loops, namely the two-loop expectation value of the four- and
six-sided loop.

Partial amplitudes and (null polygonal) Wilson loops are {\it a
priori} unrelated quantities. It is remarkable that a relation such as
the one reviewed here can exist at all. Its origins and full
implications remain to be uncovered; in section \ref{outlook} we
collect some open questions whose answers may lead to an improved
understanding of the deep and powerful structures governing the
dynamics of ${\cal N}=4$ super-Yang-Mills theory and perhaps other
four-dimensional gauge theories.



\section{Scattering amplitudes at weak coupling \label{weak_coupling}}

On-shell scattering amplitudes are perhaps the most basic
quantities computed in any quantum field theory. The standard
textbook approaches proceed to relate them through the LSZ
reduction to Green's functions which are in turn
computed in terms of Feynman diagrams. Each diagram evaluated
separately is typically more complicated that the complete
amplitude; the reason may be traced to Feynman diagrams not
exhibiting and taking advantage of the symmetries of the theory --
neither local nor global. The first instance where this shows up
is for tree level amplitudes, where one notices major
simplifications as all diagrams are added together.

Indeed, besides the scattering of physical polarizations, off-shell
scattering amplitudes also describe the scattering of (unphysical)
longitudinal polarizations of vector fields. On-shell, the equations
of motion (or, more generally, Ward identities) guarantee the
decoupling of such states. One may expose this decoupling at the
Lagrangian level by choosing a physical gauge. The resulting
gauge-fixed action does not, however, have a transparent use at the
quantum level. As usual, in an off-shell covariant and renormalizable
approach to loop corrections to scattering amplitudes, Faddeev-Popov
ghosts are needed to cancel the contribution of unphysical fields
propagating in loops.


The (generalized) unitarity-based method provides means of eliminating
the appearance of unphysical degrees of freedom, while preserving
all on-shell symmetries of the theory and avoiding the proliferation
of Feynman diagrams.
It allows the analytic construction of loop
amplitudes in terms of tree-level amplitudes. Thus, it manifestly
incorporates most (if not all) simplifying consequences of gauge
invariance and symmetries. Simplicity of loop level amplitudes is to a
large extent a consequence of simplicity of tree-level amplitudes.

In addition to the use of Feynman diagrams, there are several
methods for computing tree-level scattering amplitudes: the
Berends-Giele (off-shell) recursion relations
\cite{Berends_Giele}, MHV vertex rules \cite{CSWrules} \footnote{The
MHV vertex rules have been successfully extended to loop level in \cite{BST}.}
and the BCFW recursion relations \cite{BCFW1,BCFW2}. We will not
review them in detail and refer the reader to the original literature
and existing reviews  \cite{Dixon_TASI,Bern:2007dw,Cachazo_Svrcek}. Instead, we will be
focusing on the construction of loop amplitudes, assuming that the
tree-level amplitudes are known. After setting up the convenient
notation and describing some of the general properties of scattering
amplitudes, we will review the factorization of infrared divergences,
discuss the unitarity method and illustrate it with several
examples. We will then describe the BDS conjecture for the all-loop
resummation of $n$-point MHV amplitudes, the potential corrections and
the fact that such corrections indeed appear starting with the
six-point two-loop amplitude. We will also describe the emergence of
dual conformal invariance from the explicit expressions of amplitudes.

\subsection{Organization, presentation and general properties}

A good notation as well as an efficient organization of the
calculation and result are indispensable ingredients for the
calculation of scattering amplitudes, whether with Feynman diagrams or
by other means. They are provided, respectively, by the spinor
helicity method (for massless theories) and by color ordering, which
we now review. These methods allow the decomposition of amplitudes in
smaller, gauge-invariant pieces with transparent properties. An
enlightening discussion of these topics may be found in
\cite{Dixon_TASI}.

\subsubsection{Spinor helicity and color ordering}

In a massless theory, solutions of the chiral Dirac equation
provide an excellent parametrization of momenta and polarization
vectors which allows, among other things, the construction of
physical polarization vectors without fixing noncovariant gauges.
The main observation is that the sum over polarizations of a
direct product of a Dirac spinor and its conjugate is
\be
\sum_{s=\pm} u_s(k) {\bar u}_s(k)=-k\llap/~~.
\ee
Upon projecting
onto the chiral components one immediately finds that
\be
u_-(k){\bar u}_-(k)=-k_\mu {\bar \sigma}^\mu~~,
\label{p.sigma}
\ee
where as usual ${\bar \sigma}=(\id,-\boldsymbol{\sigma})$ are the
Pauli matrices. The decomposition of a massless four-dimensional
vector as a direct product of two 2-component commuting ``spinors''
follows also more formally from the fact that $p^2=\det(p_\mu {\bar
\sigma}^\mu)$, implying that the mass-shell condition requires that
$p_\mu {\bar \sigma}^\mu$ has unit rank, i.e.
\be (k_\mu{\bar \sigma}^\mu)_{\alpha{\dot\alpha}} =
\lambda_\alpha{\tilde{\lambda}_{\dot\alpha}} ~~~~~~\lambda\equiv
u_-(k)~~~~~~~{\tilde{\lambda}}={\bar u}_-(k)~~;
\label{split} \ee
the multiplication of spinors follows from Lorentz invariance:
\be
\langle ij\rangle =
\epsilon^{\alpha\beta}\lambda_{i\alpha}\lambda_{j\beta} ~~~~~~~~
[ij] = -\epsilon^{{\dot \alpha}{\dot \beta}} {\tilde
\lambda}_{i{\dot \alpha}}{\tilde \lambda}_{j{\dot \beta}}
\ee
In Minkowski signature $\lambda$ and ${\tilde \lambda}$ are complex
conjugate of each other.  It is useful to promote momenta to
(holomorphic) complex variables and the Lorentz group to
$SL(2,\IC)\times SL(2,\IC)$. Then, $\lambda$ and ${\tilde
\lambda}$ are independent complex variables and the decomposition
(\ref{split}) exhibits a scaling invariance
\be
\lambda\mapsto
S\lambda~~~~~~~~~~{\tilde\lambda}\mapsto
\frac{1}{S}{\tilde\lambda}
\ee
where $S$ is an arbitrary constant. We will shortly see that
scattering amplitudes have definite scaling properties under this
transformation.\footnote{For a Minkowski signature metric $S$ is a
pure phase.}

This parametrization of four-dimensional momenta allows the
construction of simple expressions for the physical polarizations of
massless vector fields. In general, gauge invariance requires that
they be transverse, and that shifts by the momentum of the
corresponding field should not change their form and
properties. Moreover, in the frame in which the vector field
propagates along a specified axis, they should take the standard form
of circular polarization vectors.

A solution to these constraints can be constructed in terms of an
arbitrary null (reference) vector $\xi$
($\xi_\mu\sigma^\mu_{\alpha{\dot\alpha}}=\xi_\alpha{\tilde\xi}_{\dot\alpha}$):
\be
\begin{array}{l}
\epsilon^+_\mu(k, \xi)=\hphantom{-}\,\displaystyle{\frac{\langle\xi|\gamma_\mu|k]}
{\sqrt{2}\langle\xi k\rangle}}
\\[10pt]
\epsilon^-_\mu(k, \xi)=-\,\displaystyle{\frac{[\xi|\gamma_\mu|k\rangle}
{\sqrt{2}[\xi k]}}
\end{array}
~~~~~~~~~~
\begin{array}{l}
\epsilon^+_{\alpha\dot\alpha}(k, \xi)=
\hphantom{-}\sqrt{2}\,
\displaystyle{\frac{\xi_\alpha{\tilde \lambda}_{\dot\alpha}}{\langle\xi k\rangle}}
\\[10pt]
\epsilon^-_{\alpha\dot\alpha}(k, \xi)=
-\sqrt{2}\,\displaystyle{\frac{\lambda_\alpha {\tilde\xi}_{\dot\alpha}}{[\xi k]}}
\end{array}
~~~~~~~~~~
\label{pol_vector}
\ee
The reference vector may be changed by a gauge transformation. Indeed,
the transformation $\epsilon(p)\mapsto \epsilon(p) + A\, k$ for some
$A$ can be realized as a change of the reference vector:
\be
\xi_\alpha\mapsto \xi_\alpha+A\,\langle \xi k\rangle \lambda_\alpha
~~~~~~~~
{\tilde \xi}_{\dot \alpha}\mapsto {\tilde \xi}_{\dot \alpha}
-A\,[\xi k ] {\tilde \lambda}_{\dot\alpha}~~.
\ee
This freedom of choosing independently the reference vector for each
of the gluons participating in the scattering process is a very
convenient tool for simplifying (somewhat effortlessly) the
expressions for (tree-level) scattering amplitudes.

\

The loop expansion of scattering amplitudes is defined in the
usual way:
\be
A=\sum_lg^{2l}A^{(l)}~~.
\ee
A clean organization of scattering amplitudes is a second useful
ingredient in the calculation of scattering amplitudes at any fixed loop
order $L$.
Besides the organization following the helicity of external states
implied by spinor helicity, at each loop order $l$ an organization
following the color structure is also possible and desirable, if only
because amplitudes are separated in at least $(n-1)!$ gauge invariant
pieces (here $n$ is the number of external legs). For an $SU(N)$ gauge
theory with gauge group generators denoted by $T^a$, it is possible to
show that any $L$-loop amplitude may be decomposed as follows:
\be
A^{(L)}=N^L\sum_{\rho\in S_n/\IZ_n}
\Tr[T^{a_{\rho(1)}}\dots T^{a_{\rho(n)}}]A^{(L)}(k_{\rho(1)}\dots
k_{\rho(n)},N) +{\rm multi{\scriptstyle -}traces}
\label{color_ordering}
\ee
where the sum extends over all
%
%
non-cyclic permutations $\rho$ of $(1\dots n)$. This is equivalent
to fixing one leg -- say the first -- and summing over all
permutations of the other legs. The coefficients
$A(k_{\rho(1)}\dots k_{\rho(n)},N)$ are called color-ordered
amplitudes. The multi-trace terms left unspecified in the equation
above do not appear in the planar (large $N$) limit, which will be
our main focus. We shall therefore ignore them in the following.
In the same limit the $N$ dependence of the partial amplitudes
drops out: \be A(k_{\rho(1)}\dots k_{\rho(n)},N)
\stackrel{N\rightarrow\infty}{\longrightarrow} A(k_{\rho(1)}\dots
k_{\rho(n)})~~. \ee The latter are the so called planar partial
amplitudes, while the subleading terms in the $1/N$ expansion as
well as the multi-trace terms in (\ref{color_ordering}) are called
non-planar partial amplitudes.

\begin{figure}[t]
\centerline{\epsfxsize 2.5 truein \epsfbox{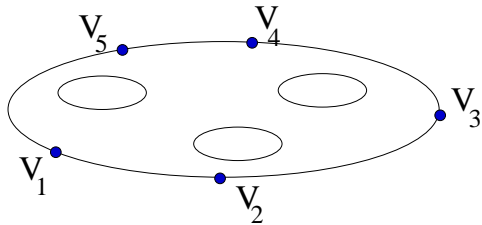}}
\caption[a]{\small The planar three-loop open string diagram
contributing to the five-gluon scattering. The single-trace
structure is manifest.}
\label{fig:string_loops}
\end{figure}

It is possible to argue for this presentation of amplitudes by
inspecting the Feynman rules and noting that their color
dependence separates from their momentum dependence. Perhaps the
cleanest argument however is in terms of string theory diagrams
\cite{Bern_Kosower_1991}.  Indeed, in string theory, gluon
scattering amplitudes are computed in terms of Riemann surfaces
with boundaries. Vertex operators carrying Chan-Paton factor are
inserted on their boundaries, with color indices contracted along
boundaries (see figure \ref{fig:string_loops}).  As one integrates
over the insertion points one sweeps over all possible orders of
inserting the operators. The cyclic permutations however are
naturally excluded because the boundaries in question are closed
curves. The boundaries carrying no vertex operators contribute the
explicit factors of $N$ in equation (\ref{color_ordering}).

\subsubsection{General properties of color ordered amplitudes}

The general properties of color-ordered amplitudes follow from their
construction in terms of Feynman diagrams (or string diagrams). The
results of other constructions must obey the same properties.  Some of
them -- such as the analytic structure -- impose powerful constraints
and in some cases uniquely determine the (tree-level) amplitudes. We
collect here some of the more important properties
\cite{Mangano_Parke_Xu}:

\begin{itemize}

\item
cyclicity (this is a consequence of the cyclic symmetry of traces)
\be
A(1,\dots n)=A(2,\dots, n, 1)
\ee

\item reflection (this is a consequence of the fact that 3-point
vertices pick up a sign under such a reflection and that an
amplitude with $n$ external legs has $(n+2L-2)$ three-point
vertices)
\be A(1,\dots n)=(-)^nA(n\dots 1) \ee

\item photon decoupling: In a theory with only adjoint fields, the diagonal $U(1)$ does not
interact with anyone. Thus, all amplitudes involving this field
identically vanish. At tree-level this property may be captured by a Ward
identity: fixing one of the external legs ($n$ below) and summing
over cyclic permutations $C(1,\ldots,n-1)$ of the remaining
$(n-1)$ legs leads to a vanishing result: \be
\sum_{C(1,\ldots,n-1)} A(1,2,3,\ldots,n) =0~~. \ee In string
theory language this is a consequence of the structure of the
operator product expansion of vertex operators. At loop level this
identity is modified and relates planar and nonplanar partial
amplitudes \cite{Bern_Kosower_1991}.

\item parity invariance (a color-ordered amplitude containing all choices
of helicities of external legs is invariant if all helicities are
reversed and simultaneously all spinors $\lambda$ are replaced by the
spinors ${\tilde\lambda}$ and vice-versa). This operation may be
expressed as a fermionic Fourier-transform \cite{RSV_2}
\be
A(\lambda_i, {\tilde\lambda}_i, \eta_{i A}) = \int d^{4 n} \psi\
\exp\left[ i \sum_{i=1}^n \eta_{i A} \psi_i^A \right]
A({\tilde\lambda}_i,\lambda_i,\psi_i^A).
\ee

\item soft (momentum) limit:
in the limit in which one momentum becomes soft the amplitude
universally factorizes as follows
\be
A^{\rm tree}(1^+,2,\ldots,n) \longrightarrow
\frac{\langle n\,2 \rangle}{\langle n\,1 \rangle\langle 1\,2 \rangle}
A^{\rm tree}(2,\ldots,n)
\ee

\item collinear limit: in the limit in which two adjacent momenta
become collinear $k_{n-1}\cdot k_n\rightarrow 0$ an $L$-loop amplitude factorizes as
\be
A^{(L)}_n(1 \dots (n-1)^{h_{n-1}},n^{h_n})\mapsto
\sum_{l=0}^L\sum_{h}
A^{(L-l)}_{n-1}(1\dots k^h){\rm Split}^{(l)}_{-h}((n-1)^{h_{n-1}},n^{h_n})~~,
\label{collinear_limit}
\ee
where $h_i$ denotes the helicity of the $i$-th gluon.
For a given gauge theory, the $l$-loop splitting amplitudes ${\rm
Split}^{(l)}_{-h}((n-1)^{h_{n-1}},n^{h_n})$ are universal functions
\cite{Kosower_splitting} of the helicities of the collinear particles,
the helicity of the external leg of the resulting amplitude and of the
momentum fraction $z$ defined as
\be
z=\frac{\xi\cdot k_{n-1}}{\xi\cdot (k_{n-1}+k_n)}~~.
\label{momentum_fraction}
\ee
In the strict collinear limit one may also use $k_{n-1} \to z k$ and
$k_{n} \to (1-z) k$ with $k^2=(k_{n-1}+k_n)^2=0$. For example, the
tree-level splitting amplitudes are:
\be
&&{\rm Split}^{(0)}_{-}(1^{+},2^+)=\frac{1}{\sqrt{z(1-z)}} \frac{1}{\langle 1\,2 \rangle}
\\
&&{\rm Split}^{(0)}_{-}(1^{+},2^-)=\frac{z^2 }{ \sqrt{z(1-z)}} \frac{1 }{ [1\,2]}
~~~~~~~~
{\rm Split}^{(0)}_{+}(1^{+},2^-)=\frac{(1-z)^2 }{ \sqrt{z(1-z)}} \frac{1}{\langle 1\,2 \rangle}
\nonumber
\ee
In ${\cal N}=4$ SYM theory Ward identities imply that all splitting
amplitudes rescaled by their tree-level expressions are the same.

Scattering amplitudes have similar factorization properties when more
than two adjacent momenta become simultaneously collinear
\cite{Kosower_splitting}.

\item multi-particle factorization:
color ordered amplitudes exhibit poles if the square of the sum of
some adjacent momenta vanishes. At tree-level this pole
corresponds to some propagator going on-shell. At higher loops,
the amplitude decomposes into a completely factorized part given
by the sum of products of lower loop amplitudes and a
non-factorized part, given in terms of additional universal
functions.  At one-loop level and in the limit $k_{1,m}^2\equiv
(k_1+\dots k_m)^2\rightarrow 0$ one finds \cite{Bern_Chalmers}
\be
A^{\rm 1\,loop}_n(1,\ldots,n)
\!\!\! \!\!\! \! \!&& \!\!\!
\longrightarrow\\
\sum_{h_p = \pm}\Big[
\!\!\!\!\!\!\!\!\!\!\!\!\! && A^{\rm
tree}_{m+1}(1,\ldots,m,k^{h_k}) \frac{i}{k_{1,m}^2} A^{\rm
1\,loop}_{n-m+1}((-k)^{-h_k},m+1,\ldots,n)
\cr
&+&\!\!\!A^{\rm 1\,loop}_{m+1}(1,\ldots,m,k^{h_k})
\frac{i}{k_{1,m}^2} A^{\rm tree}_{n-m+1}((-k)^{-{h_k}},m+1,\ldots,n)
\vphantom{\Big|}
\cr
&+&\!\!\!A^{\rm tree}_{m+1}(1,\ldots,m,k^{h_k})
\frac{i {\cal F}(1\dots n)}{k_{1,m}^2} A^{\rm tree}_{n-m+1}
((-k)^{-{h_k}}, m+1,\ldots,n)
\Big]
\nonumber
\ee

\end{itemize}

\noindent
While color ordering (\ref{color_ordering}) in the planar theory
implies that complete amplitudes may be reconstructed from $(n-1)!$
gauge invariant partial amplitudes, the first four properties listed
above imply that only a much smaller number is in fact necessary.

\subsubsection{Some simple examples}

Besides color ordering, scattering amplitudes can be organized
following the number of negative helicity gluons. One can easily
see that the amplitude with only positive helicity gluons as well
as the amplitude with a single negative helicity gluons vanish
identically at tree level in any gauge theory.
This is realized by choosing the same reference vectors for all
gluons with the same helicity and equal to the momentum of the
negative helicity gluon. In absence of supersymmetry, quantum
corrections spoil this conclusion. In the presence of
supersymmetry, its Ward identities imply that this vanishing
result is protected to all orders in perturbation theory. Indeed,
the supersymmetry transformation rules  are
\bea
&&[Q^a(\eta),\,g^{\pm}(k)]=\mp \Gamma^\pm(k,\eta)
\lambda^{a\pm}(k) \nonumber \\ &&[Q^b(\eta),\,\lambda^{b\pm}(k)]=
\mp \Gamma^\pm(k,\eta) g^{\pm}(k)\delta^{ab}\mp
i\Gamma^\pm(k,\eta)\phi_{\pm}^{ab}\epsilon^{ab} \\
&&\Gamma(k,\eta)^+=\theta [\eta,\,k],~~~~~~~~
\Gamma(k,\eta)^-=\theta \langle \eta,\, k\rangle \nonumber
\eea
where $\eta$ is a reference spinor. Acting with them on the vanishing
matrix element $\langle 0|\lambda^{a+} g^\pm g^+\dots g^+|0\rangle$
and using the fact that fermions have only helicity-conserving
interactions, it immediately follows that the all-plus amplitude
vanishes. Similarly, using the vanishing of $\langle 0|\lambda^{a+}
g^- g^+\dots g^+|0\rangle$ and making judicious choices for the
reference spinor leads to the vanishing of the amplitude with a
single negative helicity gluon \cite{Grisaru:1977px} \footnote{To
spell out the details, we use an ${\cal N}=1$ part of the
${\cal N}=4$ supersymmetry algebra and denote by
$\Lambda$ the gaugino related to the gluon by these transformations:
\be 0=\langle 0|[Q(\eta(q)),\Lambda^+ g^+
g^+\dots g^+]|0\rangle
=-\Gamma^+(q, k_1) A(g^+ g^+\dots g^+)
- \sum_i \Gamma^-(q, k_i) A(\Lambda^+ g^+\dots \Lambda_i^+ g^+)
\nonumber
\ee
implies the vanishing of the all-plus amplitude while
\be
0&=&\langle 0|[Q(\eta(q),\Lambda^+ g^- g^+\dots g^+]|0\rangle\cr
&=&-\Gamma^+(q, k_1) A(g^+ g^-\dots g^+) +
    \Gamma^-(q, k_2) A(\Lambda^+ \Lambda^- g^+\dots  g^+)-
   \sum_i \Gamma^+(q, k_i) A(\Lambda^+ g^- g^+\dots \Lambda_i^+ g^+)
\nonumber
\ee
immediately implies, after choosing $q=k_2$, the vanishing of the
amplitude with a single negative helicity gluon.  }
\be
A^{\rm tree}(g^+\dots g^+)=0~~~~~~~~~~A^{\rm tree}(g^-g^+\dots
g^+)=0~~.
\label{vanishing_amplitudes}
\ee
In the following
we will focus mainly on ${\cal N}=4$ SYM.

The first nonvanishing amplitude, having two negative helicity
gluons, takes the form \cite{Parke_Taylor1,Parke_Taylor2}
\be
A^{\rm tree}_{MHV}{\scriptstyle (1^+\dots i^-\dots j^-\dots n)}
=\frac{\langle ij\rangle^4}{\prod_{k=1}^n \langle k, k+1
\rangle}~~,
\ee
where $k$ is a cyclic index (i.e. $n+1\equiv 1$) and $i$ and $j$ are
the labels of the negative helicity gluons.  The fact that in ${\cal
N}=4$ SYM the two gluon helicity states are related by supersymmetry
makes it possible to show \cite{Bern:1996ja} that, to all loop orders,
$n$-point MHV amplitudes are cyclicly symmetric, up to an overall
factor of $\langle ij\rangle^4$ where $i$ and $j$ label, as above, the
negative helicity gluons. Indeed, using supersymmetric Ward identities
it is possible to relate the $n$-gluon amplitude to the two scalar,
$(n-2)$-gluon amplitude.  After interchanging the position of the two
scalars, which does not affect the amplitude, one may use the same
identities to obtain an amplitude with one of the two negative
helicity gluons displaced to any position. It thus follows that, to
any loop order $L$,
\be
A_{MHV}^{(L)}=A^{\rm tree}_{MHV}\,{\cal
M}^{(L)}(s_{i,i+1}, s_{i\dots i+2},\dots)~~,
\label{LloopMHV}
\ee
where ${\cal M}^{(L)}(s_{i,i+1}, s_{i\dots i+2},\dots)$ is a cyclicly symmetric
function of momenta and $s_{i...j}=(k_i+k_{i+1}+...+k_j)^2$. This
factorization of the tree-level amplitude also holds for the
infrared-singular terms of all amplitudes in all massless gauge
theories. A similar expression holds in ${\cal N}=4$ SYM also for
collinear splitting amplitudes introduced in (\ref{collinear_limit}):
\be {\rm Split}_\lambda^{(L)}(a^{h_a},b^{h_b})={\rm
Split}_\lambda^{\rm tree}(a^{h_a},b^{h_b}) \;r_S^{(L)}(z,s_{ab})
\ee
where the momentum fraction $z$ is defined in equation
(\ref{momentum_fraction}). A direct argument follows closely the one
for MHV amplitudes. Alternatively, one may extract it by simply
comparing the collinear limit of (\ref{LloopMHV}) and the expected
behavior (\ref{collinear_limit}).

\subsubsection{Soft/Collinear factorization \label{sec:soft_collinear}}

A general feature of massless gauge theories in four dimensions is the
existence of infrared singularities.\footnote{Ultraviolet divergences
may of course be present as well; as previously mentioned, our focus
is ${\cal N}=4$ SYM theory, which is free of such divergences.} Unlike
ultraviolet divergences they cannot be renormalized away, but rather
should cancel once gluon scattering amplitudes are combined to compute
infrared-safe quantities. Their structure has been thoroughly studied
and understood (see e.g.
\cite{IR_paper1,IR_paper2,IR_paper3,IR_paper4,IR_paper5,IR_paper6,
IR_paper7,IR_paper8,IR_paper9,IR_paper10,IR_paper11,IR_paper12,AM2}).
Here we briefly review some of the results specializing them,
following \cite{Bern:2005iz}, to the case of ${\cal N}=4$ SYM in
the planar limit.

In a gauge theory, infrared singularities of scattering amplitudes
come from two sources: the small energy region of some virtual
particle \be \int \frac{d\omega}{\omega^{1+\epsilon}}\propto
\frac{1}{\epsilon} \label{soft} \ee and the region in which some
virtual particle is collinear with some external one \be \int
\frac{d k_T}{k_T^{1+\epsilon}}\propto \frac{1}{\epsilon}~~.
\label{collinear} \ee Since they can occur simultaneously, at
$L$-loops the infrared singularities lead to an $1/\epsilon^{2L}$
pole.

The structure of soft and collinear singularities in a massless gauge
theory in four dimensions has been extensively studied. The
realization that soft and virtual collinear effects can be factorized
in a universal way, together with the fact
\cite{Collins_Soper_Sterman} that the soft radiation can be further
factorized from the (harder) collinear one led to a three-factor
structure for gauge theory scattering amplitudes \cite{Sen:1982bt,
Sterman:2002qn, MertAybat:2006mz}:
\be
{\cal M}_n=
    \left[\prod_{i=1}^nJ_i(\frac{Q}{\mu},\alpha_s(\mu),\epsilon)\right]
    \times S(k, \frac{Q}{\mu},\alpha_s(\mu),\epsilon)
    \times h_n(k, \frac{Q}{\mu},\alpha_s(\mu),\epsilon)~~.
\label{factorization}
\ee
Here the product runs over all the external lines. $Q$ is the
factorization scale, separating soft and collinear momenta, $\mu$
is the renormalization scale and $\alpha_s(\mu)=\frac{g(\mu)^2}{4\pi}$
is the running coupling at scale $\mu$. Both $h_n(k,
\frac{Q}{\mu},\alpha_s(\mu),\epsilon)$ and the rescaled amplitude
${\cal M}_n$ are vectors in the space of color configurations
available for the scattering process. The soft function $S(k,
\frac{Q}{\mu},\alpha_s(\mu),\epsilon)$ is a matrix acting on this
space and it is defined up to a multiple of the identity matrix.  It
captures the soft gluon radiation and it is responsible for the purely
infrared poles. For this reason it can be computed in the eikonal
approximation in which the hard partonic lines are replaced by Wilson
lines. The ``jet'' functions
$J_i(\frac{Q}{\mu},\alpha_s(\mu),\epsilon)$ do not alter the color
flow and contain the complete information on collinear dynamics of
virtual particles. Finally, $h_n(k,
\frac{Q}{\mu},\alpha_s(\mu),\epsilon)$ contains
the effects of highly virtual fields and is finite as
$\epsilon\rightarrow 0$. The jet and soft functions can be
independently defined in terms of specific matrix elements.

The factorization scale $Q$ is arbitrary (within some physical
limits); it is simply used to construct the equation
(\ref{factorization}). While it enters in each of the three factors on
the right hand side, the (rescaled) amplitude ${\cal M}_n$ is
independent of it. This independence, akin to the independence on the
renormalization scale $\mu$, leads to a evolution equation for the
soft function.

\begin{figure}[t]
\centerline{\epsfxsize 5. truein \epsfbox{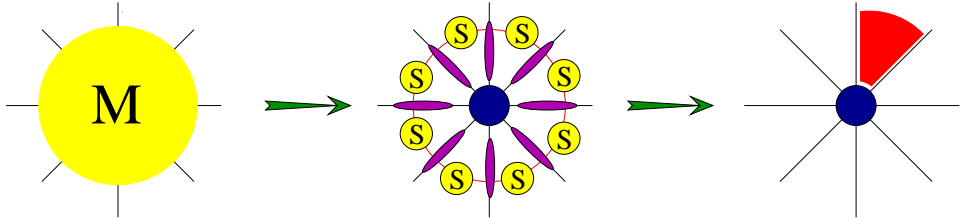}}
\caption[a]{\small Soft/Collinear factorization and its planar
limit.} \label{fig:sudakov}
\end{figure}

In the planar limit the soft/collinear factorization formulae
simplify significantly. Since in this limit there is a single
color structure, all color-space vectors reduce to a single
component.  The fact that the soft function is defined only up to
an overall function implies that, in the planar limit, it can be
completely absorbed in the jet functions $J_i$. The planar limit
implies that all interactions included in the thus redefined jet
functions are confined to adjacent gluons. In this limit it is
then instructive to consider a two-gluon process -- simply the
decay of a color-singlet state into two gluons. Direct application
of the factorization equation identifies then the square of the
jet function with the amplitude of this process which is, by
definition, the Sudakov form factor ${\cal M}^{gg\rightarrow
1}(\lambda(s_{i,i+1}/\mu),s_{i,i+1},\epsilon)$ if the two gluons
have momenta $k_i$ and $k_{i+1}$. It therefore follows that, in
the planar limit, a generic $n$-point scattering amplitude
factorizes as \be {\cal M}_n=\left[\prod_{i=1}^n{\cal
M}^{gg\rightarrow 1}
(\frac{s_{i,i+1}}{\mu},\lambda(\mu),\epsilon)\right]^{1/2}h_n~~,
\label{factorization_sudakov} \ee where $\lambda(\mu)=g(\mu)^2N$
is the 't~Hooft coupling. As before, here ${\cal M}_n$ denotes a
generic resummed amplitude, rescaled by the corresponding
tree-level amplitude.

Similarly to the soft and jet functions, the factorization
(\ref{factorization_sudakov}) implies an evolution equation and a
renormalization group equation for the factors ${\cal
M}^{[gg\rightarrow 1]}{\left(
\frac{Q^2}{\mu^2},\lambda(\mu),\epsilon\right)}$. The same
equations follow independently from the gauge invariance and the
properties of the form factor. They read
\be
\frac{d}{d\ln Q^2}{\cal M}^{[gg\rightarrow 1]}{\left(
\frac{Q^2}{\mu^2},\lambda,\epsilon\right)}=\frac{1}{2}\left[K(\epsilon,
\lambda)+G\left(\frac{Q^2}{\mu^2},\lambda,\epsilon\right)\right]
{\cal M}^{[gg\rightarrow 1]}{\left(
\frac{Q^2}{\mu^2},\lambda,\epsilon\right)}~,~~
\label{sudakov_Neq4}
\ee
where the function $K$ contains only
poles and no scale dependence.  The functions $K$ and $G$
themselves obey renormalization group equations
\cite{IR_paper2,IR_paper3,IR_paper4,{Ivanov:1985np},{Korchemsky:1985xj}}
\be
\left(\frac{d}{d\ln\mu}+\beta(\lambda)\frac{d}{dg}\right)(K+G)=0
~~~~~~~
\left(\frac{d}{d\ln\mu}+\beta(\lambda)\frac{d}{dg}\right)K(\epsilon,\lambda)
=-\gamma_K(\lambda)~~.
\ee
In ${\cal N}=4$ SYM they may be solved exactly and explicitly in terms
of the expansion coefficients of the cusp anomalous dimension
\be
f(\lambda)\equiv
\gamma_K(\lambda)=\sum_{l} a^l \gamma_K^{(l)}
\label{DefCusp}
\ee
and another set of coefficients defining the expansion of $G$:
\be
G\left(\frac{Q^2}{\mu^2},\lambda,\epsilon\right)=\sum_l {\cal G}_0^{(l)} a^l
\left(\frac{Q^2}{\mu^2}\right)^{l\epsilon}
\label{Gfunction}
\ee
where $a=\frac{\lambda}{8\pi^2}(4\pi e^{-\gamma})^\epsilon$ the
coupling constant customarily used in higher loop calculations.
%
%
An important ingredient in solving these equations is that in the
dimensionally-regularized ${\cal N}=4$ SYM theory the beta function is
\be
\beta(\lambda)=-2\epsilon \lambda~~,
\label{Neq4beta}
\ee
i.e. in the presence of the dimensional regulator the theory is
infrared-free. The solution for $K$ and $G$ may then be used to
reconstruct the Sudakov form factor (\ref{sudakov_Neq4}) which, in
turn, leads to the following expression for the factorized amplitude
\cite{Bern:2005iz}:
\be
{\cal M}_n&=& \exp\left[-\frac{1}{8}\sum_{l=1}^\infty
a^l \left(\frac{\gamma_K^{(l)}}{(l\epsilon)^2}
     +\frac{2{\cal G}_0^{(l)}}{l\epsilon}\right)\sum_{i=1}^n
\left(\frac{\mu^2}{-s_{i,i+1}}\right)^{l\epsilon}\right]\;\times\;h_n~\cr
             &=&\exp\left[
\sum_{l=1}^\infty a^l \left(\frac{1}{4}\gamma_K^{(l)}+\frac{l}{2}{\cal
G}_0^{(l)} \right){\hat I}_n^{(1)}(l\epsilon)\right]\;\times\;h_n~~.
\label{soft_collinear}
\ee
The definition of  ${\hat I}_n^{(1)}(\epsilon)$ may be easily seen
to be
\be
{\hat I}_n^{(1)}=-\frac{1}{\epsilon^2}\sum_{i=1}^n
\left(\frac{\mu^2}{-s_{i,i+1}}\right)^{\epsilon}~~;
\ee
This function captures the divergences of the planar one-loop
$n$-point amplitudes in ${\cal N}=4$ SYM.

The first few coefficients in the weak
coupling expansion of the cusp anomalous dimension and $G$
function (\ref{Gfunction}) have been evaluated directly
\cite{CuspWeak, Belitsky:2003ys,Bern:2005iz,Kotikov_1,Kotikov_2,
BCDKS,CSVcusp,CSVcollinear} with the result
\begin{eqnarray}
\label{weak_coupling_cusp}
f(\lambda) &=& \frac{\lambda}{2\pi^2}\left( 1-\frac{\lambda}{48}
+\frac{11\,\lambda^2}{11520}
-\left(\frac{73}{1290240}+\frac{\zeta_3^2}{512 \pi^6}\right)\lambda^3+ \cdots
\right) \,,
\\
G(\lambda) &=& -\zeta_3 \biggl({\lambda\over8\pi^2}\biggr)^2
+  \bigl( 6 \zeta_5 + 5 \zeta_2\zeta_3 \bigr)
\biggl({\lambda\over8\pi^2}\biggr)^3
- 2(77.56 \pm 0.02) \biggl({\lambda\over8\pi^2}\biggr)^4 + \cdots \,,
\end{eqnarray}
Using the integrability of the gauge theory dilatation
operator \cite{BES} constructed an integral equation whose solution is
the universal scaling function (conjecturally equal to the cusp
anomalous dimension) to all orders in perturbation theory. This
equation was solved in a weak coupling expansion \cite{BES} and also
in a strong coupling expansion \cite{BKK1,BKK2,BKK3}. Using the AdS/CFT
correspondence the first few coefficients in the strong coupling
expansion were evaluated in \cite{CuspStrongCoupling1,
CuspStrongCoupling2,CuspStrongCoupling3}. The leading
term in the strong coupling expansion of $G$ was computed in
\cite{AM1}:
 \begin{eqnarray}
f(\lambda)&=&\frac{\sqrt{\lambda}}{\pi}\left(1-\frac{3\ln
2}{\sqrt{\lambda}}-\frac{\rm K}{\lambda}+\cdots\right) \,,
\hskip1.0cm \lambda\to\infty\,,\\
G(\lambda)&=& (1 - \ln 2) { \sqrt{\lambda} \over 8\pi } + \cdots \,,
\hskip2.7cm \lambda\to\infty\,;
\end{eqnarray}
here ${\rm K}=\sum_{n\ge 0}\frac{(-1)^n}{(2n+1)^2}\simeq0.9159656\ldots$
is the Catalan constant.

The properties of  the collinear anomalous dimension $G$ were discussed in detail in
\cite{Dixon_Magnea_Sterman} where this function was identified with
the sum of the first subleading term in the large spin expansion of
the anomalous dimension of twist-2 operators and the coefficient of
the subleading pole in the expectation value of the cusp Wilson line
with edges of finite length.
%
%

\subsection{Loop amplitudes; generalized unitarity-based method}

Having discussed general properties of scattering amplitudes, we
now proceed to describe methods for their construction at loop
level. The goal will be to use only on-shell information for this
purpose.  we will be assuming (quite accurately) that tree-level
amplitudes are known. As we will see, the fact that Feynman
diagramatics underlies the calculation of scattering amplitudes is
a very important and useful guide. The properties of color ordered
amplitudes discussed previously will serve as a useful guide for
the completeness of the result. While most arguments apply to any
(supersymmetric) gauge theory, we will be having in mind
applications to ${\cal N}=4$ SYM.

The idea that one can use only on-shell information to construct
loop-level scattering amplitudes is of course very appealing. For
starters, one would use complete lower-loop amplitudes as building
blocks of higher amplitudes and, as such, one would build in the
calculations simplifications due to symmetries and gauge invariance.

\begin{figure}[t]
\centerline{\epsfxsize 4.0 truein
\epsfbox{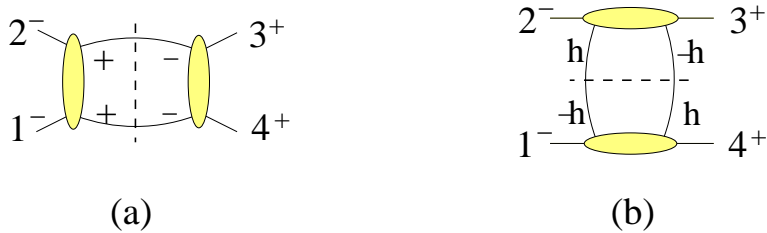}} \caption[a]{\small Singlet
and nonsinglet cuts of a one-loop four-gluon amplitude.}
\label{fig:singlet_and_nonsinglet}
\end{figure}

There is a long history associated with on-shell methods going back to
the time of the analytic S-matrix theory. The idea is that, given the
discontinuity of the amplitude in some channel -- or a cut -- one
could use a dispersion integral to reconstruct the complete
amplitude. In turn, the discontinuity of amplitudes is determined by
the unitarity condition of the scattering matrix. Indeed, separating
the interaction part of the scattering matrix
\be
S=1+iT
\ee
and requiring that $S$ is unitary $S^\dagger S=1$ implies that
\be
i(T^\dagger-T)=2\,\Im T=T^\dagger T~~.
\label{unitarity_V0}
\ee
The right hand side is the product of lower loop on-shell amplitudes;
this may be interpreted as a higher loop amplitude with some number of
Feynman propagators replaced by on-shell (or ``cut'') propagators
\be
\frac{1}{l^2+i\epsilon}\mapsto  -2\pi i \theta(l^0)\delta(l^2)~~.
\label{cut_prop}
\ee
The difference on the left hand side of equation (\ref{unitarity_V0})
is interpreted as the discontinuity in the multi-particle invariant
obtained by squaring the sum of the momenta of the cut propagators.
This interpretation is a consequence of the $i\epsilon$ prescription.
Thus, this discontinuity at $L$-loops is determined in terms of
products of lower-loop amplitudes.  There are two types of cuts:
singlet and non-singlet. In the former only one type of field
crosses the cut. In the latter several types of particles (such as
a complete multiplet in a supersymmetric theory) cross the cut.
For the one-loop four-gluon amplitude this is illustrated in
figure \ref{fig:singlet_and_nonsinglet}; in figure
\ref{fig:singlet_and_nonsinglet}(a) the tree-level amplitudes
require that only gluons can propagate along the cut propagators
while in figure \ref{fig:singlet_and_nonsinglet}(b) fields with
any helicity $h$ can cross the cut, i.e. $h=\pm 1,\pm1/2, 0$.

Having determined it, the missing (real) part of the amplitude is
constructed from a dispersion integral:
\be
\Re f(s)=\frac{1}{\pi}
P\int_{-\infty}^{\infty}dw \frac{\Im f(w)}{w-s}- C_{\infty}~~,
\ee
where $s$ is the momentum invariant flowing across the cut.  The term
coming from the contour at infinity vanishes if $f(w)\rightarrow 0$ as
$w\rightarrow\infty$. If it does not, there are subtraction
ambiguities related to terms which have no discontinuities. The first
string scattering amplitudes at one-loop were evaluated through such a
method \cite{KSV, BHS}.
%

While perfectly valid, such an approach does not make use of recent
sophisticated techniques for evaluating Feynman integrals: identities,
modern reduction techniques, differential equations, reduction to
master integrals, etc. A reinterpretation of the equation
(\ref{unitarity_V0}) leads however in this direction. Indeed, besides
representing the discontinuity of the amplitude, the right-hand-side
of that equation also represents the part of the amplitude which
contains the cut propagators. In fact, the right hand side of that
equation contains a combination of parts of the amplitude containing
two, three up to $(L+1)$ propagators.

It is not hard to see that separately each of these pieces are given
by products of on-shell lower-loop amplitudes. This conclusion may be
reached by thinking of the complete amplitude from a Feynman diagram
perspective.
Consider looking at the part of the amplitude which contains some
prescribed set of propagators such that if they are cut the amplitude
falls apart in at least two disconnected pieces.  Since the full
amplitude is a sum of Feynman diagrams, each of the resulting pieces
it itself a sum over all Feynman diagrams having as external legs
(some) of the original external legs as well as (some of) the cut lines.
%
%
Thus each of the resulting parts is itself an on-shell amplitude, with
the on-shell condition being a consequence of the cut conditions
(\ref{cut_prop}).

This observation, originally due to Bern, Dixon, Dunbar and Kosower
\cite{BDDK_npt} and improved at one-loop level in \cite{BCF_unitarity},
allows to ``cut'' more than $(L+1)$ propagators for an $L$-loop
amplitude, generalizing the unitarity relation
(\ref{unitarity_V0}). Similarly to regular cuts, generalized cuts can
be either of singlet and nonsinglet types.  These properties open the
possibility of going beyond reconstructing the amplitudes from
dispersion integrals: instead, one identifies the pieces of an
amplitude with some prescribed set of propagators. Analyzing
sufficiently many combinations of propagators one is guaranteed to be
able to reconstruct the complete amplitude.
Indeed, the fact that Feynman rules express scattering amplitudes as a
sum of terms containing propagators and vertices implies that, after
integral reduction, each term in the result contains part of the
propagators present in the initial Feynman diagrams. By analyzing all
possible generalized cuts one probes all possible combinations of
propagators and thus all possible terms originating from the Feynman
diagrams underlying the scattering process.

The argument above assumes that the (generalized) cuts are constructed
in the regularized theory -- i.e. in $d$-dimensions (perhaps with
$d=4-2\epsilon$).
 In practice however it is much simpler to start by
analyzing four-dimensional cuts, as one can saturate them with
four-dimensional helicity states and also make use of the simplifying
consequences of the supersymmetric Ward identities, such
as (\ref{vanishing_amplitudes}).
Four-dimensional cuts however potentially miss terms arising from the
$(-2\epsilon)$-dimensional components of the momenta in the
momentum-dependent vertices. Such terms must be separately accounted
for (either by considering $d$-dimensional cuts or by other means). In
supersymmetric theories one can argue
\cite{BDDK_cut_constructibility},
based on the improved power-counting of the theory, that at one-loop
level such terms do not exist through ${\cal O}(\epsilon^0)$ (in the
sense that through ${\cal O}(\epsilon^0)$ one-loop amplitudes follow
from four-dimensional cut calculations).

Let us illustrate this discussion with a simple example -- that of the
four gluon scattering amplitude in ${\cal N}=4$ SYM.  We will organize
the calculation in terms of regular, two-particle cuts reinterpreted in
the spirit of generalized unitarity-based method. There are two cuts
-- in the $s$ and in the $t$-channels. Depending upon the external
helicity configuration either one or both cuts are of non-singlet
type, with the complete ${\cal N}=4$ supermultiplet crossing it. As
discussed previously, the helicity information in any MHV amplitude
(such as this one) is carried by an overall factor of the tree-level
amplitude (\ref{LloopMHV}).  The remaining function may be thus
computed by choosing the most convenient helicity
configuration. Choosing $(1^-2^-3^+4^+)$ and evaluating the
four-dimensional $s$-channel cut (figure
\ref{fig:singlet_and_nonsinglet}(a)) one finds without difficulty that
\be
A{\scriptstyle (l_2, 1^-, 2^-, l_1)}A{\scriptstyle (-l_1, 3^+, 4^+, -l_2)}=
i s_{12}s_{23} A{\scriptstyle (1^-2^-3^+4^+)} \frac{1}{(l_2+k_1)^2(l_2-k_4)^2}~~.
\label{1loop_4pt}
\ee
Here $s_{i\dots j}=(k_i+k_{i+1}+\dots +k_j)^2$ and we have used the fact that the
cut condition allows one to write $2k_1\cdot l_2=(k_1+l_2)^2$. In the
propagator-like structures one recognizes the cut of a scalar box
integral in $\phi^3$ theory (that is, the integrand of a box integral
in $\phi^3$ theory in which two propagators have been removed and the
on-shell condition for the corresponding momenta is imposed). At this
stage one can argue based on the ultraviolet behavior of ${\cal N}=4$
SYM that the full answer is given by the box integral whose
$s$-channel cut we have just computed. Indeed, any other scalar
integral diverges in a smaller number of dimensions than ${\cal N}=4$
SYM and thus cannot appear in the final result. The conclusion of this
argument can be confirmed by the evaluation of the (nonsinglet)
$t$-channel cut (figure \ref{fig:singlet_and_nonsinglet}(b)). The
simplest way to see this is to make use again of the equation
(\ref{LloopMHV}) and note that up to the tree-level factor, the
$t$-channel cut in the configurations $(1^-2^-3^+4^+)$ and
$(1^+2^-3^-4^+)$ are the same. The latter is again a singlet cut,
being given by a relabeling of equation (\ref{1loop_4pt}). To
summarize, we find \cite{BDDK_npt} that
\be
{\cal M}_4^{(1)}=\frac{i}{\pi^{d/2}}s_{12}s_{23}\int d^dl~\frac{1}
{l^2(l-k_1)^2(l-k_{12})^2(k+k_4)^2}\equiv -\frac{1}{2}s t I_4(s,t)~~,
\label{1loop4pt}
\ee
thus reproducing the well-known result of \cite{GreenSchwarz}.

\begin{figure}[t]
\centerline{\epsfxsize 4.3 truein
\epsfbox{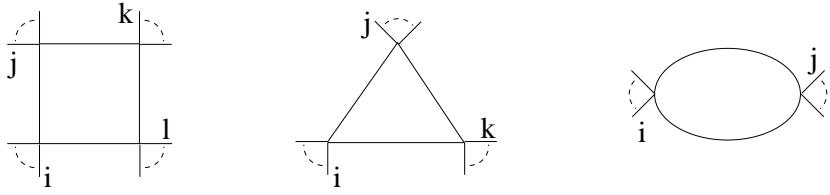}} \caption[a]{\small Box,
triangle and bubble scalar integrals. The clusters at each corner
 are constructed from color-adjacent external legs. If more than one
external leg is present at a corner, then that corner
is ``massive'' as the total momentum is no longer light-like.}
\label{fig:one_loop_integrals}
\end{figure}

The fact that a scalar box integral appears in the result of this
calculation is not surprising. On general grounds one can show
that in any four-dimensional massless theory, any one-loop
scattering amplitude may be expressed as a linear combination of
scalar box, triangle and bubble integrals (i.e. integrals with
four, three and two propagators, respectively, and no
loop-momentum factors appearing in the numerator) with rational
coefficients (see figure \ref{fig:one_loop_integrals}) and a
rational function which has no cuts in any channel. It was shown
in \cite{BDDK_npt} that in a supersymmetric theory this rational
contributions are absent and that in such theories one-loop
amplitudes are constructible using four-dimensional cuts.

\begin{figure}[t]
\centerline{\epsfxsize 4.0 truein \epsfbox{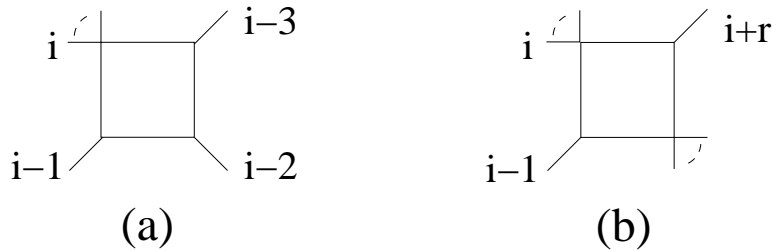}}
\caption[a]{\small The one-mass (a) and easy two-mass (b)
integrals.} \label{fig:1m2me}
\end{figure}

For one-loop amplitudes in ${\cal N}=4$ SYM one can do much better
than the above by noticing \cite{BDDK_npt} that
the one-loop amplitudes with external states belonging to the same
${\cal N}=1$ vector multiplet may be written as a sum of box
integrals.
%
%
Besides a massless box integral which occurs only for four-gluon
scattering, these integrals fall in five different classes: one-mass,
easy two-mass, hard two-mass, three- and four-mass box integrals,
depending on whether massive or massless momenta are injected at the
corner of the box. The first two classes are shown in figure
\ref{fig:1m2me}.  The box integrals are defined and given in ref.
\cite{bdk_integrals_1,bdk_integrals_2} (the four-mass boxes are from
ref. \cite{denner,davydychev_1,davydychev_2}).
Since each box integral has a unique set of four propagators
(cf. figure \ref{fig:one_loop_integrals}), a quadruple cut (i.e. the result of
eliminating four propagators and using the on-shell condition for
their momenta) isolates a unique box integral and its coefficient
\cite{BCF_unitarity}. The quadruple cut of the amplitude is, following
the previous discussion, simply given by the product of four tree
amplitudes evaluated on the solution of the on-shell conditions for
the four propagators. Thus:
\be
c_{ijkl}=\frac{1}{2}\sum_{h_{q_i}}
A{\scriptstyle (q_1,i\dots j-1,-q_2)}A{\scriptstyle (q_2,j\dots k-1,-q_3)}
A{\scriptstyle (q_3,k\dots l-1,-q_4)}A{\scriptstyle (q_4,l\dots
i-1,-q_1)}
\Big|_{q_1^2=q_2^2=q_3^2=q_4^2=0}
\label{qc_coefs}
\ee
where the labels $i,j,k,l$ are cyclic indices and label the first
external leg at each corner of the box, counting clockwise.
The sum runs over all possible helicity assignments on the internal lines.
The factor of $1/2$ above is due to the four on-shell conditions
having two solutions with equal values of the quadruple-cut box
integrals are equal. The sum over these solutions is implicit in the
sum in equation (\ref{qc_coefs}).
An implicit assumption is made in writing this expression. Any
amplitude contains at least one box integral with one three-point
corner. In Minkowski signature -- i.e. with real momenta -- the
corresponding tree-level three-point amplitude vanishes identically. A
nonvanishing result requires interpreting the loop momentum as complex.

We will later need the expression for the one-loop MHV amplitude.
As we discussed, the four-point amplitude is given by
($\ref{1loop4pt})$. For an arbitrary number of external legs
(larger than four), the result initially obtained in
\cite{BDDK_npt} (which can be reproduced using quadruple cuts and
complex momenta) reads:
\be {\cal
M}_{n=2m+1}^{(1)}&=&-\frac{1}{2}\sum_{r=2}^{m-1}\sum_{i=1}^n
(t_{i-1}^{[r+1]}t_i^{[r+1]}-t_{i}^{[r]}t_{i+r+1}^{[n-r-2]})\;I_{4;r;i}^{2me}
-\frac{1}{2} \sum_{i=1}^n t_{i-3}^{[2]}t_{i-2}^{[2]}\;I_{4;i}^{1m}
\cr {\cal
M}_{n=2m}^{(1)}&=&-\frac{1}{2}\sum_{r=2}^{m-2}\sum_{i=1}^n
(t_{i-1}^{[r+1]}t_i^{[r+1]}-t_{i}^{[r]}t_{i+r+1}^{[n-r-2]})\;I_{4;r;i}^{2me}
-\frac{1}{2} \sum_{i=1}^n t_{i-3}^{[2]}t_{i-2}^{[2]}\;I_{4;i}^{1m}
\cr &&-\frac{1}{2}\sum_{r=2}^{m-2}\sum_{i=1}^n
(t_{i-1}^{[m]}t_i^{[m]}-t_{i}^{[m-1]}t_{i+m}^{[n-m-1]})\;I_{4;m-1;i}^{2me}
\label{one_loop_full}
\ee
where $I_{4;i}^{1m}$
and $I_{4;r;i}^{2me}$ are the one-mass (figure \ref{fig:1m2me}(a))
and easy two-mass (figure \ref{fig:1m2me}(b)) integrals and
$t_i^{[r]}$ are multi-particle invariants
$t_i^{[r]}=(k_i+\dots+k_{i+r-1})^2$. \footnote{This is a more compact
notation for $s_{i\dots (i+r-1)}$.}.

\subsection{Calculations at higher loops}

Higher loop calculations in ${\cal N}=4$ SYM enjoy similar
simplifications, though to a lesser extent.  An analog of the 1-loop
integral basis is not available, in the sense that the members of all
proposed bases are in fact functionally dependent integrals
\footnote{Notable examples are the two-loop four-point integral basis
with massless external legs \cite{Smirnov_Veretin} and the two-loop
four-point integral basis with one massive external leg
\cite{Gehrmann_Remiddi}.}; moreover, not all integrals have
sufficiently many propagators such that the cut condition on all of
them does not completely freeze the integrals. It was pointed out
\cite{Buchbinder_Cachazo} that under certain circumstances, after all
propagators have been set on-shell, an additional propagator-like
structure appears which can be used to set an additional on-shell
condition. The lack of independence of the integral basis does not
allow however a straightforward identification of the resulting
product of tree amplitudes with the coefficient of the integral which
is isolated by these cuts.

Generalized cuts can nevertheless be used to great effect to
isolate parts of the full amplitude containing some prescribed set
of propagators. One needs to ensure that integrals are not
double-counted and that all cuts are consistent with each other.
The previous arguments continue to hold and imply that the
complete amplitude can be reconstructed from its $d$-dimensional
generalized cuts. A detailed, general algorithm for assembling the
amplitude was described in \cite{QCD_splitting_amplitude}. In a
nutshell, starting from one (generalized) cut, one corrects it
iteratively such that all the other cuts are correctly reproduced.
%

While fundamentally all cuts have equal importance, some of them
exhibit more structure, which makes them ideal starting points for
the reconstruction of the amplitude. Such are the iterated
two-particle cuts, defined as a sequence of two-particle cuts
which at each stage reduces the number of loops by one
unit.\footnote{It is fairly clear that {\it a priori} there exist
integrals which do not exhibit any two-particle cuts. Such
contributions to the amplitude are not captured in this way. An
example is provided by the four-loop four-gluon planar amplitude
\cite{BCDKS}.} Their importance stems from the fact that
two-particle cuts with MHV amplitudes on both sides are naturally
proportional to another MHV tree amplitude: \be &&A^{\rm
tree}(l_2^+ 1^+,\dots ,m_1^-, \dots ,m_j^-,\dots,c_2^+,l_1^+)
A^{\rm tree}(-l_1^-,(c_2+1)^+,\dots, n^+, -l_2^-)\cr
&&~~~~~~~~~~\propto A^{\rm tree}(1^+,\dots ,m_1^-, \dots
,m_j^-,\dots,n^+)~~. \ee The proportionality coefficient can be
partial-fractioned into a sum of terms recognizable as cuts of box
integrals with polynomial coefficients in external invariants.
Repeatedly sewing an MHV tree amplitude onto such a construct
yields another MHV tree amplitude as natural common factor.

\begin{figure}[t]
\centerline{\epsfxsize 4.5 truein \epsfbox{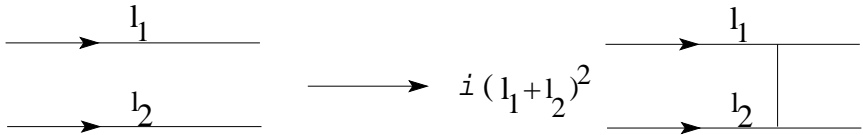}}
\caption[a]{\small The rung rule.}
\label{fig:rung_rule}
\end{figure}

For a four-particle amplitude the iteration of two-particle cuts can
be explicitly solved and yields the so-called rung rule
\cite{Bern:1997nh}. It
states that the $L$-loop integrals which follow from iterated
two-particle cuts can be obtained from the $(L-1)$-loop amplitudes by
adding a rung in all possible (planar) ways and in the process
multiplying the numerator by $i$ times the invariant constructed from
the momenta of the lines connected by the rung. This rule is
illustrated in figure
\ref{fig:rung_rule}.

For higher multiplicity amplitudes the rung rule is less effective
and it is necessary to explicitly evaluate the relevant iterated
cuts. The strategy discussed in this section can be used to
compute quite high loop amplitudes in ${\cal N}=4$ SYM. In the
next section some explicit results obtained in this way will be
discussed. It is important to keep in mind that, in contrast to
one-loop calculations, four-dimensional cut calculations are not
necessarily sufficient. Indeed, ${\cal O}(\epsilon)$ terms at
one-loop order may combine with singular terms from other loops to
yield pole terms and/or finite terms at higher loops. Besides the
obvious one-loop ${\cal O}(\epsilon)$ arising from integrals whose
integrand manifestly exhibit $d$-dimensional Lorentz-invariance,
such terms may also arise from integrals containing explicitly the
$(-2\epsilon)$ components of the loop momenta.  Usually called
``$\mu$-integrals'', at higher loops they contain the
$(-2\epsilon)$ components of any number of the loop
momenta.\footnote{Such appear already at one-loop level if one is
interested in expressions valid to all orders in $\epsilon$. An
example is provided by a parity-odd ${\cal O}(\epsilon)$ term in
the one-loop five-point amplitude \cite{Bern:1996ja}: \be {\cal
M}^{(1)\mu}_5\propto \int \frac{d^4pd^{-2\epsilon}\mu}{(2\pi)^d}
\frac{\epsilon(k_1,k_2,k_3,k_4)\;\;\mu^2}{(p^2-\mu^2)((p-k_1)^2-\mu^2)
((p-k_{12})^2-\mu^2)((p+k_{45})^2-\mu^2)((p+k_{5})^2-\mu^2)}~~.
\nonumber \ee Similar integrals occur in all higher-multiplicity
one-loop amplitudes. Two-loop analogs of such integrals will
appear in section \ref{6gamp}.  }
One may decide whether such terms, not constructible from
four-dimensional cuts, are present in the amplitude by comparing the
infrared divergences emerging from a four-dimensional cut calculation
with the expected structure implied by the soft and collinear
factorization theorem.


An apparently alternative method for determining the four-dimensional
cut-constructible part of scattering amplitudes was suggested in
\cite{cachazo_leading}. The basic idea is based on the observation
that an amplitude possess singularities for specific momentum
configurations, determined by their construction in terms of Feynman
diagrams. These singularities must be correctly reproduced by any
presentation of the amplitude in terms of simpler integrals. Moreover,
singularities exhibited by these simpler integrals but not present
in the sum of Feynman diagrams are spurious and should cancel out.
The identification of the leading singularities of amplitudes proceeds
by cutting the largest possible number of propagators and matching the
result against a judicious choice of a(n overcomplete) basis of
integral functions. At $L$-loops, integrals with $4L$ propagators are
completely localized. Integrals with fewer propagators are however
not. Additional propagator-like structures appear sometimes due to
Jacobians coming from solving the cut conditions which are
manifest. ``Cutting'' these additional ``propagators'' leads to a
complete localization of the integrals and expresses the result in
terms of product of tree-level amplitudes.
This proposal has been tested for all the amplitudes constructed by
independent means and appears to correctly reproduce the
four-dimensional cut-constructible part of the amplitude. The odd part
of the two-loop six-point amplitude was constructed only through this
method \cite{CSV_odd_six_pt}.

\subsection{Some explicit higher loop results at
low multiplicity \label{examples_HL}}

Using generalized unitarity, a number of higher loop amplitudes have
been explicitly computed and their properties analyzed. Due to the
increase in the number of kinematic invariants with the number of
external particles, the complexity of the analysis increases as the
number of external legs is increased.
Here we discuss some of the available results, in increasing order of
their complexity.  First we will discuss the four-point amplitudes at
two- and three-loops. As we have seen previously, splitting amplitudes
provide a link between the lower and higher-point amplitudes; we will
review them next and then proceed to the five-point amplitude.

\begin{figure}[t]
\centerline{\epsfxsize 4.3 truein \epsfbox{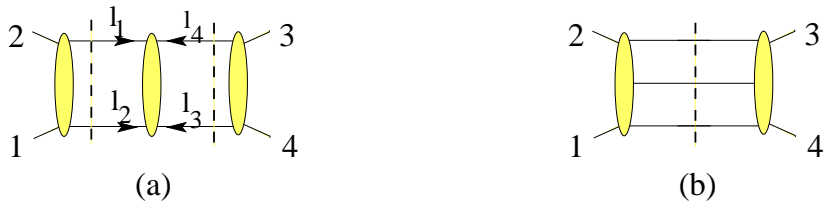}}
\caption[a]{\small The double two-particle and the three-particle cuts
determining the four-gluon scattering amplitude in ${\cal N}=4$ SYM at
two-loop level.}
\label{double2pc4pt}
\end{figure}

The integrand of the four-point scattering amplitude at two-loops
was found in \cite{Bern:1997nh} and evaluated in
\cite{Anastasiou:2003kj} using the results of
\cite{smirnov_2loops}. It can be evaluated by a considering a
double two-particle cut as in figure \ref{double2pc4pt}a.
As previously mentioned, they are correctly captured by the rung
rule. It is instructive to follow the details of the calculation in
this relatively simple case and in the process also have an explicit
example of the rung rule; they may in fact be constructed by
iteratively using the equation (\ref{1loop_4pt}). For the purpose of
the calculation one needs to pick some helicity assignment; we will
choose $(1^-2^-3^+4^+)$; thus, we need to evaluate
\be
A_4^{\rm tree}{\scriptstyle (l_2, k_1^-, k_2^-, l_1)}
A_4^{\rm tree}{\scriptstyle (-l_1, -l_4, -l_3, -l_2)}
A_4^{\rm tree}{\scriptstyle (l_4, k_3^+, k_4^+, l_3)}~~,
\label{eg_4pt_2pc}
\ee
where the helicities of the cut lines are fixed by the requirement
that the tree-level amplitudes are nonvanishing. The product of the
first two tree-level amplitudes may be easily reorganized following
equation (\ref{1loop_4pt}) to be
\be
A_4^{\rm tree}{\scriptstyle (l_2, k_1^-, k_2^-, l_1)}
A_4^{\rm tree}{\scriptstyle (-l_1, -l_4, -l_3, -l_2)}
=i s_{12}(k_2-l_4)^2 A{\scriptstyle (-l_3,1^-,2^-,-l_4)}
\frac{1}{(l_2-k_1)^2(l_2+l_3)^2}~~.
\ee
Further application of equation (\ref{1loop_4pt}) leads to a final
expression for the product in equation (\ref{eg_4pt_2pc}):
\be
&&A_4^{\rm tree}{\scriptstyle (l_2, k_1^-, k_2^-, l_1)}
  A_4^{\rm tree}{\scriptstyle (-l_1, -l_4, -l_3, -l_2)}
  A_4^{\rm tree}{\scriptstyle (l_4, k_3^+, k_4^+, l_3)}\cr
&&~~~~
=i s_{12}(k_2-l_4)^2 \frac{1}{(l_2-k_1)^2(l_2+l_3)^2}A_4^{\rm tree}{\scriptstyle (-l_3,1^-,2^-,-l_4)}
                                                     A_4^{\rm tree}{\scriptstyle (l_4, k_3^+, k_4^+, l_3)}\cr
&&~~~~
=A^{\rm tree}_4{\scriptstyle (k_1^-,k_2^-,k_3^+,k_4^+)}
   \left[is_{12}(k_2-l_4)^2 \frac{1}{(l_2-k_1)^2(l_2+l_3)^2}\right]
   \left[is_{12}s_{23}\frac{1}{(l_3-k_1)^2(l_3+k_4)^2}\right]\cr
&&~~~~
=-s_{12}^2s_{23}A^{\rm tree}_4{\scriptstyle (k_1^-,k_2^-,k_3^+,k_4^+)}\pic{15}{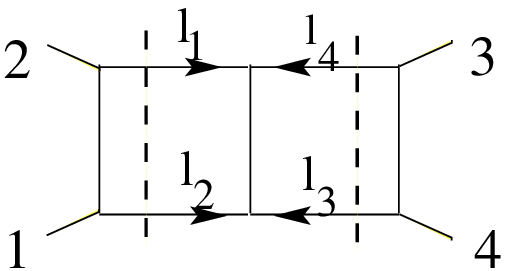}{.5}~~,
\ee
where we have used again the cut conditions to organize the result in
terms of propagators.  One notes without difficulty that momentum
conservation implies the cancellation of the numerator factor
$(k_2-l_4)^2$ against the denominator factor $(l_3-k_1)^2$ in the last
equality. This cancellation is crucial for having a Feynman integral
interpretation for the generalized cut in equation
(\ref{eg_4pt_2pc}). The conclusion of this calculation is that the
two-loop four-gluon amplitude contains the double-box integral whose
cut appears in the equation above.  This calculation also illustrates
the application of the rung rule (cf. fig.\ref{fig:rung_rule}).

The other double two-particle cuts are obtained by simple relabeling of the
previous calculation. Thus, one finds that they imply that the two-loop
four-gluon amplitude in ${\cal N}=4$ SYM is (for any choice of helicity
assignment) given by \cite{Bern:1997nh}
\be
{\cal M}^{(2)}_4{\scriptstyle (k_1,k_2,k_3,k_4)}
=-\frac{1}{4}
s_{12}s_{23}\left\{~s_{12} \pic{16}{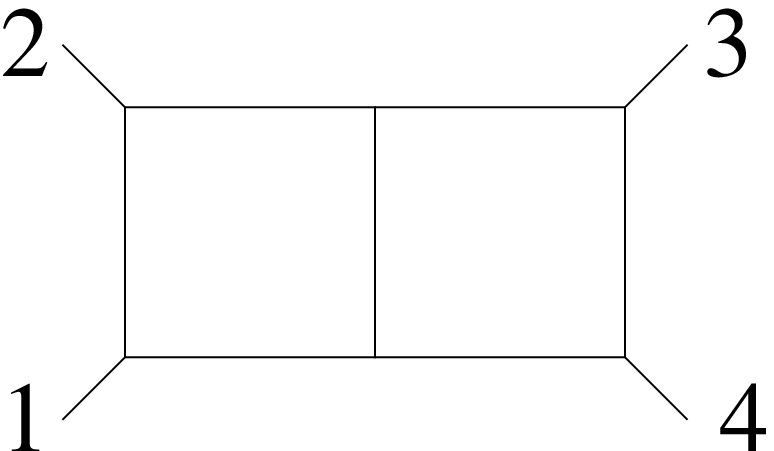}{.25}~+\;s_{23}
\pic{28}{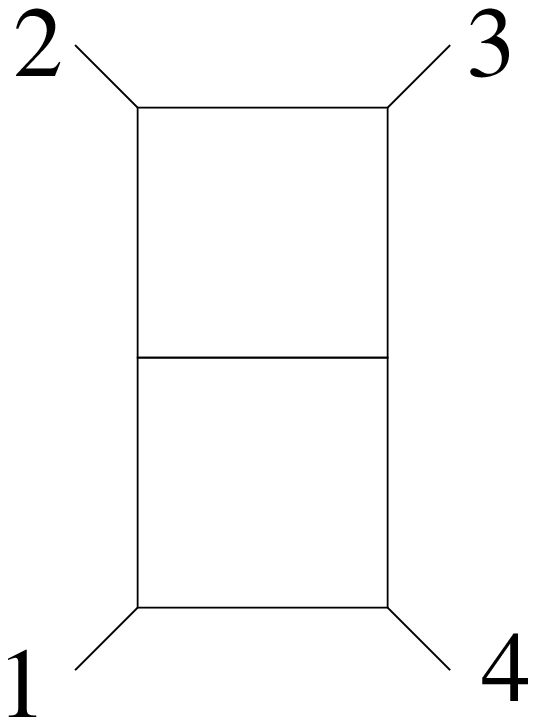}{.25}~\right\}~~.
\label{2loop_full}
\ee
The ultraviolet behavior of ${\cal N}=4$ SYM suggests\footnote{
Superspace arguments \cite{Howe_Stelle_YM} imply that at two-loops,
${\cal N}=4$ SYM is logarithmically-divergent in seven
dimensions. This is however only suggestive of (\ref{2loop_full})
being the full answer, as there may exist more divergent contributions
whose leading ultraviolet behavior cancels out.}  that this is indeed
the complete amplitude, a fact confirmed by the evaluation of the
three-particle cut.

Similar (though somewhat lengthier) manipulations or repeated
application of the rung rule leads to the three-loop four-gluon amplitude
\cite{Bern:1997nh,Bern:2005iz}. The notable fact is that, unlike the
two-loop amplitude, the three-loop integrand retains some dependence
of the loop momentum in its numerator.

\begin{figure}[t]
\centerline{\epsfxsize 5. truein \epsfbox{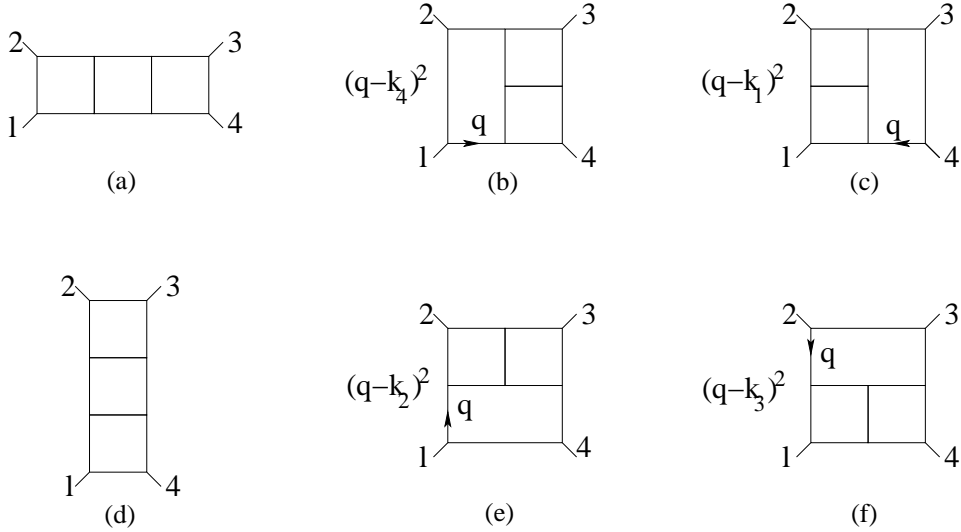}}
\caption[a]{\small Integrals featuring in the three-loop 4-gluon amplitude.}
\label{fig:threeloop}
\end{figure}

\be
{\cal M}^{(3)}_4{\scriptstyle (k_1,k_2,k_3,k_4)}&=&-\frac{1}{8}s_{12}s_{23}
\left[~s_{12}^2 I_4^{(3)a}(s_{12}, s_{23})
     +2s_{12}   I_4^{(3)b}(s_{12}, s_{23})\right.\cr
&&~~~~~~~~~~~~\left.+ s_{23}^2 I_4^{(3)d}(s_{12}, s_{23})
                    +2s_{23}I_4^{(3)e}(s_{12}, s_{23})~\right]
\label{3loops4g}
\ee
where the integrals $I_4^{(3)a,b,d,e}$ are shown in figure
\ref{fig:threeloop}. The second and third integrals on each row of
that figure are equal and also $I_4^{(3)d}(s_{12},
s_{23})=I_4^{(3)a}(s_{23}, s_{12})$.

A link between lower and higher point amplitudes at any number of
loops is provided by the splitting amplitudes introduced in equation
(\ref{collinear_limit}). A unitarity-based all-order proof of those
equations as well as a means of directly evaluating the splitting
amplitudes was discussed in \cite{Kosower_splitting} for arbitrary
gauge theories. Similar to scattering amplitudes, they are determined
by tree-level information up to the appropriate treatment of the
intermediate momentum $p$ in equation (\ref{collinear_limit})
 which must be kept massive throughout the
calculation.  The one-loop splitting amplitudes can be obtained without
difficulty either by considering collinear limits of higher loop
amplitudes \cite{Bern_Chalmers} or by direct evaluation
\cite{Kosower_Uwer_1}.  The two-loop splitting amplitude in ${\cal N}=4$
SYM theory have been computed and their properties have been analyzed in
\cite{QCD_splitting_amplitude, Anastasiou:2003kj}.

\subsubsection{A possible integral basis at higher loops;
               Conformal integrals \label{conf_ints}}


${\cal N}=4$ SYM is a conformal field theory at the quantum level;
conformal invariance may be observed in correlation functions of
operators of definite (anomalous) dimension. In the context of the
AdS/CFT correspondence this symmetry is related to the existence of an
exact $SO(2,4)$ isometry of the anti-de-Sitter space. At the level of
on-shell scattering amplitudes however (super)conformal invariance is
obscured beyond tree level; after removing the effects of the
(infrared) regulator which explicitly breaks it, the momentum space
realization of the generators of the conformal group still exhibits
anomalies analogous to the holomorphic anomaly of collinear operators
\cite{CSWanomaly}.

It was observed in \cite{Drummond:2006rz} by explicitly inspecting the
known results for the one-, two- and three-loop four-gluon amplitudes
that the integrals appearing in the rescaled amplitude ${\cal M}_4$
exhibit, if regularized by keeping the external legs off-shell, (in a
sense we will describe below) an $SO(2,4)$ symmetry apparently
unrelated to the four-dimensional conformal group. To expose this
symmetry one solves the momentum conservation constraint at each
vertex by writing each momentum as a difference of two variables
\be
p_i=x_i-x_{i+1}~~.
\ee
We use the notation $p_i$ here to denote generically both external momenta
as well as loop momenta.  These variables define the position $x_i$ of
the vertices of the dual graph.
\begin{figure}[t]
\centerline{\epsfxsize 6. truein \epsfbox{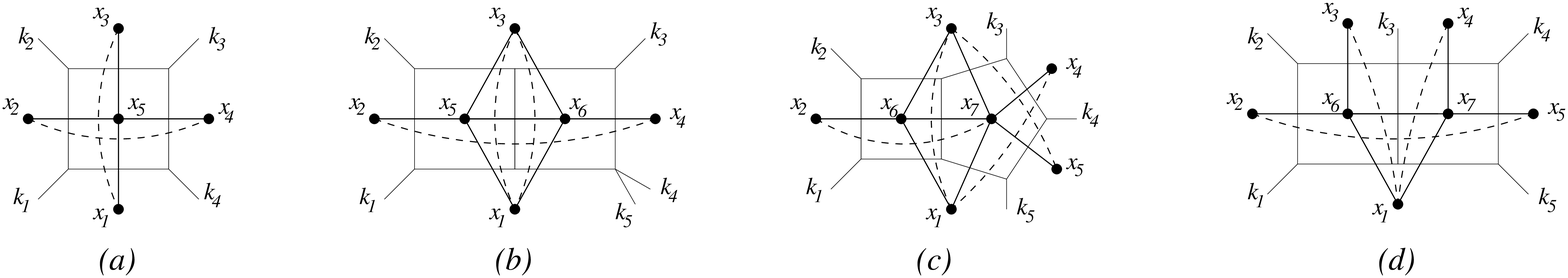}}
\caption[a]{\small Examples of pseudoconformal integrals. Points $x_i$ label
the dual graph, a solid line connecting two points $x_i$ and $x_j$
corresponds to a factor of $1/x_{ij}^2$ while a dashed line
corresponds to a factor of $x_{ij}^2$. The integral is pseudoconformal
if the difference between the number of solid lines and dashed lines
at a vertex equals $4$ if the vertex is inside the loops of the
original graph and zero for all other points $x_i$. The graphs $(b)$,
$(c)$ and $(d)$ show that the integrals appearing in the even part of
the five-point two-loop amplitude \cite{CSV_5pt, BCKRS} are pseudoconformal.}
\label{ConformalIntegralsFigure}
\end{figure}
This way the momentum conservation constraint is replaced by an
invariance under uniform shifts of the dual coordinates
$x_i$. Moreover, their Lorentz transformation properties are identical
to those of the momenta.  Since the dual variables are unconstrained
one may also define an inversion operator
\begin{eqnarray}
I=\sum_i I_i~~~~~~~I:x_i^\mu\mapsto \frac{x_i^\mu}{x_i^2}~~.
\label{inversion}
\end{eqnarray}

An off-shell regularization of infrared divergences allows the
construction of planar loop integrals which are invariant under such a
transformation. Indeed, properties of dual graphs imply that in any
planar integral all inverse propagators can be written as the square
of a difference of two $x_i$-s. Thus, propagators transform
homogeneously (with weight $(+1)$ in each of the two $x_i$-s) under
the transformation (\ref{inversion}). The weight of each $x_i$ in the
transformation of all propagators defining the integral equals twice
the number of propagators containing this variable.
The four-dimensional loop integration measure transforms homogeneously
(with weight $(-4)$). It therefore follows that a numerator factor
transforming homogeneously with the appropriate weight would render
the integral invariant under simultaneous inversion of all dual
variables $x_i$.
Simple graphical rules capturing the transformation under inversion of
an integral are illustrated in figure
\ref{ConformalIntegralsFigure}. Let us illustrate the details by
discussing a simple example -- the one-loop box integral shown in
figure \ref{ConformalIntegralsFigure}a. Up to numerator factors, the
relevant integral is
\be
I_a=\int d^4 x_5 \frac{1}{x_{51}^2x_{52}^2 x_{53}^2 x_{54}^2}~~;
\ee
each of the propagators present is denoted by a solid line in figure
\ref{ConformalIntegralsFigure}a. As mentioned, under inversion this
integral transforms as
\be
I:I_a\mapsto \int \frac{d^4 x_5}{(x_5^2)^4}
\frac{(x_5^2x_1^2)(x_5^2x_2^2)(x_5^2x_3^2)(x_5^2x_4^2)}
{x_{51}^2x_{52}^2 x_{53}^2 x_{54}^2}=x_1^2x_2^2x_3^2x_4^2\,I_a~~,
\ee
i.e. it transforms homogeneously with weight $(+2)$ for each of the
coordinates $x_i$ unrelated to the loop momentum. If the external
momenta are massless -- $k_i^2=0$ -- then the only way to compensate
for this transformation is by adding a factor of
$s_{12}s_{23}=x_{13}^2x_{24}^2$ since
\be
I:x_{13}^2x_{24}^2\mapsto \frac{x_{13}^2x_{24}^2}{x_1^2x_2^2x_3^2x_4^2}
\ee
and thus $s_{12}s_{23}I_a$ is invariant. If two opposite external legs
are massive -- say $k_1^2\ne 0$ and $k_3^2\ne 0$ -- a further
numerator factor is possible since $k_1^2 k_3^2=x_{12}^2 x_{34}^2$ no
longer vanishes and transforms as
\be
I:x_{12}^2 x_{34}^2\mapsto \frac{x_{12}^2 x_{34}^2}{x_1^2x_2^2x_3^2x_4^2}~~.
\ee
Further possibilities occur if more of the external legs are
massive. A similar discussion may be carried out at higher loops
\cite{Drummond:2006rz}.

One can also define a dilatation generator, under which all integrals
transform homogeneously and carry the same weight as under rescaling
of momentum variables. Together with translations of the dual
variables and their inversion this generate an $SO(2,4)$ algebra
called {\it dual conformal symmetry}.

It turns out that all integrals which appear in the four-gluon
amplitude through three-loops are invariant under dual conformal
transformations if they are regularized with an off-shell
regulator. The amplitudes are however constructed assuming dimensional
regularization; due to the change in the dimension of the integration
measure this regularization breaks the inversion invariance.
Dimensionally-regularized integrals which, if regularized with an
off-shell regulator are invariant under dual conformal transformations
are called {\it pseudo-conformal integrals} \cite{DHKS4_2loop}.
It is interesting to note that by this definition $\mu$-integrals are
also pseudo-conformal. Indeed, with an off-shell regulator the
integrand is treated as four-dimensional and thus vanishes identically
for these integrals.

The appearance of pseudo-conformal integrals is not limited to
four-point amplitudes in ${\cal N}=4$ SYM; they also generate the
scalar factor of $n$-point one-loop MHV amplitudes, the even part of
the two-loop five-point amplitude (cf. Figure
\ref{ConformalIntegralsFigure}) and the even part of the two-loop
six-point amplitude \cite{BDKRSVV} which we shall review shortly.

It is not clear what is the underlying reason for the appearance of
dual conformal invariance at weak coupling. It is moreover not clear
whether its appearance persists to all loop orders (perhaps up to
integrals whose integrands vanish identically in four dimensions
\cite{BDKRSVV}). It is nevertheless a useful guide in organizing
higher loop calculations. If it indeed survives to all orders in
perturbation theory it provides a general (though nevertheless
overcomplete at higher loops) basis of integrals organizing parts of
higher loop amplitudes in ${\cal N}=4$ SYM.

\subsection{The BDS conjecture and potential departures form it
\label{BDS_and_departures}}

In section \ref{examples_HL} we discussed, following
\cite{Bern:1997nh, Bern:2005iz}, higher loop corrections to the
four-gluon amplitude. The direct evaluation of the integrals in
the two-loop four-gluon amplitude \cite{Anastasiou:2003kj} reveals
a surprising structure: up to terms of order $\epsilon$, \be {\cal
M}_4^{(2)}(\epsilon)&=& \frac{1}{2}\left({\cal
M}_4^{(1)}(\epsilon)\right)^2 +f^{(2)}(\epsilon){\cal
M}_4^{(1)}(2\epsilon)+C^{(2)}+{\cal O}(\epsilon)~~.
\label{2loop4ptEval} \ee Equally surprisingly, the same expression
holds for the two-loop splitting amplitude
\cite{Anastasiou:2003kj}.
Such an iterative behavior is to be expected for the infrared-singular part
of the amplitudes; indeed, it is only a consequence of the
soft/collinear factorization theorem discussed previously
(cf. eq. (\ref{soft_collinear})). The surprising fact is that this
structure extends to the finite part of the amplitude, in particular
that $C^{(2)}$ is a constant.

The fact that splitting amplitudes provide a link between higher
and lower-point amplitudes at fixed loop order suggests a
generalization of the iteration relation above to arbitrary number
of external legs.  Indeed, an ansatz which correctly captures the
behavior of the amplitude in collinear limits as well as its
infrared singularities is
\be
{\cal M}_n^{(2)}(\epsilon)&=&
\frac{1}{2}\left({\cal M}_n^{(1)}(\epsilon)\right)^2
+f^{(2)}(\epsilon){\cal M}_n^{(1)}(2\epsilon)+C^{(2)}+{\cal
O}(\epsilon)~~. \label{ABDK}
\ee
Similarly to the explicit
calculation of the four-point amplitude, the main feature of this
ansatz is that $C^{(2)}$ and $f^{(2)}(\epsilon)$ are independent
of the external momenta and also of the number of external legs.
The five-gluon amplitude at two-loops obeys this ansatz; the same
cannot be said however about the six-gluon amplitude, as we shall
discuss in section \ref{6gamp}.

A similarly surprising result followed \cite{Bern:2005iz} from the evaluation
of the three-gluon amplitude (\ref{3loops4g}); throughout the finite
part, it obeys an iterative structure similar to that if the two-loop
amplitude.
\be
{\cal M}_4^{(3)}(\epsilon)&=&
- \frac{1}{3}\left({\cal M}_4^{(1)}(\epsilon)\right)^3
+ {\cal M}_4^{(2)}(\epsilon){\cal M}_4^{(1)}(\epsilon)
+ f^{(1)}(\epsilon){\cal M}_4^{(1)}(3\epsilon)
+ C^{(3)} + {\cal O}(\epsilon)~~;
\nonumber
\ee
This equation as well as (\ref{2loop4ptEval}) are consistent with the
resummed amplitude taking an exponential form with the exponent given
in terms of the one-loop four-gluon amplitude. Assuming that the same
is true for the splitting amplitude, Bern, Dixon and Smirnov
\cite{Bern:2005iz} suggested that, to all loop orders, the rescaled $n$-point
MHV amplitude is given by
\be
{\cal M}_n=
\exp\left[\sum_{l=1}^\infty \, a^l f^{(l)}(\epsilon){\cal M}^{(1)}_n(l\epsilon)+C^{(l)}
+{\cal O}(\epsilon)\right]
\label{BDS_conjecture}
\ee
where the coefficients \be f^{(l)}(\epsilon)=f_0^{(l)}+\epsilon
f_1^{(l)}+\epsilon^2f_2^{(l)} \ee are independent of the number of
external legs. The $\epsilon$-independent part, $f_0^{(l)}$, are
the Taylor coefficients of the cusp anomalous dimension or
universal scaling function (\ref{DefCusp}) \be
f(\lambda)={4}\sum_{l=0}^\infty \,a^l f_0^{(l)}~~. \ee The
appearance of the cusp anomalous dimension is of course dictated
by the infrared structure of the amplitude.  Similarly,
$f_1^{(l)}$ and $f_2^{(l)}$ define the functions \be g(\lambda)
 = 2 \sum_{l=2}^\infty \frac{a^l}{l} f^{(l)}_1 \equiv
2\int \frac{d \lambda}{\lambda} \,G(\lambda)~~~~~~~~~~ k(\lambda)
 = -\frac{1}{2} \sum_{l=2}^\infty \frac{a^l}{l^2} f^{(l)}_2 \,;
\label{small_g}
\ee
the former being twice the first logarithmic integral of $G$ entering in the
Sudakov form factor (\ref{Gfunction}).

In the construction of (\ref{BDS_conjecture}) it was assumed that
the splitting amplitude obeys an all-order exponentiation similar
to the infrared-singular part of the amplitude:
\be
r_S=
\exp\left[\sum_{l=1}^\infty a^l
f^{(l)}(\epsilon)r_S^{(1)}(l\epsilon)\right]~~.
\label{splitting_all_orders}
\ee
This relation may be justified using dual conformal
invariance. Indeed, as we will see in some detain in section
\ref{constraints_WL_vev} following \cite{Drummond:2007au}, the four-
and five-point amplitudes are uniquely fixed by requiring that this
symmetry, observed in explicit calculations, exists to all orders in
perturbation theory. Then, taking the collinear limit of the
five-point amplitude immediately yields (\ref{splitting_all_orders}).
%
%

The infrared poles are apparent in the equation (\ref{BDS_conjecture})
and, using equation (\ref{soft_collinear}), may be readily
isolated together with the associated dependence on the two-particle
invariants:
\begin{equation}
{\rm Div}_n =
-\sum_{i=1}^n \Biggl[
      \frac{1}{8 \epsilon^2} f^{(-2)}\biggl(\frac{\lambda\mu_{IR}^{2\epsilon}}{(-s_{i,i+1})^\epsilon}\biggr)
    + \frac{1}{4 \epsilon}   g^{(-1)}\biggl(\frac{\lambda\mu_{IR}^{2\epsilon}}{(-s_{i,i+1})^\epsilon}\biggr)
              \Biggr]\,,
\label{divergence_general}
\end{equation}
where the invariants $s_{i,i+1}$ are assumed to be negative. The
functions $f^{(-2)}$ and $g^{(-1)}$ are respectively the second
and first logarithmic integrals of the functions $f(\lambda)$ and
$G(\lambda)$ . Extracting this divergent part defines the finite
remainder $F^{(1)}_n(0)$.
\begin{equation}
\ln{\cal M}_n = {\rm Div}_n + \frac{f(\lambda)}{4} F^{(1)}_n(0) +
n k(\lambda) + C(\lambda) \, \label{BDS_Vn}
\end{equation}
with $C(\lambda)=\sum_{l=1}^\infty C^{(l)}a^l$. In the simplest
case of the four-gluon amplitude the finite remainder
$F^{(1)}_n(0)$ takes the form \be
F^{(1)}_4(0)=\frac{1}{2}\left(\ln\frac{s_{12}}{s_{23}}\right)^2+4\zeta_2~~.
\label{4pt_1loop_amp} \ee For more than four external legs the
finite remainders $F^{(1)}_n(0)$ have a more complicated form:
\begin{equation}
F_n^{(1)}(0) = \frac{1}{2} \sum_{i=1}^n g_{n,i} \,,
\label{OneLoopFiniteRemainder}
\end{equation}
where
\begin{eqnarray}
g_{n,i} &=&
-\sum_{r=2}^{\lfloor n/2 \rfloor -1}
  \ln \Biggl(\frac{ -t_i^{[r]}}{ -t_{i}^{[r+1]}  }\Biggr)
  \ln \Biggl(\frac{ -t_{i+1}^{[r]}}{ -t_{i}^{[r+1]}  }\Biggr) +
D_{n,i} + L_{n,i} + \frac{3}{2} \zeta_2 \,,
\label{UniversalFunci}
\end{eqnarray}
in which $\lfloor x \rfloor$ is the greatest integer less than or equal
to $x$ and, as in (\ref{one_loop_full}), $t_i^{[r]} = (k_i + \cdots + k_{i+r-1})^2$
are momentum invariants.
(All indices are understood to be $\mod n$.)
The form of $D_{n,i}$ and $L_{n,i}$ depends upon whether $n$ is odd or even.
For the even case ($n=2m$) these quantities are given by
\begin{eqnarray}
D_{2m,i} &=&
  -\sum_{r=2}^{m-2}
\Li_2 \Biggl( 1-   \frac{t_i^{[r]} t_{i-1}^{[r+2]}}
                        {t_i^{[r+1]}t_{i-1}^{[r+1]}}   \Biggr)
- \frac{1}{2} \Li_2 \Biggl( 1- \frac{t_{i}^{[m-1]} t_{i-1}^{[m+1]}}
                                    {t_i^{[m]} t_{i-1}^{[m]}}
 \Biggr)
 \,,
\nn \\
L_{2m,i} &=&
   \frac{1}{4}
  \ln^2 \Biggl(\frac{-t_{i}^{[m]}}{-t_{i+1}^{[m]}} \Biggr) \,.
\label{DLeven}
\end{eqnarray}
In the odd case ($n=2m+1$), we have,
\begin{eqnarray}
D_{2m+1,i} &=&
  -\sum_{r=2}^{m-1}
\Li_2 \Biggl( 1- \frac{t_i^{[r]} t_{i-1}^{[r+2]}}
                      {t_{i}^{[r+1]} t_{i-1}^{[r+1]}}
  \Biggr)
\,,
\nn \\
L_{2m+1,i} &=&
  - \frac{1}{2}
  \ln \Biggl(\frac{-t_i^{[m-1]}}{-t_i^{[m+1]}} \Biggr)
  \ln \Biggl(\frac{-t_{i+1}^{[m]}}{-t_{i-1}^{[m+1]}} \Biggr) \,.
\label{DLodd}
\end{eqnarray}
These expressions for $D_{n,i}$ and $L_{n,i}$ are found
\cite{BDDK_npt} by inserting the explicit values of the box
integrals into equation (\ref{one_loop_full}).

By construction, the BDS conjecture captures the correct infrared
singularities as well as the correct behavior under collinear
limits. Thus, departures from it should contain no infrared
singularities and moreover should have vanishing collinear limits
in all channels.

Additional constraints may be found if one assumes that dual conformal
invariance is a property of MHV amplitudes to all orders in
perturbation theory \cite{Drummond:2007au}. While it is a plausible assumption especially in
light of the strong coupling prescription for the calculation of
scattering amplitudes \cite{AM1} which we will discuss shortly, this
assumption needs to be verified on a case by case basis. Nevertheless,
if this assumption holds, it leads to the conclusion that departures
from the BDS ansatz must be exhibit dual conformal invariance as a
consequence of their finiteness. Thus, similarly to two-dimensional
conformal field theories, such corrections must be functions of
invariants under the inversion transformations (\ref{inversion}). Dual
conformal invariants can be constructed for any kinematics with at
least six momenta.  As we will see in more detail in section
\ref{constraints_WL_vev}, in this case they are\footnote{
Parity-odd dual conformal invariants can also be constructed. Explicit
calculations \cite{CSV_odd_six_pt} show that, at least at
two-loop order, all parity-odd terms in the six-point amplitude
exponentiate following the BDS ansatz.}
\be
u_1=
      \frac{s_{12}s_{45}}{s_{123}s_{345}}~~~~
u_2=
      \frac{s_{23}s_{56}}{s_{234}s_{123}}~~~~
u_3=
      \frac{s_{34}s_{61}}{s_{345}s_{234}}~~.
\label{SixPtConformalCrossRatios} \ee
The number of such cross-ratios -- i.e. $x_{ij}x_{kl}/x_{ik}x_{jl}$
with the difference between any two labels of at least two units --
grows with the number of external points. Clearly, dual conformal
invariance would imply a reduction in the number of independent
arguments of (the finite part of) MHV amplitudes. This point will be
further discussed in section \ref{constraints_WL_vev}.

To probe the structure of amplitudes it is useful to define the
{\it remainder function} $R_A$: \be
\Remainder_{An}(a)=\ln(1+\sum_l a^l {\cal M}_n^{(l)})-
\left(\sum_l a^l f_l(\epsilon){\cal
M}_n^{(1)}(l\epsilon)+C(a)\right)~~. \label{def_remainder_n} \ee
This is a finite, dual conformally invariant function of the
coupling constant which encodes the departure of the $n$-point MHV
rescaled amplitude from the BDS ansatz. The ${\cal O}(a^2)$ part
may be readily extracted and reads
\begin{equation}
\Remainder^{(2)}_{An} \equiv \lim_{\epsilon \to 0} \left[ M_n^{(2)}(\epsilon)
- \left( \frac{1}{2} \bigl(M_n^{(1)}(\epsilon) \bigr)^2
+  f^{(2)}(\epsilon) \, M_n^{(1)}(2\epsilon) + C^{(2)}\right)
\right]\,.
\label{def_remainder_6}
\end{equation}
Note that the terms in parenthesis are just the ABDK ansatz
(\ref{ABDK}) for the 2-loop MHV amplitude with arbitrary multiplicity.

\subsection{The six-point MHV amplitude at two-loops and the BDS ansatz   \label{6gamp}}

As previously mentioned, the ABDK/BDS ansatz was constructed based on
explicit calculations of four gluon amplitudes at two- and three-loop
orders as well as of the collinear splitting amplitudes at two-loop
order and was subsequently tested through the calculation of the
five-point amplitude at two-loops. Assuming that dual conformal
invariance holds to all loop orders, the fact that no conformal
cross-ratios can be constructed for four- and five-point kinematics
suggests that these amplitudes are determined to all orders by their
infrared singularities.  In later sections we will discuss to what
extent this interpretation is accurate; the ABDK/BDS ansatz will obey
an anomalous Ward identity for dual conformal transformations which
has a unique solution for four- and five-point kinematics. Thus, in
these cases, the ABDK/BDS ansatz necessarily reproduces the
scattering amplitudes.

For higher-point kinematics scattering amplitudes may contain in
principle additional information beyond that related to its infrared
divergences, which is captured by finite functions of conformal ratios
and vanishes in all collinear limits.

Similarly to the five-point MHV amplitude at two-loops, the two-loop
six-point MHV amplitude contains an even and an odd part. The even
part was evaluated in \cite{BDKRSVV} and information on the odd part
was found in \cite{CSV_odd_six_pt}. Projecting the ABDK ansatz
(\ref{ABDK}) onto parity-even and parity-odd components, it follows
quickly that similar iteration relations should hold separately for
the even and odd parts of rescaled amplitudes. It turns out that,
while the odd part of the amplitude obeys the iteration relation
\cite{CSV_odd_six_pt}, the even part does not, signaling departures
from the ABDK/BDS ansatz.

\begin{figure}[ht]
\centering
\includegraphics[scale=0.65]{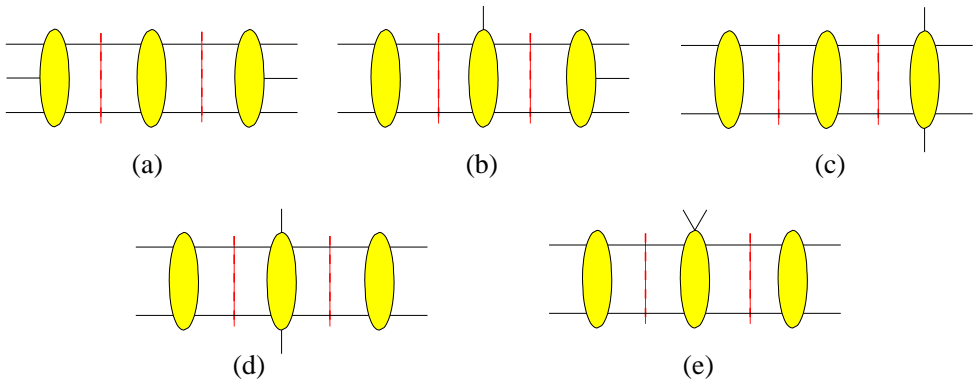}
\caption{Cuts capturing the complete structure of the six gluon
amplitude at two loops. } \label{fig:allcuts_6pt}
\end{figure}

Following the discussion in previous sections, to reconstruct the
amplitude one should analyze all of its generalized cuts. The known
ultraviolet behavior of the theory however implies certain
simplifications.  As mentioned before, this is quite analogous to the
observation that one-loop amplitudes can be written solely in terms of
box integrals. The analogous statement for two-loop amplitudes of
arbitrary multiplicity is that they are completely determined by
($d$-dimensional) iterated two-particle cuts. The relevant topologies
are listed in figure \ref{fig:allcuts_6pt}.

Here, as well as for more general amplitudes, it is useful to
organize the result as the sum of the part constructible from
four-dimensional cuts $M_6^{(2), D=4} (\epsilon)$ and the part
accessible only through some (partial) $d$-dimensional cuts
$M_6^{(2), \mu} (\epsilon)$
\begin{equation}
M_6^{(2), D=4-2\ep}(\epsilon) = M_6^{(2), D=4} (\epsilon) +
                                M_6^{(2), \mu} (\epsilon)~~.
\label{TwoLoopAssembly}
\end{equation}
The latter terms are built from $\mu$-integrals and their
four-dimensional generalized cuts vanish identically (in the sense
that all cut propagators are considered four-dimensional).

\begin{figure}[ht]
\centering
\includegraphics[scale=0.9]{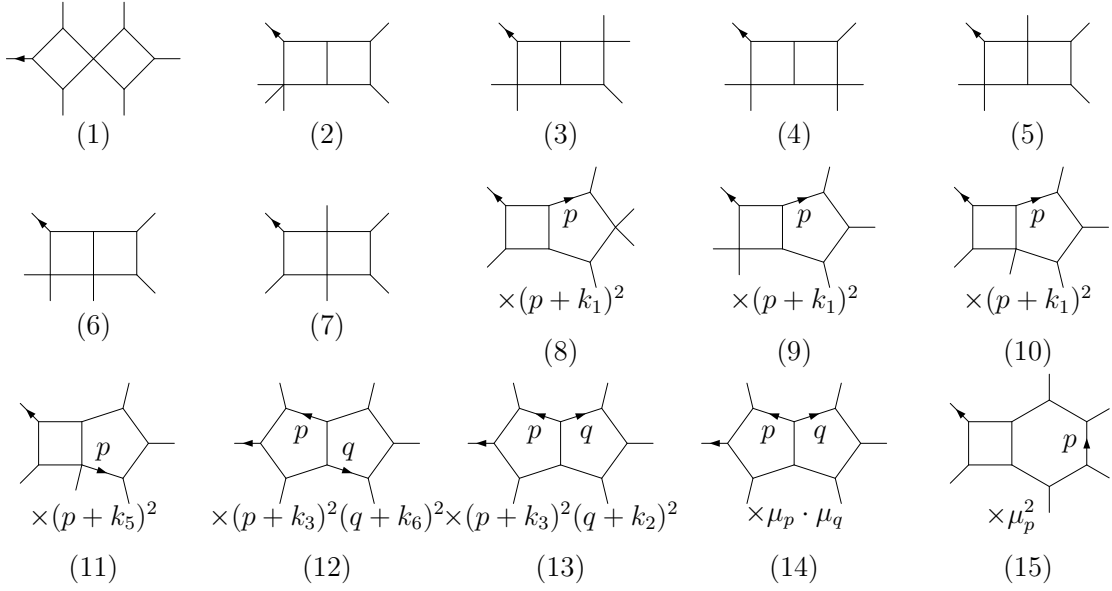}
\caption{Integral topologies appearing in the even part of the six
gluon amplitude at two loops. All momenta are considered to be
outgoing and the arrow on an external line denotes the line
carrying momentum $k_1$. As before, $\mu_p$ and $\mu_q$ denote the
$(-2\epsilon)$-dimensional part of the loop momenta.}
\label{fig:topologies_6pt}
\end{figure}

The four-dimensional cut-constructible parity-even part of the
amplitude is given entirely in terms of pseudo-conformal integrals
\cite{BDKRSVV};
\begin{eqnarray}
M_6^{(2),D=4}(\epsilon) &=& \frac{1}{16} \sum_{12~{\rm perms.}}
\Bigg[
\frac{1}{4} c_1 I^{(1)}(\epsilon)
+ c_2 I^{(2)}(\epsilon)
+ \frac{1}{2} c_3 I^{(3)}(\epsilon)
+ \frac{1}{2} c_4 I^{(4)}(\epsilon)
+ c_5 I^{(5)}(\epsilon)
\cr
&&\qquad\qquad\quad\null
+ c_6 I^{(6)}(\epsilon)
+ \frac{1}{4} c_7 I^{(7)}(\epsilon)
+ \frac{1}{2} c_8 I^{(8)}(\epsilon)
+ c_9 I^{(9)}(\epsilon)
+ c_{10} I^{(10)}(\epsilon)
\cr
&&\qquad\qquad\quad\null
+ c_{11} I^{(11)}(\epsilon)
+ \frac{1}{2} c_{12} I^{(12)}(\epsilon)
+ \frac{1}{2} c_{13} I^{(13)}(\epsilon)
\Bigg] \,.
\label{TwoLoopAssemblyA}
\end{eqnarray}
The integrals $I_i$ are listed in figure \ref{fig:topologies_6pt}
and the corresponding coefficients for the $(1,2,3,4,5,6)$
permutation are:
\begin{eqnarray}
c_1 &=&
s_{61} s_{34} s_{123} s_{345} + s_{12} s_{45} s_{234} s_{345} +
s_{345}^2 (s_{23} s_{56} - s_{123} s_{234})\,,\cr
c_2 &=&
2 s_{12} s_{23}^2\,,\cr
c_3 &=&
s_{234} (s_{123} s_{234} - s_{23} s_{56})\,,\cr
c_4 &=&
s_{12} s_{234}^2\,,\cr
c_5 &=&
s_{34} (s_{123} s_{234} - 2 s_{23} s_{56})\,,\cr
c_6 &=&
- s_{12} s_{23} s_{234}\,,\cr
c_7 &=&
2 s_{123} s_{234} s_{345} - 4 s_{61} s_{34} s_{123} - s_{12} s_{45}
s_{234} - s_{23} s_{56} s_{345}\,,\cr
c_8 &=&
2 s_{61} (s_{234} s_{345} - s_{61} s_{34})\,,\cr
c_9 &=&
s_{23} s_{34} s_{234}\,,\cr
c_{10} &=&
s_{23} (2 s_{61} s_{34} - s_{234} s_{345})\,,\cr
c_{11} &=&
s_{12} s_{23} s_{234}\,,\cr
c_{12} &=&
s_{345} (s_{234} s_{345} - s_{61} s_{34})\,,\cr
c_{13} &=&
- s_{345}^2 s_{56}\,.
\label{IntegralCoefficients_Deq4}
\end{eqnarray}
It is not hard to check that all terms appearing in
(\ref{TwoLoopAssemblyA}) are indeed pseudo-conformal integrals.
Their relative coefficients are $0,\pm1,\pm2$ and $\pm4$, which
represents a chance of patterns from the four- and five-point
amplitudes where they were only $0$ and $\pm1$. It is currently
unclear what is the origin of this change.

The remaining parity-even part of the amplitude, which may be
determined by performing generalized cuts with at least one
two-particle $d$-dimensional cut is \cite{BDKRSVV}
\begin{eqnarray}
M_6^{(2),\mu}(\epsilon) &=& \frac{1}{16} \sum_{12~{\rm perms.}}
\Bigg[
  \frac{1}{4} c_{14} I^{(14)}(\epsilon)
+ \frac{1}{2} c_{15} I^{(15)}(\epsilon)
\Bigg] \,;
\label{TwoLoopAssemblyB}
\end{eqnarray}
the coefficients for the identity permutation $(1,2,3,4,5,6)$ are
\begin{eqnarray}
c_{14} &=&
-2 s_{345} (s_{123} s_{234} s_{345} - s_{61} s_{34} s_{123} - s_{12}
  s_{45} s_{234} - s_{23} s_{56} s_{345})\,,\cr
c_{15} &=&~~\,
2 s_{61} ( s_{123} s_{234} s_{345} - s_{61} s_{34} s_{123} - s_{12}
   s_{45} s_{234} - s_{23} s_{56} s_{345})\,.
\label{IntegralCoefficientsDneq4}
\end{eqnarray}
Their dual conformal properties are somewhat nontransparent; following
the definition given in section \ref{conf_ints} they may be
interpreted as pseudo-conformal as their integrand vanishes
identically in four dimensions. Explicit calculation \cite{BDKRSVV}
shows that they do not contribute to the remainder function
(\ref{def_remainder_6}); thus, their presence may be ascribed to the
infrared structure of the amplitude, interpretation strengthened by
the fact that they either integrate to ${\cal O}(\epsilon)$ ($I_{14}$)
or they exhibit infrared poles ($I_{15}$).

The analytic evaluation of the integrals appearing in the
six-point amplitude remains a difficult open problem, with
potential applications beyond ${\cal N}=4$ SYM. To test for the
structure and the conformal properties of the remainder function,
\cite{BDKRSVV} evaluated the amplitude at a variety of kinematic
points $K^{(0)}$ through $K^{(5)}$. For the state of the art in
the evaluation of Feynman integrals we refer the reader to
\cite{Smirnov_book,
mathematica_implementation1,mathematica_implementation2,
mathematica_implementation3}.
Two of the kinematic points, $K^{(0)}$ and $K^{(1)}$, were chosen to
have the same cross-ratios while the momentum invariants are
different. The results \cite{BDKRSVV} of the evaluation of the
remainder function $\Remainder_{A6}$ are shown in table \ref{RemainderTable}.

\begin{table}

\begin{center}
\begin{tabular}{||c|c||c||}
\hline
\hline
kinematic point & $(u_1, u_2, u_3)$ & $\Remainder_{A6}$    \\
\hline
\hline
$K^{(0)}$ & $(1/4, 1/4, 1/4)$  & $ 1.0937 \pm  0.0057$   \\
\hline
$K^{(1)}$ &$(1/4, 1/4, 1/4)$  & $  1.076 \pm 0.022$  \\
\hline
$K^{(2)}$ &  $\, (0.547253,\, 0.203822,\, 0.881270) \,$
                            & $ -1.659 \pm  0.014$   \\
\hline
$K^{(3)}$ &($28/17, 16/5, 112/85)$  & $ -3.6508 \pm 0.0032\,$  \\
\hline
$K^{(4)}$ & $(1/9, 1/9, 1/9)$ & $  5.21  \pm   0.10$  \\
\hline
$K^{(5)}$ & $(4/81, 4/81, 4/81)$ &  $ 11.09 \pm 0.50 $ \\
\hline
\end{tabular}
\end{center}

\caption[a]{\label{RemainderTable} The numerical remainder compared
with the ABDK ansatz~(\ref{ABDK}) for various kinematic points. The
second column gives the conformal cross-ratios introduced in
(\ref{SixPtConformalCrossRatios}).}

\end{table}


The main observation is that the remainder function is nonzero to a
high level of confidence thus suggesting that the ABDK/BDS ansatz
captures only part of the amplitude. It is also important to note that
the remainder functions at kinematic points $K^{(0)}$ and $K^{(1)}$
are equal within within the errors.  This strongly suggests that
$\Remainder_{A6}$ is indeed a function of only conformal cross-ratios --
{\it i.e.} is invariant under dual conformal transformations.

The conclusion of this calculation is thus that the BDS ansatz
should be modified for six point amplitudes and beyond. For this
purpose it is instructive to identify the origin of the remainder
function within the arguments that led to this ansatz.  In short,
the full structure of collinear limits for $n$-point amplitudes with
$n\ge 6$ is somewhat more involved. One may consider limits in
which more than two particles are simultaneously collinear:
\be
k_{i}=z_i k~~{\rm for}~i=1\dots m~~~~~\sum_{i=1}^m z_i=1~~~~z_i\le 1
~~~~~k^2\rightarrow 0~~.
\ee
For the six-point amplitude only a triple-collinear limit (i.e. $m=3$
above) exists.
%
%
While vanishing in all double-collinear limits, the remainder function
for the six-point amplitude has a nontrivial triple-collinear limit,
which in fact allows its complete reconstruction. We refer the reader
to \cite{BDKRSVV} for more detailed discussions in this direction.

\section{Scattering amplitudes at strong coupling through the
AdS/CFT duality
\label{strong_coupling}}

The AdS/CFT correspondence \cite{Maldacena:1997re, Gubser:1998bc,
Witten:1998qj} provides the only direct access to the strong coupling
regime of the ${\cal N}=4$ SYM; it relates four-dimensional ${\cal
N}=4$ SYM theory and type IIB string theory on $AdS_5
\times S^5$ space through the identification of string states and
gauge-invariant operators. The two gauge theory parameters -- the
't~Hooft coupling $\lambda$ and the rank of the gauge group $N$ -- are
expressed in terms of the radius of curvature of the space and the
string coupling by the well-known relations
\begin{equation}
\label{parameter_relations}
\sqrt{\lambda} \equiv \sqrt{g^2_{YM} N}
=\frac{ R^2 }{ \alpha' }
~,~~~~~~~~~~~~
\frac{1}{N} \sim g_s~~.
\end{equation}
Thus, in the limit of a large number of colors the splitting and
joining of strings is suppressed and in the limit of large 't
Hooft coupling, the string theory lives on a weakly curved space.
In this regime the string theory is completely described by a
weakly-coupled worldsheet sigma-model.

By appending an open string sector to closed string theory in
$AdS_5\times S^5$ gluon scattering amplitudes could in principle
be directly computed, on the string side of the AdS/CFT correspondence,
in terms of integrated correlation functions of vertex operators.
Due to the presence of color factors it is, however, unclear how
much of the resulting structure can be captured entirely in terms
of closed string data. We will argue, following \cite{AM1,AM3},
that the color-stripped partial amplitudes do have such a
description, to leading order in the strong coupling expansion.

As extensively discussed in the previous section, a further property
of on-shell scattering amplitudes is the need of an infrared regulator
due to the presence of infrared divergences. Dimensional
regularization and variants thereof remain the preferred gauge theory
regularization scheme.
As we will see, a choice of regularization is also needed to
define the scattering on the string theory side of the AdS/CFT
correspondence. We will discuss two such choices: first, to set up
the calculation, we will use as a regulator a $D$-brane cutting
off the infrared part of $AdS_5$. After formulating and
understanding the prescription for computing scattering amplitudes
at strong coupling we will modify the regulator to one akin to the
gauge theory dimensional regulator; this will allow a direct
comparison with the strong coupling limit of
(\ref{BDS_conjecture}).

\subsection{The general construction \label{generalpresc}}

In the presence of an open string sector on the string theory side of
the AdS/CFT correspondence, the calculation of open string scattering
amplitudes in the Poincar\'e patch is, in principle, conceptually
straightforward: one simply computes the transition amplitude between
some in and out asymptotic states. To describe scattering amplitudes
of ${\cal N}=4$ SYM fields, these states must be located at spatial
infinity in the directions parallel to the boundary of the $AdS$
space.  As usual, two-dimensional conformal invariance on the string
worldsheet allows a description of the asymptotic states in terms of
local vertex operators inserted on the boundary of the worldsheet.

Perhaps the natural place for the worldsheet boundary and for the
vertex operators is the AdS boundary located at $z=0$ in the
coordinates\footnote{Notice that we interchanged the notation for
original and dual variables with respect to the one used in \cite{AM1}.}
\begin{equation}
\label{ads5met}
ds^2 =R^2 \frac{d{\bf y}_{3+1}^2+dz^2}{z^2} ~~.
\end{equation}
It is however not hard to see that such a choice is not allowed:
indeed, scattering amplitudes are expected to be divergent and as
such must be evaluated in the presence of a regulator. As
discussed at length in section \ref{weak_coupling}, the expected
divergences arise from low energy modes; thus, a natural regulator
eliminating these modes is a D3-brane placed at some fixed and
large value of the radial coordinate $z=z_{IR}\gg R $ and
extending along the four boundary directions ${\bf y}_{3+1}$.

\begin{figure}[ht]
\centering
\includegraphics[scale=0.35]{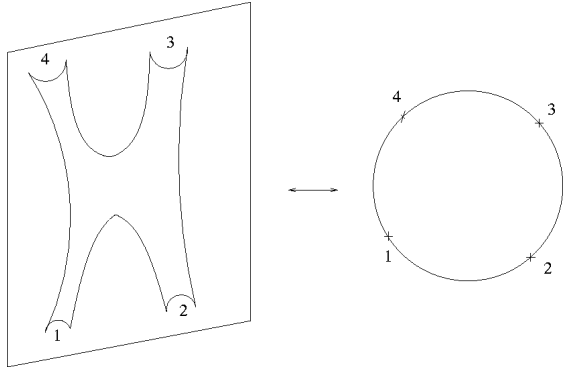}
\caption{worldsheet corresponding to the scattering of four open
strings.} \label{branescattering}
\end{figure}

An interesting property of the Poincar\'e patch is that the spatial
infinity of the regulator brane introduced above coincides with the
spatial infinity of the AdS boundary.\footnote{It is perhaps
interesting to note that, regardless of the coordinate system, a
regulator brane excising the part of AdS space describing the low
energy modes of ${\cal N}=4$ SYM theory always intersects the
boundary. Consider for example the global coordinates; the global time
is dual to SYM energy scale. Eliminating low energy modes amounts to
placing a D3-brane at on some surface of fixed time; it is not hard to
see that this D3-brane will also intersect the boundary of AdS
space. } Thus, the asymptotic states of the scattering process are
simultaneously defined on the regulator brane as well as on the
boundary of $AdS$ space. Two-dimensional conformal invariance may then
be used to describe them through vertex operators located either on
the boundary or on the regulator brane (see
fig. \ref{branescattering}).

The momenta carried by vertex operators in these two
representations of the asymptotic states are, however, different
due to the fact that the anti-de-Sitter space has a nontrivial
metric. Indeed, if the vertex operator carried momentum $k$ when
placed on the boundary, it carries the corresponding proper
momentum $k_{pr}$ when placed on the regulator brane \be
\label{kpr_vs_k} k_{pr}=k \frac{z_{IR}}{R}~~. \ee Here $k$ is the
momentum conjugate to the boundary coordinates ${\bf y}_{3+1}$ and
must be kept fixed\footnote{$k$ plays the role of gauge theory
momentum} as the infrared regulator is removed -- i.e. $z_{IR} \to
\infty$.
Thus, when described in terms of vertex operators placed on the
regulator D3-brane, the scattering process occurs at arbitrarily high
(proper) momenta and fixed angle.

In flat space this regime was studied by Gross, Mende and Manes
\cite{Gross:1987kza,Gross:1989ge}. The key result of their
analysis is that, to leading order in the large momentum expansion,
the calculation of the scattering amplitude is dominated by a saddle
point of the classical action, deformed by the insertion of vertex
operators. At string tree-level, the worldsheet has the topology of a
disk and the vertex operators corresponding to the scattering states
are inserted on its boundary. Perturbative corrections to the
saddle-point contribution are a series in inverse-momentum invariants
(e.g. $(\alpha' s)^{-1}$ and $(\alpha' t)^{-1}$ for a four-point
amplitude) and yield the large energy expansion of the amplitude.

The $\alpha'$ expansion of scattering amplitudes is thus correlated to
the energy expansion; this can be understood on dimensional grounds
and it is a feature of the free worldsheet theory for strings in flat
space. In curved spaces the situation is different and scattering
amplitudes are a double-series in energy $(\alpha's)$ and in (inverse)
curvature $(\alpha'{\cal R})\sim (\alpha'/R^2)$.\footnote{Here
${\cal R}$ denotes the curvature and $R$ denotes the curvature radius.}

One may repeat the arguments of \cite{Gross:1987kza,Gross:1989ge}
for the gluon scattering amplitude described by vertex operators
placed on the regulator brane in AdS space. The fact that the
vertex operators are necessarily proportional to
$\exp(ik_{pr}\cdot y)$ and that $k_{pr}$ is large (cf.
(\ref{kpr_vs_k})) guarantees that, as the regulator is removed
($z_{IR}\rightarrow\infty$), the conclusion that the scattering
amplitude calculation is dominated by a saddle-point continues to
hold, i.e. \be A\propto e^{iS}~~, \ee where $S$ is the value of
the classical worldsheet action at the saddle-point.\footnote{For
sufficiently many particles it is known (in flat space) that there
exist multiple saddle-points. The same may happen in AdS space as
well. Presumably, for a fixed configuration, a single saddle-point
yields the dominant contribution to the amplitude. It would be
very interesting to clarify this issue.}

As discussed above, this result may receive corrections of two
types: $(a)$ corresponding to the large energy expansion and $(b)$
corresponding to the curvature expansion. Corrections of type
$(a)$ must then be a series in $(\alpha's_{pr})^{-1}$ where
$s_{pr}$ denotes some Mandelstam invariant on the regulator brane.
Due to (\ref{kpr_vs_k}), this is also a series in $z_{IR}^{-2}$
and thus they are irrelevant as $z_{IR}\rightarrow\infty$, for any
finite, not necessarily large, value of the field theory
invariants $s$. Corrections of type $(b)$ are governed by the
curvature radius of AdS space. From (\ref{parameter_relations}) it
follows that the curvature expansion is a series in
$\frac{1}{\sqrt{\lambda}}$ which should therefore reproduce the
strong coupling expansion of scattering amplitudes.

To summarize this (long) argument, the leading term in the strong
coupling expansion of a scattering amplitude of states in ${\cal N}=4$
SYM theory is exactly given by the value of the classical worldsheet
action at a saddle-point. The corresponding worldsheet has the
topology of a disk with boundary placed on the regulator brane and
containing the vertex operators describing the scattering
states. Corrections around this saddle-point yield the strong coupling
expansion of the corresponding amplitude. This analysis captures both
the color and the kinematic dependence of amplitudes.

While this conclusion \cite{AM1} is very encouraging, the boundary
conditions for the construction of the saddle-point cannot be easily
formulated due to a lack of sufficiently explicit expressions for open
string vertex operators in AdS space. It turns out however that the
boundary conditions for the construction of the saddle-point are
easier to formulate in terms of the T-dual boundary
coordinates. Indeed, consider a metric of the form
\be
ds^2=h(z)^2(dy_\mu dy^\mu+dz^2)~~.
\ee
On an Euclidean worldsheet, the T-dual coordinates $x^\mu$ are defined
by\footnote{This transformation, while identical to T-duality for
compact coordinates, should be interpreted here as a formal
sigma model duality transformation. Apart from their use here
\cite{AM1}, such transformations have been also used
\cite{Kallosh:1998ji} to simplify the action of the Green-Schwarz
superstring in $AdS_5\times S^5$.}
\be
\partial_\alpha x^\mu = i h^2(z) \epsilon_{\alpha \beta}
\partial_\beta y^\mu~~.
\label{T_duality_mock_up}
\ee
For the specific $h(z)$ in (\ref{ads5met}) this transformation,
combined with the redefinition $r=R^2/z$ has no apparent effect on the metric
\cite{Kallosh:1998ji}. Indeed, after the first step the metric becomes
\begin{equation}
ds^2=\frac{z^2}{R^2} dx_\mu dx^\mu+\frac{R^2}{z^2}dz^2~~,
\end{equation}
which is just another copy of AdS$_5$. The important point however
is that, unlike the metric (\ref{ads5met}), its boundary is
located at $z=\infty$. Further introducing the radial coordinate
\be r=\frac{R^2}{z} \label{coord_transf} \ee leads to the metric
\begin{equation}
\label{dualmetric} ds^2=\frac{R^2}{r^2} \left(\,dx_\mu
dx^\mu+dr^2\,\right)~~,
\end{equation}
which is identical to (\ref{ads5met}) except that, as implied by the
coordinate transformation (\ref{coord_transf}), the boundary of this
space should be {\it defined} to be at $r=0$.

The sequence of transformations has important effects on vertex
operators and on the regulator brane:

\noindent $(i)$ the regulator brane is located at
$r_{IR}=\frac{R^2}{z_{IR}}\rightarrow 0$ which is located close to the
{\it boundary} of the resulting space.

\noindent $(ii)$
T-duality transformation of a compact coordinate interchanges momentum
and winding states. The transformation (\ref{T_duality_mock_up}) has a
similar effect: the zero-mode of the field $y$ corresponding to the
momentum $k^\mu$ (and described by a local vertex operator) is
replaced by a ``winding'' mode of the field $x$ implying that the
difference between the two endpoints of the string obeys
\begin{equation}
\Delta x^\mu=2 \pi k^\mu~~,
\label{winding}
\end{equation}
as may be seen by integrating (\ref{T_duality_mock_up}) over the
space-like worldsheet coordinate ($\Delta x\equiv
x(\sigma=2\pi)-x(\sigma=0)$).  Thus, each vertex operator is replaced
by a line segment connecting two points whose coordinate difference is
a multiple of the momentum carried by the vertex operator. Moreover,
since the regulator brane is located near the boundary, the momentum
carried by these vertex operators is that of the gauge theory
scattering states.

To summarize, the boundary conditions defining the saddle-point
generating the leading term in the strong coupling expansion of
scattering amplitudes (see fig. \ref{originaltdual}) imply that
the boundary of the worldsheet is a line constructed as follows:

\begin{figure}[ht]
\centering
\includegraphics[scale=0.28]{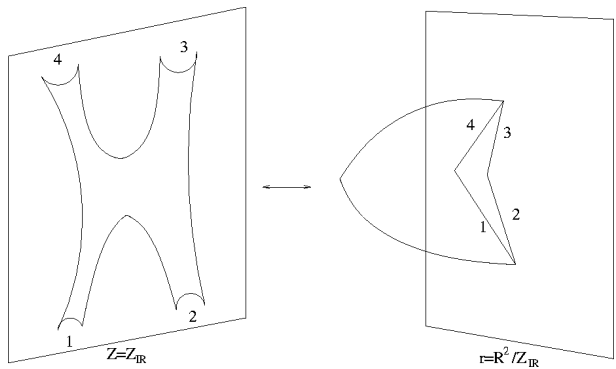}
\caption{Comparison of the worldsheet in original and T-dual
coordinates.} \label{originaltdual}
\end{figure}

\begin{itemize}

\item For each particle of momentum $k^\mu$, draw a
segment joining two points separated by $\Delta x^\mu=2 \pi
k^\mu$.

\item Concatenate the segments according to the
insertions on the disk (corresponding to a particular color
ordering or to a particular ordering of vertex operators on the
original boundary)

\item As gluons are massless, the segments are
light-like. Due to momentum conservation, the diagram is closed.

\end{itemize}

\begin{figure}[ht]
\centering
\includegraphics[scale=0.5]{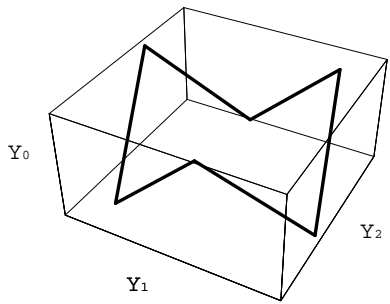}
\caption{Polygon of light-like segments corresponding to the
momenta of the external particles.} \label{6lines}
\end{figure}

\noindent
Figure \ref{6lines} shows an example of such boundary conditions,
corresponding to the scattering of six particles.

As the infrared regulator is removed, i.e. as $z_{IR} \rightarrow
\infty$ in the original coordinates, the boundary of the worldsheet
moves towards the boundary of the T-dual metric, at $r=0$. To leading
order in the strong coupling expansion the computation of scattering
amplitudes becomes formally equivalent to that of the expectation
value of a Wilson loop given by a sequence of light-like segments.

The standard prescription \cite{Rey:1998ik, Maldacena:1998im} implies
that the leading exponential behavior of the $n-$point scattering
amplitude is given by the area $A$ of the minimal surface that ends on
a sequence of light-like segments on the boundary
\begin{equation}
\label{finalpresc}
{\cal A}_n \sim e^{-\frac{\sqrt{\lambda}}{2 \pi}A(k_1,...,k_n)}~~.
\end{equation}
The area $A$ contains the relevant kinematic information through its
boundary conditions.

It is important to stress the following two points:

\noindent$(i)$ In implementing the duality transformation
(\ref{T_duality_mock_up}) at the level of vertex operators most of
their structure -- in particular the polarization of the particle
-- was ignored, as we were interested in the leading strong
coupling regime. Thus, such information is not directly available
at the level of the light-like polygon describing the boundary
conditions. To access this information (which may, in some sense,
be considered as subleading in the $1/\sqrt\lambda$ expansion) it
appears necessary to return to the vertex operator picture for the
scattering process. This should expose, for example, that all-plus
helicity amplitudes vanish identically (\ref{vanishing_amplitudes}).

\noindent$(ii)$
The map (\ref{winding}) between vertex operators and light-like
segments operates independently of the color structure of the original
amplitude. Thus, one finds one light-like polygon for each
color-ordered amplitude. The corresponding color factor may be
reconstructed from the order of momenta of particles.

In the following, we will show in detail how this construction works
for the scattering of four gluons and compare our results with field
theory expectations.

\subsection{Four gluon scattering}

\begin{figure}[ht]
\centering
\includegraphics[scale=0.4]{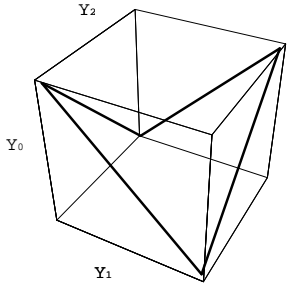}
\caption{Polygon corresponding to the scattering of four gluons.}
\label{4lines}
\end{figure}

The simplest scattering process involves four particles and
is characterized by the usual Mandelstam invariants
\begin{equation}
s=-(k_1+k_2)^2
~,~~~~
t=-(k_2+k_3)^2
\end{equation}
The discussion in the previous section suggests that, in the strongly
coupled ${\cal N}=4$ SYM theory, the amplitude for this process is
governed by the minimal surface ending on the light-like polygon shown
in figure \ref{4lines}.

As implied by the equation (\ref{winding}),
the difference between the coordinates of each of the corners of the
polygon (cusps) are, up to a factor of $(2\pi)$,
the momenta of the corresponding gluon.
In drawing figure \ref{4lines}, it was assumed that the third
component of the momentum (described by the boundary
coordinate $y_3$) vanishes. This choice imposes no kinematical
restrictions, as it does not imply any relations between the two
Mandelstam invariants.

The area of a surface embedded in a higher dimensional space is
simply given by the integral of the induced metric \be A=\int
d\sigma d\tau \sqrt{-\det \partial_\alpha x^\mu \partial_\beta
x^\nu g_{\mu\nu}(x)}~~. \label{NGaction} \ee Finding minimal
surfaces amounts simply to treating this area as an action (the
Nambu-Goto action) and solving the classical equations of motion,
subject to the desired boundary conditions. Then, the area is
obtained by evaluating (\ref{NGaction}) on the resulting
configuration.

\subsubsection{The single cusp solution}
\label{singlecusp}

As a warm up exercise let us discuss the solution near the cusp
where two of the light-like lines meet. This problem was
originally considered in \cite{Kruczenski:2002fb} and it will
prove useful for generating the solution relevant for the
four-gluon scattering. The surface can be embedded into an $AdS_3$
subspace of $AdS_5$
\begin{equation}
ds^2=\frac{-dx_0^2+dx_1^2+dr^2}{r^2}~~.
\end{equation}
We are interested in finding the surface ending on a light-like
Wilson line which is along $x^1=\pm x^0$, $x^0>0$ \footnote{One
can also consider Wilson loops along $x^0=\pm x^1, ~x^1>0$. The
basic difference with the ones considered here is that their
worldsheet is Lorentzian and $z$ is imaginary.} (see figure
\ref{cusp}). This configuration has both boost and scale symmetry,
which are made manifest by the following ansatz:

\begin{figure}[ht]
\centering
\includegraphics[scale=0.4]{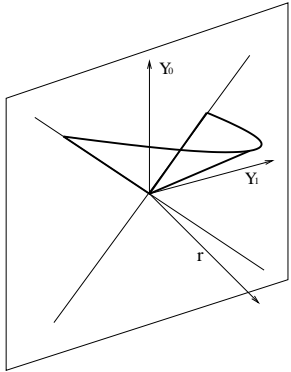}
\caption{Single cusp solution.} \label{cusp}
\end{figure}

\begin{equation}
x_0=e^\tau \cosh \sigma,\hspace{0.3in}x_1=e^\tau \sinh
\sigma,\hspace{0.3in}r=e^\tau w(\tau) ~~.
\label{cusp_ansatz}
\end{equation}
This ansatz establishes a close relation between the affine
coordinates $(\tau,\sigma)$ and the target space coordinates
$(x^0,x^1)$; boosts in the $(0,1)$ plane and scale transformations
are simply shifts of $\sigma$ and $\tau$, respectively.

Equations for the remaining function $w(\tau)$ may be found by
evaluating the equations of motion on the ansatz
(\ref{cusp_ansatz}). Alternatively, one may simply evaluate the
Nambu-Goto action on the ansatz (\ref{cusp_ansatz}) and derive an
equation for $w(\tau)$ by varying the result with respect to it.
Choosing the second path, the Nambu-Goto action becomes
\begin{equation}
S_{NG}=\frac{R^2}{2\pi\alpha'}\int d\sigma d\tau
\frac{\sqrt{1-\left(w(\tau)+w'(\tau)\right)^2}}{w(\tau)^2}~~.
\end{equation}
One can then explicitly check that $w(\tau)=\sqrt{2}$ solves the equations
of motion and has the correct boundary conditions. Hence the surface
is given by
\begin{equation}
\label{eq:singlecusp} r=\sqrt{2}\sqrt{x_0^2-x_1^2}~~.
\end{equation}
Notice that the surface lies entirely outside the light-cone of the
origin, hence it is Euclidean.

\subsubsection{Four cusps solution  \label{four_cusp_solution}}

The four cusps solution is closely related to the single cusp
solution discussed above. The relevant solution of the Nambu Goto
action can be embedded in a $AdS_4$ subspace of $AdS_5$,
parametrized by $(r,x_0,x_1,x_2,x_3=0)$. It is moreover convenient
to fix reparametrization invariance by choosing
$(\sigma_1,\sigma_0)=(x_1,x_2)$. With these choices, the
Nambu-Goto action describes the dynamics of two fields, $r$ and
$x_0$, living in the space parametrized by $x_1$ and $x_2$
\begin{equation}
S = \frac{R^2 }{ 2 \pi \alpha'} \int dx_1 dx_2 \frac{ \sqrt{ 1 + (
\partial_i r)^2 - (\partial_i x_0)^2 - ( \partial_1 r
\partial_2x_0 - \partial_2 r \partial_1 x_0 )^2 } }{ r^{ 2 } }~~.
\end{equation}

\begin{figure}[ht]
\centering
\includegraphics[scale=0.7]{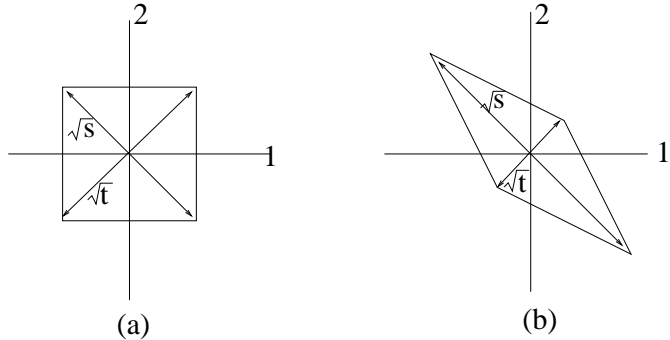}
\caption{Projection to the plane $(x_1,x_2)$ of the surface for
the cases $s=t$ and $s \neq t$.}
\label{rombosr}
\end{figure}

 The classical equations of motion should then be supplemented by the
 appropriate boundary conditions. We consider first the case with
 $s=t$, where the projection of the Wilson lines is a square. By scale
 invariance, we can choose the edges of the square to be at
 $x_1,x_2=\pm 1$. The boundary conditions can be easily seen to be
\begin{equation}
r(\pm 1,x_2)=r(x_1,\pm 1)=0,~~~~x_0(\pm 1,x_2)=\pm
x_2,~~~y_0(x_1,\pm 1)=\pm x_1 ~~.
\end{equation}
The form of the solution near each of the cusps can be obtained by
rotations and boosts from the single cusp solution
(\ref{eq:singlecusp}). The following field configuration satisfies the
boundary conditions and has the correct properties near each of the
cusps
\begin{equation}
\label{squaresol}
x_0(x_1,x_2)=x_1x_2,~~~~~r(x_1,x_2)=\sqrt{(1-x_1^2)(1-x_2^2)}~~.
\end{equation}
 Remarkably it turns out to be a solution of the equations of
 motion. However, in order to capture the kinematical dependence of
 the area \footnote{On dimensional grounds the area, if finite,
 should be a function of the form $f(s/t)$.} we need to consider more
 general solutions with $s \neq t$. In this case the projection of the
 surface to the $(x_1,x_2)$ plane will not be an square but a rhombus,
 with $s$ and $t$ given by the square of the distance between opposite
 vertices, as shown in figure \ref{rombosr}.

The symmetry generators of anti-de-Sitter space act nonlinearly on the
Poincar\'e patch coordinates (\ref{dualmetric}). They are however
useful for generating new and interesting worldsheet configurations
from known ones, since they can change the Mandelstam variables $s$
and $t$; it would therefore be useful to linearize their action.
This is realized by passing to the so-called embedding coordinates, in
which $AdS_5$ is viewed as a hypersurface
\begin{equation}
-X_{-1}^2-X_{0}^2+X_1^2+X_2^2+X_3^2+X^2_4=-1~~.
\label{embedding_eq}
\end{equation}
embedded in $\IR^{2,4}$. Clearly, this constraint equation is
manifestly invariant under the $SO(2,4)$ symmetry group of $AdS_5$,
which is also the Lorentz group of the embedding space.
The Poincar\'e coordinates $({\bf x},r)$ are but a particular solution
of the constraint equation (\ref{embedding_eq}):
\begin{eqnarray}
X^\mu&=&\frac{x^\mu}{r},~~~~~~\mu=0,...,3 \\
\nonumber
X_{-1}+X_4&=&\frac{1}{r},\hspace{0.3in}X_{-1}-X_4=\frac{r^2+x_\mu
x^\mu}{r}~~.
\end{eqnarray}
Direct use of the solution  (\ref{squaresol}) implies that, in the
embedding coordinates, the
minimal surface describing the scattering of four gluons with
the $s=t$ kinematics  is
\begin{equation}
 X_0 X_{-1}=X_1 X_2,~~~~~~X_3=X_4=0~~.
\end{equation}
In this form it is then easy to use the $SO(2,4)$ to generate from
(\ref{squaresol}) new minimal surfaces, corresponding to $s\ne t$
gauge theory kinematics.

Through the AdS/CFT correspondence, the $SO(2,4)$ of the dual AdS
space should have a gauge theory counterpart. Due to the T-duality
transformation relating the boundary coordinates of this space with
gauge theory momenta this symmetry is necessarily different from the
usual position space conformal invariance of ${\cal N}=4$ SYM theory.
As discussed in section \ref{conf_ints}, the action of a ``dual
conformal group'' can be identified at the level of perturbative
scattering amplitudes. While {\it a priori} these two symmetries are
unrelated (as e.g. the former exists at strong coupling while the
latter at weak coupling), it is nevertheless tempting to interpret the
latter as the weak coupling version of the former. This interpretation
is strengthened by the relation between MHV scattering amplitudes and
Wilson loops, which will be discussed in section \ref{Amp_vs_WL}.


 Solutions with $s \neq t$ can be obtained by starting from
 (\ref{squaresol}) and performing a boost in the $(0,4)$ plane. In this
 way we change the distance between opposite vertices of the square.
 \begin{equation}X_0 X_{-1}=X_1 X_2,~~~X_4=0~~~\rightarrow~~~X_4-v
 X_0=0,~~~\sqrt{1-v^2}X_0 X_{-1}=X_1X_2\end{equation} After the boost,
 we end up with a two-parameter solution, one related to the size of
 the initial square and another related to the boost parameter. The
 solution can be conveniently written as
\begin{eqnarray}
\label{solution_st} r={a \over \cosh u_1 \cosh u_2+b \sinh u_1
\sinh u_2},~~~~ x_0= {a \sqrt{1+b^2} \sinh u_1 \sinh u_2 \over
\cosh u_1 \cosh u_2+b \sinh u_1 \sinh u_2} \\ x_1={a \sinh u_1
\cosh u_2 \over \cosh u_1 \cosh u_2+b \sinh u_1 \sinh u_2},~~~~
x_2={a \cosh u_1 \sinh u_2 \over \cosh u_1 \cosh u_2+b \sinh u_1
\sinh u_2}
\end{eqnarray}
where we have written the surface as a solution to the equations of
motion in conformal gauge
\begin{equation}
iS=-{R^2 \over 2 \pi \alpha'} \int {\cal L} =-{R^2 \over 2 \pi }
\int du_1 du_2  { 1 \over 2 }
 { \left( \partial r\partial r +  \partial x_\mu \partial x^\mu
 \right)\over r^2 }\end{equation}
$a$ and $b$ encode the kinematical information of the scattering
as follows
\begin{equation}
-s(2 \pi)^2 = {8 a^2 \over (1-b)^2},~~~~~
-t (2 \pi)^2 ={8 a^2 \over (1+b)^2},
 ~~~~~{ s \over t } = \frac{ (1+b)^2 }{ (1-b)^2 }
\end{equation}
To obtain the four point scattering amplitude at strong coupling it
should suffice, following the discussion in section
\ref{generalpresc}, to evaluate the classical action on the solution
(\ref{solution_st}). However, in doing so, one finds a divergent
answer. That is of course the case, since we have ignored the infrared
regulator. In order to obtain a finite answer we need to reintroduce a
regulator.

\subsubsection{Dimensional regularization at strong coupling
\label{Dimregstrong}  }

Gauge theory amplitudes are regularized by considering the theory
in $D=4-2\epsilon$ dimensions. More precisely (see discussion in
section 2), one starts with ${\cal N}=1$ in ten dimensions and
then dimensionally reduces to $4-2\epsilon$ dimensions. For
integer $2\epsilon$ this is precisely the low energy theory living
on a $Dp-$brane, where $p=3-2\epsilon$. We regularize the
amplitudes at strong coupling by considering the gravity dual of
these theories. \footnote{See section \ref{generalizations} for a
brief discussion of the subtleties of sigma-model actions in these
backgrounds and the calculations of $1/\sqrt{\lambda}$ corrections
to the classical area of the minimal surface.}  The string frame
metric is (see e.g. \cite{Duff:1994an,Itzhaki:1998dd})
\begin{eqnarray}
ds^2=f^{-1/2}dx_{4-2\epsilon}^2+f^{1/2}\left[ dr^2 +r^2
d\Omega^2_{5+2\epsilon} \right],~~~~~ f=(4 \pi^2 e^\gamma)^\epsilon
\Gamma(2+\epsilon) \mu^{2\epsilon} \frac{\lambda}{r^{4+2\epsilon}}
\end{eqnarray}
Following the steps described above, one is led to the following
action
\begin{equation}
S=\frac{\sqrt{c_\epsilon \lambda} \mu^\epsilon }{2\pi}
\int \frac{{\cal L}_{\epsilon=0}}{r^\epsilon}\end{equation}

Where ${\cal L}_{\epsilon=0}$ is the Lagrangian density in the
absence of the regulator. The presence of the factor $r^\epsilon$,
which arises from the conformal factor of the induced metric, will
have two important effects. On one hand, previously divergent
integrals will now converge. On the other hand, the equations of
motion will now depend on $\epsilon$ and it turns out to be very
difficult to find the solution for general $\epsilon$. However, we
are interested in computing the amplitude only up to finite terms
as we take $\epsilon \rightarrow 0$. In that case, it turns out to
be sufficient to plug the original solution into the
$\epsilon$-deformed action \footnote{Up to a contribution from the
regions close to the cusps that add an unimportant additional
constant term.}. The evaluation of integrals leads to \cite{AM1}
\begin{equation}
S \approx \sqrt{\lambda} \frac{\mu^\epsilon}{a^\epsilon} ~ _2
F_1\left( \frac{1}{2},-\frac{\epsilon}{2},\frac{1-\epsilon}{2};b^2
\right)~~.
\end{equation}
Finally, expanding in powers of $\epsilon$ yields the final answer
\begin{eqnarray}
\label{finalfour}
{\cal A}&=&e^{i S}=\exp \left[i
S_{div}+\frac{\sqrt{\lambda}}{8\pi}\left(\log{\frac{s}{t}}
\right)^2+\tilde{C} \right] ~~,\\
S_{div}&=&2S_{div,s}+2S_{div,t}~~,\\
iS_{div,s}&=&-\frac{1}{\epsilon^2}\frac{1}{2\pi}\sqrt{\frac{\lambda
\mu^{2\epsilon}}{(-s)^\epsilon}}
-\frac{1}{\epsilon}\frac{1}{4\pi}(1-\log 2) \sqrt{\frac{\lambda
\mu^{2\epsilon}}{(-s)^\epsilon}}~~.
\end{eqnarray}
This should be compared with the field theory expectations, equations
(\ref{divergence_general}) and (\ref{BDS_Vn}), specialized to the case
$n=4$:
\begin{eqnarray}
\label{final_four_decomposition}
{\cal A}&\sim& \left({\cal A}_{div,s} \right)^2
\left({\cal A}_{div,t} \right)^2 \exp \left\{\frac{f(\lambda)}{8}
(\ln s/t)^2 +{\rm const.} \right\} \\
{\cal A}_{div,s}&=&\exp \left\{-\frac{1}{8 \epsilon^2}f^{(-2)}
\left(\frac{\lambda \mu^{2\epsilon}}{s^\epsilon} \right)
-\frac{1}{4 \epsilon}g^{(-1)}
\left(\frac{\lambda \mu^{2\epsilon}}{s^\epsilon} \right) \right\}~~.
\end{eqnarray}
It is important to notice that the general structure is in perfect
agreement with the general structure of infrared divergences in ${\cal
N}=4$ SYM theory.  It is not hard to see that the leading divergence
has the correct coefficient, given by the strong coupling limit of the
cusp anomalous dimension \cite{Kruczenski:2002fb}
\footnote{The appearance of the cusp anomalous dimension in the equations
(\ref{final_four_decomposition}) may appear surprising at first
sight. Indeed, by analogy with weak coupling arguments based on
finiteness of physical quantities constructed from gluon scattering
amplitudes, the natural quantity entering
(\ref{final_four_decomposition}) should be the large spin limit of the
anomalous dimension of twist-2 operators. It was however shown in
\cite{Kruczenski:2007cy}
that worldsheet calculations of the cusp anomaly and of the large
spin limit of the anomalous dimension of twist-2 operators are related
by an analytic continuation and and target space symmetry
transformations. Thus, similarly to the weak coupling result of
\cite{cusp_vs_twist2_1, cusp_vs_twist2_2, Bassetto:1993xd},
the cusp anomaly equals the large spin limit of the anomalous
dimension of twist-2 operators to all orders in the $1/\sqrt{\lambda}$
expansion. }
\begin{equation}
f(\lambda)=\frac{\sqrt{\lambda}}{\pi}~~.
\end{equation}
Moreover, from (\ref{finalfour}) one could extract the strong coupling
behavior of the function $g(\lambda)$ introduced in equation
(\ref{small_g}):
\begin{equation}
g(\lambda)=\frac{\sqrt{\lambda}}{2\pi}(1-\log 2)~~.
\end{equation}
Notice that due to the scheme dependence of $g(\lambda)$, it
should be computed using the same regularization as in
perturbative computations, that of course is not the case for
$f(\lambda)$. Finally, the kinematic dependence of the finite term
(\ref{final_four_decomposition}) reproduces the strong coupling
limit of the BDS prediction (\ref{BDS_Vn})-(\ref{4pt_1loop_amp})
for the four gluon scattering amplitude.

\subsubsection{Radial Cut-off}

A more common regularization scheme for computing minimal areas in
$AdS$ is to introduce a cut-off in the radial direction. The correct
procedure would be to impose the boundary conditions at some small
$r=r_c$. It turns out, however, that in order to compute the finite
piece as $r_c \rightarrow 0$ it suffices to use the original solution
and cut the integral giving the area at $r=r_{c}$\footnote{This finite
part arises in a similar way when using the conjectured string theory
version of gauge theory dimensional regularization.}

In order to compute the regularized area for the scattering of
four gluons it is convenient to work in conformal gauge. In this
case, the problem reduces to the calculation of the area enclosed by the
curve

\begin{equation}\frac{a}{\cosh u_1 \cosh u_2+b \sinh u_1 \sinh
u_2}=r_{c}\end{equation}
 One way to compute the area is
by expanding the integrand in power series of $r_c / a$ and
integrating term by term. Equivalently, one can use Green's
theorem and express the area as a one dimensional integral over
the boundary of the worldsheet. The result is

\begin{equation}i S
=-\frac{\sqrt{\lambda}}{2\pi}A,~~~~~
A=\frac{1}{4}\left(\log\left(\frac{r_c^2}{-8
\pi^2 s }\right)
\right)^2+\frac{1}{4}\left(\log\left(\frac{r_c^2}{-8 \pi^2 t
}\right) \right)^2-\frac{1}{4}\log^2(\frac{s}{t})+{\rm const.}
\end{equation}
Several comments are in order. First, notice that the structure of
infrared divergences is of the form $\log^2(r^2_c/s)$ \footnote{Notice
that a very similar structure appears when using off-shell
regularization, as done in \cite{Drummond:2007aua}.}, and the
coefficient in front of double logs and the finite piece is
proportional to the cusp anomalous dimension at leading strong
coupling, as in the case of dimensional regularization. Second, single
logs have been absorbed into the double logs. Finally, the finite term
reproduces, up to an additive constant, the results of dimensional
regularization. Hence, the computation of amplitudes at strong
coupling does not need to be done by using dimensional regularization,
unless we are interested in computing the function $g(\lambda)$ and
the constant $C(\lambda)$ entering in equations
(\ref{divergence_general}) and (\ref{BDS_Vn}), respectively, and computed
by using dimensional regularization.

\subsubsection{Structure of infrared poles at strong coupling}

Even if the relevant solutions for minimal surfaces for the cases
$n>4$ are presently unknown, the IR structure of amplitudes at strong
coupling for the general case of $n-$point amplitudes can easily be
understood.

Given the
cusp formed by a pair of neighboring gluons with momenta $k_i$ and
$k_{i+1}$ we associate the kinematical invariant
$s_{i,i+1}=(k_i+k_{i+1})^2$. We expect the following structure for the
infrared-divergent part of the action

\begin{equation}i S_{div}=-\frac{\sqrt{\lambda}}{2\pi}\sum_i
I(\frac{r_c^2}{s_{i,i+1}})\end{equation} where
$I(\frac{r_c^2}{s_{i,i+1}})$ can be computed following
\cite{Buchbinder:2007hm},
either by using dimensional regularization or a radial cut-off. For
later applications the later scheme will be more useful to us, in this
case

\begin{equation}4 I = \int^1_\xi \int^1_\frac{\xi}{X^-} \frac{1}{X^-
X^+}=\frac{1}{2}\log^2 \xi,~~~~~~~\xi=\frac{r_c^2}{-8 \pi^2
s_{i,i+1}}\end{equation} Hence, when using a radial cut-off as
regulator, we expect the following structure for scattering
amplitudes at strong coupling

\begin{equation}i S_n=-\frac{\sqrt{\lambda}}{16\pi}\sum_{i=1}^n
\log^2\left(\frac{r_c^2}{-8 \pi^2 s_{i,i+1}} \right)+\text{Fin}(k_i)
\end{equation} It is easy to check that the general form of the amplitude
for the case $n=4$ is consistent with this general expression.

For the discussion of the next subsection, it will be important to consider a
radial cut-off that depends on the point at the boundary we are
approaching, {\it i.e.} $r_c(x)$. In that case, the structure of
the amplitude turns out to be as follows

\begin{equation}i S_n=-\frac{\sqrt{\lambda}}{16\pi}\sum_{i=1}^n
\log^2\left(\frac{r_c^2(x_i)}{-8 \pi^2 s_{i,i+1}}
\right)+\text{Fin}(k_i)+\sum_{i=1}^n E_{edge}^i(r_c) \end{equation} The
last sum in this expression corresponds to finite extra
contributions coming from the edges

\begin{equation}
\label{general}
E_{edge}^i={\sqrt{\lambda} \over 2\pi}\int_0^1 {ds \over s}
\log \left(\frac{r_c(s)r_c(1-s)}{r_c(0) r_c(1)} \right)
\end{equation} where $s$ running from zero to one parametrizes the
$ith$ edge, namely $x^\mu (s)=x^\mu_i+s(x^\mu_{i+1}-x^\mu_i)$ and
$r_c(s)$ is a shorthand notation for $r_c(x(s))$. For instance, a
simple example is that of a cut-off that takes the value
$r_c(x_i)$ at the $ith$ cusp and varies linearly between cusp and
cusp, in this case

\begin{equation}\label{simp}E_{edge}^i
={\sqrt{\lambda} \over 4 \pi} \log^2{r_c(x_i)
\over r_c(x_{i+1})} \end{equation}

\subsection{A conformal Ward Identity \label{CWI_strong_coupling}}

An important ingredient in the construction of the minimal surface
governing the four-gluon scattering amplitude was the existence of a
dual $SO(2,4)$ symmetry.\footnote{In principle this symmetry is
unrelated to the original conformal symmetry. It has been suggested
\cite{RTW} that, at the level of the worldsheet sigma-model, the symmetries
of the dual $AdS$ space are in fact part of the hidden (non-local)
symmetries of the original $AdS$ space sigma-model \cite{BPR}. } This
symmetry allowed the construction of new solutions and fixed the
finite piece of the scattering amplitude.  Naively, this conformal
symmetry would imply that the amplitude is independent of $s$ and $t$,
since they can be sent to arbitrary values by a dual conformal
symmetry.  The whole dependence on $s$ and $t$ arises due to the
necessity of introducing an infrared regulator. However, we will see
that, after keeping track of the dependence on the infrared regulator,
the amplitude is still determined by the dual conformal symmetry.

At quantum level, classical symmetries manifest themselves through the
existence of (potentially anomalous) Ward identities, the simplest is
the quantum version of the conservation of the Noether current. More
complicated Ward identities describe the action of symmetry generators
on gauge-invariant quantities and potentially constrain their quantum
expressions. In this direction it is possible to construct Ward
identities for the dual $SO(2,4)$ symmetry and study the constraints
they impose on scattering amplitudes. For this purpose it is
convenient to regularize the amplitude calculation with a radial
cut-off.

Given the momenta $k_i$ of the external gluons, the boundary of the
worldsheet contains cusps located at $x_i$, with $2\pi
k_i=x_i-x_{i+1}$. Now imagine that we regularize the area by choosing
a cut-off $r_c$. Moreover, we would like this cut-off to depend on the
point at the boundary we are approaching, {\it i.e.}  $r_c \rightarrow
r_c(x)$. From the discussion above we expect the regulated area to
have the general form
\begin{equation}
A^{reg}_n=f(\lambda)\; \sum_{i=1}^n \log^2 \left(\frac{r_c^2(x_i)}{-2
x_{i-1,i+1}^2} \right)+\text{Fin}(x_i)~~,
\end{equation}
where we have ignored extra terms coming from the edges of the contour
as they can be seen not to affect the following argument
\footnote{One can convince oneself, for instance, by considering the
simplified case (\ref{simp}) and applying the generator of special
conformal transformations to such extra terms. It is also instructive
to apply the generator of dual conformal transformations, whose
relevant piece is of the form $\int ds x^\mu(s)
r_c(s)\frac{\delta}{\delta r_c(s)}$, to the general extra terms
(\ref{general}) and compare this expression to eq. (34) of
\cite{Drummond:2007au} .}. $SO(2,4)$ transformations will then
act on the points $x_i$ and $r_c(x_i)$. By requiring the area to
be invariant under the action of special conformal transformations
generated by $\mathbb{K}^\mu$
\begin{equation}
\label{Keq}
\mathbb{K}^\mu A^{reg}_n=\left(\sum_{i=1}^n 2 x_i^\mu
(x_i \cdot \partial_{x_i}+r(x_i) \partial_{r(x_i)})-x_i^2 \partial_{x_i^\mu}
\right) A^{reg}_n=0
\end{equation}
one may derive an equation for the finite part of the amplitude.
\footnote{As the introduction of a infrared cut-off breaks (dual)
conformal invariance, one obtains terms that depend explicitly on
$r_c$ on the right hand side of (\ref{Keq}). As we take $r_c
\rightarrow 0$, conformal invariant is recovered and such terms
vanish.} At weak coupling this equation was constructed in
\cite{Drummond:2007au} from the analysis of the dual conformal
properties of Wilson loops in dimensional regularization. Its strong
coupling counterpart was constructed in\cite{Komargodski:2008wa} using
the strong coupling version of dimensional regularization
discussed in section \ref{Dimregstrong}.

Assuming that dual conformal symmetry is present beyond the strong
coupling limit, a similar argument can be extended to all values
of the coupling, {\it e.g.} by using as regulator an energy scale
$\mu(x)$ and assuming that the amplitude has divergences which
depend only on $\mu(x)$ at the cusps (or assuming that special
conformal transformations annihilate the extra pieces coming from
the edges, as happened at strong coupling.) In the next section it
will be discussed how the same Ward identity arises when studying
the conformal properties of cusped light-like Wilson loops.

It turns out, that for the case of $n=4$ and $n=5$, this equation
fixes uniquely the form of the finite piece, to be the one in the
 BDS conjecture. At this point we do not know if the dual conformal symmetry
 is an exact property of planar amplitudes. We do know, however, that it is a
 symmetry of all the weak and strong coupling computations that have been
 done so far. If we assume that it is a symmetry, then we conclude that
the BDS conjecture for four and five gluons is correct.

\subsection{Other processes}

\subsubsection{Processes involving asymptotic gluons and local operators}
%
%

It is natural to extend the discussion in section
\ref{generalpresc} to cover the decay of fields $\phi_i$ which
couple to the ${\cal N}=4$ SYM fields through some gauge-invariant
operators \footnote{Such couplings are toy models for effective
interactions arising from integrating out heavy fields.} \be {\cal
L}=L_{YM}+\phi_i O_i~~. \label{deform} \ee While these fields may
be dynamical, we will not be interested in their other
interactions focusing only on processes containing a single field
$\phi_i$.\footnote{This guarantees that the deformation
(\ref{deform}) can be treated at the linearized level and thus it
does not lead to any modifications to the $AdS_5\times S^5$
geometry.}  We have in mind processes similar to the ones that
arise when we consider $e^+ e^- \to \gamma \to $jets
\footnote{Other interesting processes are the electron-quark
elastic scattering or the Higgs boson decaying into two gluons,
the later described by the Sudakov form factor.}. In this case, we
can analyze the process to lowest order in the electro-magnetic
coupling constant $\alpha_{em}$ but to all orders in the strong
coupling constant $\alpha_{s}$ (taken here to be the 't~Hooft
coupling) by noticing that the photon couples to the
electromagnetic current of QCD and this in turn can produce
various final states. Thus the final hadronic state is produced by
acting with the QCD electromagnetic current on the vacuum. To
lowest order in $\alpha_{s}$ the state is, of course,
a quark-antiquark pair.

 We now want to consider analogous processes in ${\cal N}=4$ super
 Yang Mills at strong coupling in the planar approximation. Thus we
 consider a process where we add a local operator to the theory and we
 produce gluons. The local operator is a single trace operator with
 given momentum
\begin{equation}
{\cal O}(q) = \int d^4x e^{ i q\cdot x} {\cal O}(x)
\end{equation}
We can consider any operator of the theory. Concrete examples are the
stress tensor, the R-symmetry currents, etc.

We are interested in exclusive final states consisting of individual
gluons, or other members of the ${\cal N}=4$ supermultiplet. From now
on the word ``gluon'' will mean any element of the supermultiplet: a
gluon, fermion or scalar, all in the adjoint representation. The
asymptotic states for these colored objects are well defined
only in the presence of an infrared regulator. The simplest one is dimensional
regularization, which consists in going to $4-2\epsilon$
dimensions. Then the theory is free in the infrared (\ref{Neq4beta})
and gluons are good asymptotic states. On the gravity side this can be
done by considering the near horizon metrics of D-$p$ branes with
$p=3-2\epsilon$, as explained in sec. \ref{Dimregstrong}.

 Once we regularize, we have a worldsheet whose boundary conditions in
 the far past or future are set at $z\sim
\infty$, where the asymptotic gluons live, and the operator
conditions are set at the boundary of $AdS_5$,  $z \sim 0$, in
(\ref{ads5met}) . In the T-dual  coordinates the asymptotic states
carry winding number which is proportional to the momentum. The
gluon final states are represented as before by considering a
sequence of light-like lines at $r=0$. Each light-like segment
joints two points separated by  $ 2 \pi k^\mu_i$. As opposed to
the situation considered in \ref{generalpresc}, this sequence is
not closed. In fact we have $\sum_{i=1}^n k_i^\mu = q^\mu$ where
$q^\mu$ is the momentum of the operator. It is convenient to
formally think of the coordinate along $q^\mu$ as compact and to
consider a closed string as winding that coordinate. This is
equivalent to saying that we consider an infinite periodic
superposition of the set of momenta $\{ k_1,k_2, \cdots , k_n \}$.

We now should give a prescription for the operator. An operator
insertion  leads to a string that goes to the $AdS_5$ boundary,
$z=0$ in the coordinates (\ref{ads5met}).
 This implies that it should
go to $r=\infty$ in the dual coordinates.
 Thus we consider a string stretching along
the direction $q^\mu$ that goes to $r=\infty$.

As a simple example, consider a two gluon
 state and an operator insertion.
 The two gluon momenta obey  $k^\mu_1 + k^\mu_2 = q^\mu$.
Let us consider the case where the momentum $q^\mu$ is spacelike
and $k_1$ is incoming and $k_2$ outgoing. By performing a boost we
can choose the momenta as
\begin{equation}
\label{momconf} 2\pi
k_1^\mu=(\kappa/2,\kappa/2,0),\hspace{0.3in}2\pi
k_2^\mu=(-\kappa/2,\kappa/2,0),\hspace{0.3in}2\pi
q^\mu=(0,\kappa,0)
\end{equation}
It is convenient to view the direction $y^1$ as a compact
direction with period $y^1  \sim y^1 +  \kappa$ so that the total
winding number of the string corresponds to an allowed closed
string. This closed string has to end on the boundary of the
original $AdS_5$ space (\ref{ads5met}) at $z=0$. In terms of the
dual metric (\ref{dualmetric}) it should go to $r=\infty$.

\begin{figure}[ht]
\centering
\includegraphics[scale=0.8]{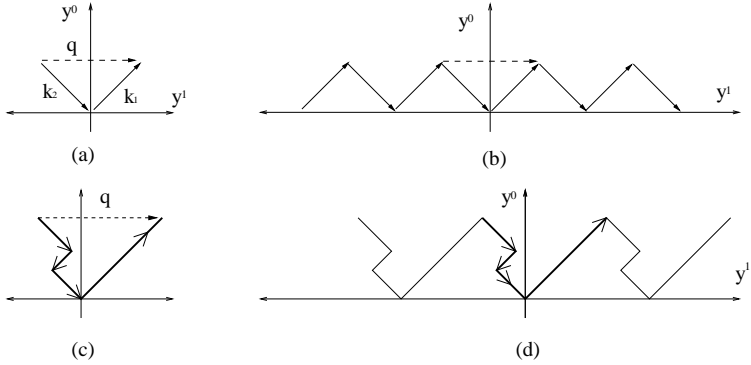}
\caption{In (a) we consider the configuration of light-like lines
corresponding to the initial and final state gluons under
consideration. In (b) we consider an infinite repetition of the
configuration. In (c) a more general configuration with four
gluons is considered and in (d) we draw the corresponding periodic
version.} \label{manygluons}
\end{figure}

In order to find the surface it is convenient to consider an infinite
repetition of these momenta, which are following a zig-zag path in the
$y^0,y^1$ plane as shown in figure (\ref{manygluons}).
%

\begin{figure}[ht]
\centering
\includegraphics[scale=0.65]{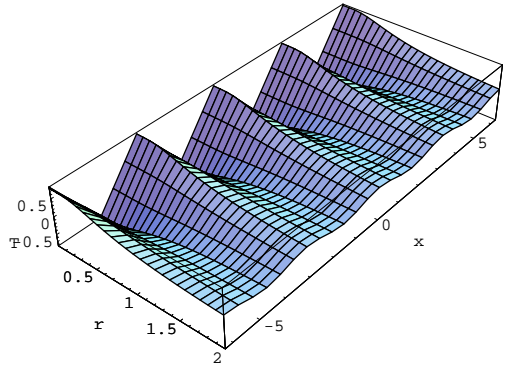}
\caption{Approximate form of the solution. Ar $r=0$ the surface
ends on a zig-zag, while for large $r$, $t$ decays exponentially.}
\label{approxzigzag}
\end{figure}

We  look for a worldsheet which is extended in the radial $AdS_5$
direction, from $r=0$, where it ends on the  contour displayed in
picture (b), and extends all the way to $r\to \infty$. As we go to
large $r$ the surface is extended in the $y^1$ spatial direction
but is localized in time. See figure (\ref{approxzigzag}) for a
picture of the expected surface.

The amplitude is  given by computing the area over one period of
the resulting surface, since this is the region of the worldsheet that corresponds to the scattering process.  Let us point out some features of the
solution, which we have not found explicitly.
  First, one can write the Nambu-Goto action
by choosing $r$ and $y$ as worldsheet coordinates so that $t(r,y)$
is the unknown function. The action is

\begin{equation}
\label{actng} i S=-\frac{R^2}{2\pi \alpha'}\int dy dr
\frac{\sqrt{1-(\partial_y t)^2-(\partial_r t)^2}}{r^2}
\end{equation}
The equations of motion coming from this action should be supplemented
with the appropriate boundary conditions
\begin{equation}
\label{boundcon}
t(r=\infty,y)=0,\hspace{0.2in}t(r=0,y)=y~~\mbox{for}~ |y|\leq
\frac{\kappa}{4},\hspace{0.2in}t(r=0,y)=\frac{\kappa}{2}-y~~\mbox{for}~\frac{\kappa}{4}
\leq y \leq \frac{3}{4}\kappa
\end{equation}
and extended in a periodic way outside this range, $t(r,y + \kappa
)=t(r,y)$. This equation could, in principle, be solved numerically.
We expect that for distances much bigger than the size of the zig-zag,
{\it i.e.} $r \gg \kappa $, $t(r,y)$ is very small and satisfies a
linear equation obtained by expanding (\ref{actng}) for small $t$.
Expanding $t$ in Fourier modes, $t(r,y)=\sum_n t_{n}(r)e^{i k_n y}$,
with $k_n=2 \pi n/\kappa$, we obtain the following equation for
$t_{n}(r)$
\begin{equation}
 -k_n^2 t_{n}(r)+r^2 \partial_r
\left[ \frac{1}{r^2} \partial_r t_{n}(r)\right]=0
~~~~\rightarrow~~~~
t_{n}(r)= c_{n}  e^{-k_n r}(k_n r+1)
\end{equation}
where we have kept only the decaying solution (for positive
$k_n$). Note that due to the exponential decay, already when $r$ is a
few times bigger than $|\kappa|$, the above solution will be a good
approximation. The coefficients $c_{n}$ are determined by imposing the
boundary condition at $r=0$ (\ref{boundcon}) , but we should recall
that we cannot use the linearized equation in that region.

The problem has a scaling symmetry that implies that we can scale
out $\kappa$ so that the solution is
\begin{equation}
t(r,y)=\kappa \hat{t}(\frac{r}{\kappa},\frac{y}{\kappa})~~.
\end{equation}
It is then easy to see, that the value of the classical action
(\ref{actng}) on this solution is formally independent of the scale
$\kappa$, as expected from scaling symmetry. However, since there is a
divergence, an explicit $\kappa$ dependence is introduced when we
subtract the divergence. Let us understand the divergences. Let us
first consider the large $r$ region.  The integral in the region of
large $r$ converges since $t\to 0$ so that we are simply integrating
$dr/r^2$. This might seem a bit surprising since we expected to obtain
terms of the form $\log r$ that are related to the anomalous dimension
of the operator.  Notice, however, that a logarithmic term in the
classical area would have implied an anomalous dimension of order
$\sqrt{\lambda}$. Thus, the boundary conditions we considered
correspond to operators whose anomalous dimension vanishes at this
order.  For a protected operator such as the stress tensor, whose
dimension equal to four, this is indeed the case. We expect to obtain
logarithmic terms when we go to higher order in the $1/\sqrt{\lambda}$
expansion.

We can now consider the small $r$ region. The analysis of this
region is the same as the analysis in the small $r$ region for the
gluon scattering amplitudes discussed at the beginning of this
section. One can dimensionally regularize the problem by going to
$d = 4 - 2 \epsilon$ dimensions. Then the Lagrangian becomes
\begin{equation}\label{newlag} L = { \sqrt{ \lambda} \over 2 \pi} c_\epsilon
\kappa^{-\epsilon} \int \frac{d \hat r d \hat y}
{ {\hat r}^{\epsilon} }
{\cal L}_0[\hat t(\hat y, \hat r)]
\end{equation}
where we have rescaled all variables so that the only dependence on
$\kappa$ is in the overall factor. In (\ref{newlag}) $c_\epsilon$ is a
function of only $\epsilon$.  The divergences arise from the region
near the cusps connecting the momenta of two adjacent gluons and they
can be computed using the single cusp solution considered in section
\ref{singlecusp}. The value of the action is given by integrating only
over one period in $y$. It evaluates to a function of the form
\begin{equation} \label{actmeson}i S =
-\frac{ \sqrt{ \lambda}}{2 \pi}
{ \mu^\epsilon \over
( 2 \pi \kappa) ^{\epsilon} }  \left[ 2 \left({ 1  \over
\epsilon^2 } + { 1- \log 2   \over 2  \epsilon} \right)  + C
\right] \end{equation}
The coefficients of the divergent terms are locally determined and are
the same as in \ref{Dimregstrong}, so that we would only need the
solution to compute the constant $C$. For the simplest case of two
gluons, the solution does not depend on any kinematical variable. As
we consider configurations with more gluons the solution, and the
value of the amplitude, will start depending on the kinematic
invariants.

\subsubsection{Processes involving a mesonic operator and final
quark and antiquarks}

In this subsection we consider a small variant of the configuration
considered above. We consider a large $N$ theory with flavors and we
insert a mesonic operator, which contains a quark and an antiquark
field.\footnote{Here and elsewhere quarks denote fields in the
fundamental representation of the gauge group. For $SU(N)$ their color
charge if $C_F=(N^2-1)/2N$ which, in the planar limit, is half of the
charge of an adjoint field $C_A=N$.}
Flavors correspond to adding D-branes in the bulk
\cite{Aharony:1998xz,{Karch:2002sh},Kruczenski:2003be}.
A mesonic operator corresponds
to an open string mode on the D-brane that is extended over $AdS_5$.
For example, we could consider the insertion of a flavor symmetry
current which couples to a $q, ~\bar q$ pair. This is analogous to the
electromagnetic current in QCD. Other amplitudes involving quarks have
been considered at strong coupling in
\cite{Komargodski:2007er,{McGreevy:2007kt}} and will be reviewed in
the next subsection.

\begin{figure}[ht]
\centering
\includegraphics[scale=0.4]{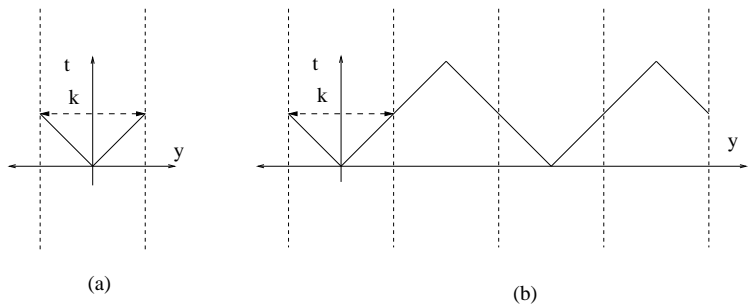}
\caption{Once we extend the solution outside the strip as shown in
the figure, it reduces to the zig-zag solution, with twice the
period.} \label{fqq}
\end{figure}

In the presence of the infrared regulator, the quarks correspond to
open strings that are attached to the D-brane and sit at $z\sim
\infty$ or $r\sim 0$. The discussion is very similar to the one for
closed strings.  One difference is that now we do not require the
configuration to be periodic. However, since we obey Neumann boundary
conditions on the boundary of the open string, which translate to
Dirichlet boundary conditions in the T-dual variables, we find that
the solution can be extended outside the strip into a periodic
function with a period which is twice the original width of the strip,
see figure (\ref{fqq}) .

Thus, if we consider a configuration with momenta as in
(\ref{momconf}), the solution is simply given by
\begin{equation}
t=2\kappa \hat{t}(\frac{r}{2\kappa},\frac{y}{2\kappa})
\end{equation}
where $\hat{t}(\hat{r},\hat{y})=\hat{t}(\hat{r},\hat{y}+1)$ is the
rescaled solution with period one. The action is simply half of
the action (\ref{actmeson}) but with the replacement $\kappa
\rightarrow 2 \kappa$. After we re-express it in terms of $\kappa$
we obtain
\begin{equation}
i S = - { \sqrt{\lambda} \over 2 \pi } { \mu^\epsilon \over ( 2
\pi \kappa )^\epsilon } \left[   { 1 \over \epsilon^2 } + { 1- 3
\log 2   \over  2  \epsilon}    + \left\{ { C \over 2 } - { \log 2
\over 2} + (\log 2 )^2 \right\} \right]
\end{equation}
Thus we see that function $g(\lambda) $ which determines the
subleading infrared divergences is different for a gluon than a quark.
Namely, we have
\begin{equation}
g_{gluon}(\lambda ) = { \sqrt{\lambda} \over 2 \pi } (1 - \log 2)
~,~~~~~~~ g_{quark} (\lambda ) ={ \sqrt{\lambda} \over 2 \pi } (1
- 3 \log 2) \end{equation}
where $g_{gluon}$ was computed in the previous subsection. Due to the
collinear nature of the function $g$, in the case that we have a cusp
that joins a quark and a gluon we expect to have the average of the
above two formulas \footnote{In another words, $g$ should measure the
contribution coming from the region closed to the edges joined by the
cusp.}. Similarly, we can consider asymptotic states corresponding to
a quark and a antiquark plus extra gluons.

\subsubsection{Processes involving quarks and gluons}

Scattering of quarks and gluons were considered in
\cite{Komargodski:2007er} and \cite{McGreevy:2007kt}. As already
mentioned, we can have fundamental matter by adding extra D-branes
in the bulk. More precisely, we add $N_f$ D7-branes that wrap an
$S^3\subset S^5$ in the $AdS_5 \times S^5$. Such configuration
preserves ${\cal N}=2$ SUSY. As long as $N_f \ll N$ the back
reaction of the D7-branes can be ignored and they may be treated
as probes.

\begin{figure}[ht]
\centering
\includegraphics[scale=0.5]{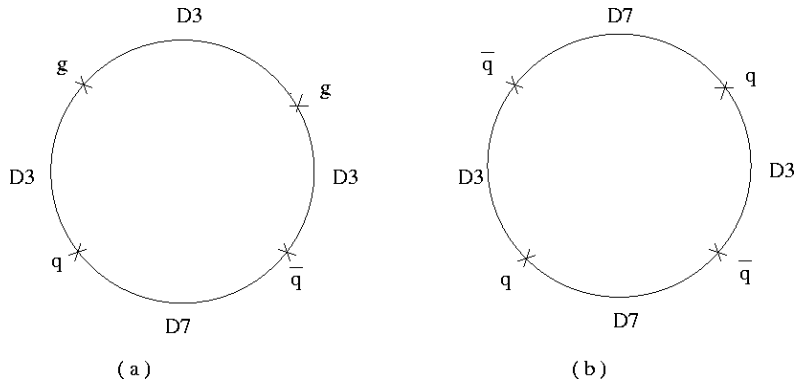}
\caption{worldsheet with the topology of a disk with vertex
operator insertions, corresponding to the $\bar{q}ggq$ amplitude (a)
and $\bar{q}q\bar{q}q$ amplitude (b)} \label{quarksgluons}
\end{figure}

The quarks will then be given by open strings between the infrared
regulator D3-brane and the D7-branes. \footnote{This is consistent
with the fact that quarks transform in the fundamental
representation of $SU(N)$, while gluons transform in the adjoint representation.}
Scattering amplitudes can then be computed by considering a
worldsheet with the topology of a disk with vertex operator
insertions, corresponding to open strings states belonging either
to the $(3,3)$ or $(3,7)$ sector; the former are gluons and the
latter are quarks. For instance, in figure \ref{quarksgluons} we
can see amplitudes corresponding to the scattering of quarks and
gluons (a) and only quarks (b).

\begin{figure}[ht]
\centering
\includegraphics[scale=0.6]{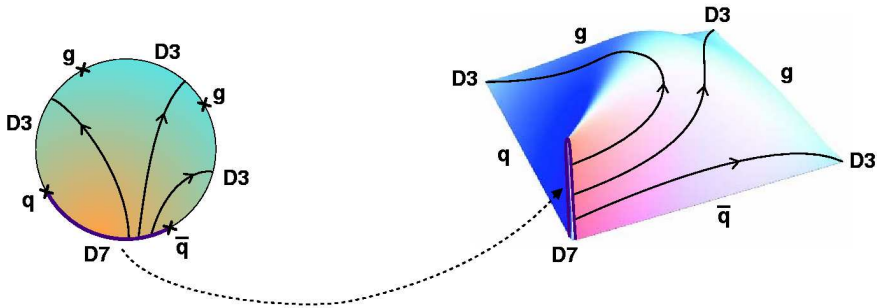}
\caption{worldsheet corresponding to $\bar{q}ggq$ scattering and
its dual version.} \label{D7-cusp}
\end{figure}

As before, it is convenient to T-dualize along the directions
longitudinal to the regulator D3-brane. Such directions are also
shared by the D7-branes, so the $x_{3,1}$ coordinates satisfy
Neumann boundary conditions on the boundary. Hence, after
performing the T-duality, the projection of the worldsheet to the
boundary of $AdS_5$ is again a polygon formed by light-like edges
given by the momenta of the particles undergoing the scattering. A
crucial difference, however, is the following: the components of
the boundary ending on the regulator D3-brane have Dirichlet
boundary conditions on the radial direction $r$. When the boundary
ends on the D7-branes, on the other hand, it satisfies Neumann
boundary conditions on the radial direction, so it can extend into
the bulk. This implies, that after T-duality, at the location of a
cusp between two quarks (or quark-antiquark), the worldsheet can
extend into the bulk, folding back into itself, see figure
\ref{D7-cusp} \footnote{We thank the authors of
\cite{McGreevy:2007kt} for providing us with this figure.}.

In \cite{McGreevy:2007kt} it was shown that amplitudes for quarks
and gluons at strong coupling can be constructed from special
gluon amplitudes. For instance, based on symmetry arguments, the
authors have argued that the minimal surface for the $({\bar q}gg
q)$ amplitude with momenta $k_1,k_2,k_3$ and $k_4$, is half of the
surface corresponding to the scattering of six gluons with momenta
$2k_1,k_2,k_3,2k_4,k_3$ and $k_2$. \footnote{More precisely, it
can be seen that the polygon corresponding to the scattering of
six gluons crosses itself at the midpoints of the lines associated
with $k_1$ and $k_4$, which we can fix at $x=0$. Then, the
worldsheet maps into itself under $R\Omega$, where $R$ denotes
reflection through the intersection point and $\Omega$ is the
worldsheet parity. The part of the worldsheet at $x=0$ is
invariant under such transformation and it can be shown that the
radial coordinate satisfies Neumann boundary conditions there.}
Using the known expression for the divergent piece of such
amplitude one obtains
\begin{equation}
\label{qggq} \log{A_{\bar{q}ggq}}_{div}=
-\frac{f(\lambda)}{8}\left(\log^2\left(\frac{\mu^2}{-2s} \right)
+\frac{1}{2}\log^2\left(\frac{\mu^2}{-t} \right)\right)
-\frac{g(\lambda)}{2}\left(\log \left(\frac{\mu^2}{-2s}\right)
+\frac{1}{2}\log \left(\frac{\mu^2}{-t} \right) \right)
\end{equation}
As the finite piece of the six gluons amplitude is at present unknown,
only the divergent part of the above amplitude can be computed. By
looking at the relevant diagrams in the double line notation,
\cite{McGreevy:2007kt} argued that the exchange of soft gluons
between quark and antiquark in the above configuration is suppressed
by a factor $1/N$, and so it vanish in the large $N$ limit. As a
result it is appropriate to rewrite the amplitude (\ref{qggq}) in the
following way
\begin{equation}
\log{A_{\bar{q}ggq}}_{div}=-\frac{f(\lambda)}{8}
\left(\log^2\left(\frac{\mu^2}{-s}
\right)+\frac{1}{2}\log^2\left(\frac{\mu^2}{-t} \right)
\right)-\frac{g(\lambda)}{4}\log \left(\frac{\mu^2}{-t}
\right)-\frac{g_{qg}(\lambda)}{2}\log \left(\frac{\mu^2}{-s}
\right)~~,
\end{equation}
with $g(\lambda)$ the collinear anomalous dimension between two
gluons and $g_{qg}(\lambda)$ the one between a gluon and a quark.
Using the known strong coupling expressions for the collinear
anomalous dimension for gluons and the cusp anomalous dimension we
obtain
\begin{equation}
g_{qg}(\lambda)=g_{gluon}(\lambda)-\frac{f(\lambda)}{2}\log{2}
=\frac{\sqrt{\lambda}}{2\pi}(1-2\log{2})~~.
\end{equation}
This result realizes the expectation, discussed in the previous
section, that the collinear anomalous dimension for a pair quark-gluon
is the average of the collinear anomalous dimension for two gluons and
the one for two quarks.

\subsection{Further generalizations}
\label{generalizations}

In the following we briefly describe many generalizations of the
results presented in this section. We refer the reader to the original
literature for the details.

According to the discussion in section \ref{generalpresc}, the
calculation of the scattering amplitudes in ${\cal N}=4$ SYM theory at
strong coupling is equivalent to finding the regularized area of a
minimal surface ending on a special polygon with light-like edges on
the boundary of anti-de-Sitter space.
While it may be possible to find their areas without actually knowing
the solution for minimal surfaces, the most direct approach requires
first constructing the classical solutions to the worldsheet
equations of motion and then evaluating their areas using perhaps one
of the regulators discussed in previous sections.
Explicit results for the strong coupling limit of scattering
amplitudes may shed light on the resummation of such amplitudes
for general number of external legs.
%

In section \ref{four_cusp_solution} we have described the solution
corresponding to the case of four light-like edges. Once a solution is
found, additional ones may be constructed through $SO(2,4)$
transformations.\footnote{Such solutions were explicitly constructed
in \cite{Ryang:2007bc}.} It should be mentioned that, while $SO(2,4)$
is a symmetry of the action in the absence of the regulator, it is no
longer so once a regulator is introduced. The value of the regularized
area will depend on whether the regulator is introduced before or
after the transformation is performed.  The construction of minimal
surfaces ending on polygons with more sides, however, appear to be a
much harder problem. Partial progress towards such a goal has appeared
in the literature:

\begin{itemize}

\item
The authors of \cite{Jevicki:2007pk,Jevicki:2007aa} showed how to apply the
dressing method in order to construct new classical solutions for
Euclidean worldsheets in $AdS$. While these solutions may be useful
in order to study certain Wilson loops by means of the AdS/CFT
duality, their boundary conditions do not correspond to the boundary
conditions relevant for studying scattering amplitudes.

\item
Some partial progress in constructing approximate solutions with
light-like polygonal boundary conditions was presented in
\cite{Itoyama:2007ue,Itoyama:2007fs,Itoyama:2008je}. Assuming the
BDS ansatz holds and making use of the demonstrated equivalence of
one-loop MHV amplitudes and light-like Wilson loops
\cite{Drummond:2007aua,BHT} which will be discussed in detail in
section \ref{Amp_vs_WL}, the prescription for computing scattering
amplitudes at strong coupling can be written in purely geometrical
terms, of the schematic form
\begin{equation}
\oint_\Pi \frac{dx^\mu dx'_\mu}{(x-x')^2}+\frac{4C(a)}{\gamma_K(a)}
=A_\Pi^{min}~~.
\end{equation}
Here $C(a)$ is a function solely of the coupling constant, $\Pi$ is a
light-like polygon, the left hand side of the above equation is, up to
a factor of $\gamma_K/4$, the one loop MHV amplitude associated to
$\Pi$ and $A_\Pi^{min}$ is the area of the minimal surface ending on
$\Pi$.
Conversely, departures from the above relation imply departures
from the BDS ansatz at strong coupling. Such departures can be studied
for some particular examples by considering approximate solutions.
While such approach seems interesting, unfortunately it has not lead to
new solutions yet.

\item
Solutions corresponding to the scattering of six and eight gluons were
discussed in \cite{Astefanesei:2007bk}. In particular, the authors
construct new solutions by cutting and gluing the known
solutions. However, these solutions appear to satisfy extra boundary
conditions and are not clearly related to the relevant solutions.

\item
In \cite{Dobashi:2008ia}, a method was developed in order to construct
numerically new solutions relevant to scattering amplitudes. The
method consists in solving the discretized equations of motion with a
given boundary condition. A difficulty in the numerical evaluation of
the area, is the need of a regulator. This causes large errors in the
computation of the finite piece of the area.  However, given the
difficulties in constructing analytical solutions, a numerical
approach may be appropriate.  Besides, the study of numerical
solutions can give a hint about their analytical properties.

\end{itemize}

The $S^5$ part of the bulk geometry as well as the fermionic
sector of the theory did not play any role in the strong coupling
arguments relating scattering amplitudes and cusped light-like
Wilson loops described in section \ref{generalpresc}. This
suggests, at least in the strong coupling limit, a certain degree
of universality of the results, extending beyond the planar ${\cal
N}=4$ SYM theory. The string theory analysis was carried out for
several other theories, for some of which the analog of the BDS
ansatz is also available.

\begin{itemize}
\item
In \cite{Nastase_finite_T} the prescription was extended to the case
of ${\cal N}=4$ SYM at finite temperature. In this case, the gravity
dual is described by the black hole $AdS$ metric, of the form
\begin{equation}
ds^2=\frac{R^2}{z^2}\left(-h(z) dy_0^2
+dy_3^2+\frac{dz^2}{h(z)}\right),
\hspace{0.3in}h(z)=1-\frac{r_0^4}{R^8}z^4
\end{equation}
with the temperature $T=\pi \frac{r_0}{R^2}$. After
T-dualizing and performing the change of coordinates
$r=\frac{R^2}{z}$, we are led to the action
\begin{equation}
ds^2=\frac{R^2}{r^2}\left(-\frac{dx_0^2}{h(z)}+dx_3^2
+\frac{dr^2}{h(z)}\right),\hspace{0.3in}h(z)=1-\frac{r_0^4}{r^4}
\end{equation}
The authors define gluon scattering amplitude at strong coupling and
finite temperature by a light-like Wilson loop living at the
horizon $r=r_0$ of this dual metric. Unlike the zero temperature
case, now the $T$-dual metric is different from the original
metric.
As a consequence, this is best interpreted as a way of evaluating
scattering amplitudes at finite temperature rather than as a
duality between amplitudes and Wilson loops.
Quite interestingly, both scattering amplitudes and Wilson loops in
the original geometry are related to observables of
the theory: the usual Wilson loop to the jet quenching parameter
and the scattering amplitude to the viscosity coefficient.

\item
In \cite{{OzTY}} scattering amplitudes in planar $\beta$-deformed
${\cal N}=4$ SYM were considered.\footnote{The $\beta$-deformation is
an exactly marginal deformation of ${\cal N}=4$ SYM theory. The ${\cal
N}=1$ supersymmetric version, due to Leigh and Strassler
\cite{LeighStrassler}, amounts to adding a completely symmetric term
in the superpotential. An alternative presentation of the theory is a
noncommutative deformation of ${\cal N}=4$ SYM in which the
noncommutative product is based on the R-charge of fields --
$\phi_i*\phi_j=e^{i\beta_{ab}q_i^aq_j^b} \phi_i\phi_j$ where $q_i^a$
is the charge vector of the field $\phi_i$ and $\beta_{ab}$ is a real
antisymmetric matrix. This formulation allows
for simple non-supersymmetric generalizations
\cite{frolov_beta}. The gravity dual of the ${\cal N}=1$ theory was
constructed in \cite{Lunin_Maldacena} while that of the
nonsupersymmetric deformations in \cite{frolov_beta}.} In particular,
the authors have considered deformations that break supersymmetry down
to ${\cal N}=1$ and ${\cal N}=0$.
It was known \cite{Khoze_beta} that, for real deformation parameter
$\beta$, that planar scattering amplitudes in these theories are the
same as in ${\cal N}=4$ SYM, order by order in perturbation theory.
In \cite{{OzTY}} it was found that the same conclusion holds in the
strong coupling expansion.
This is no longer true for complex deformation parameters; starting at
five-loop order, the scattering amplitudes in the deformed theory are
different \cite{Khoze_beta} from those of ${\cal N}=4$ SYM.

The planar scattering amplitudes of orbifolds of ${\cal N}=4$ SYM
theory are also known \cite{BerJoh} to be identical to those of the
parent theory, up to a trivial rescaling of the coupling constant.
The strong coupling expansion enjoys similar properties.

\item Finally, ${\cal N}=2$ theories with matter in the fundamental
representation were considered by \cite{Komargodski:2007er,
McGreevy:2007kt} and were discussed in some detail in the previous
subsection.

\end{itemize}

A further question, discussed in some detail in
\cite{Kruczenski:2007cy},
is whether the agreement between the BDS ansatz and the result of the
AdS/CFT calculation for the four-gluon scattering amplitudes
extends beyond the leading order in the strong coupling expansion.
To this end it is necessary to evaluate the quantum corrections to the
regularized area of the minimal surface, i.e. the corrections to the
worldsheet partition function in the presence of the minimal surface
background. The strategy is clear and has been extensively applied to
the calculation of the energy of extended string solutions: one simply
expands the relevant worldsheet action \cite{Kallosh:1998ji} around
the minimal surface and integrates out the fluctuations. For example,
the leading $\frac{1}{\sqrt{\lambda}}$ correction is given by the
logarithm of the ratio of determinants of the bosonic and fermionic
fluctuations.
This analysis was carried out \cite{Kruczenski:2007cy}, with some
negative conclusions: (a) it is not clear how to construct the
complete worldsheet action for D$(3-2\epsilon)$ branes; (b)
perturbative inclusion of the regulator both at the level of the
worldsheet action as well as in the minimal surface leads to a
divergent answer at one-loop level. Indeed, since the worldsheet
theory is conformal (in a two-dimensional sense),
the nontrivial conformal factor in the induced worldsheet metric,
which is the basis of the regularization discussed in section
\ref{Dimregstrong}, drops out of the calculation of one-loop
corrections. Thus, it appears to be necessary to either have an exact
minimal surface in the regularized geometry or to switch to a
different regulator.
The evaluation of $\frac{1}{\sqrt{\lambda}}$ corrections to the area of the
minimal surface describing the scattering of four gluons at string
coupling remains an interesting open question.


\section{Scattering Amplitudes vs. Wilson loops at
weak coupling \label{Amp_vs_WL}}

The discussion in the previous section connects (to leading order in
the strong coupling expansion) two apparently different quantities:
scattering amplitudes and the expectation value of a special type of
Wilson loops. Without additional specifications this relation is
restricted to MHV amplitudes which, as discussed in section
\ref{weak_coupling}, are completely defined by a single function ${\cal M}_n$.
To be specific, an MHV amplitude $A_{MHV}(k_1,...k_n)$ is
conjectured to be equal to the expectation value of a Wilson loop
$W(x_1,...,x_n)$ constructed from noncollinear light-like segments
connecting points $x_i$ defined by $k_i=x_i-x_{i+1}$, as in the
equation (\ref{winding}).\footnote{The apparent factor of $2\pi$
difference between $k_i=x_i-x_{i+1}$ and equation (\ref{winding})
may be explained away by invoking the invariance of the
expectation value of the Wilson loop under scale transformations.}

\subsection{The general statement}

While the arguments discussed in the previous section are phrased
in the string theory dual to ${\cal N}=4$ SYM theory, the final statement
%
%
appears to be independent of the coupling constant. Despite the fact
that neither scattering amplitudes nor this particular type of Wilson
loops are BPS quantities, one may conjecture \cite{Drummond:2007aua,
BHT} a weak coupling relation of a similar type as at strong coupling:
\be
\ln(1+\sum_{l=1}^\infty a^l{\cal M}_n^{(l)})=
\ln(1+\sum_{l=1}^\infty \WLa^lW_n^{(l)})+{\cal O}(\epsilon)~~{\rm
where}~~ \WLa=a|_{\epsilon=0}~~.
\label{Amp_WL_relation}
\ee
Here ${\cal M}_n^{(l)}$ is the $l$-loop
correction to the rescaled MHV amplitude (\ref{LloopMHV}) and
$W_n^{(l)}$ is the $l$-loop correction to the expectation value of the
corresponding Wilson loop.  As usual, given a contour $C_n$, a Wilson
loop is defined as
\be
\langle W_{C_n} \rangle&=&
\frac{1}{N}\langle 0 | \Tr P \exp{\left(i g \int_{C_n} d\tau
A_\mu(x(\tau))  \dot{x}^\mu(\tau)\right)} |0\rangle \cr
&\equiv&\frac{1}{N}\int {\cal D} A {\cal D} \lambda {\cal D} \phi \;
e^{i S_\epsilon} \;\Tr P \exp{\left(i g\oint_{C_n} A(x) \right)}~~,
\label{WL_general_YM}
\ee
where $A=dx^\mu A_\mu^a T^a$, $T^a$ are
the $SU(N)$ generators in the fundamental representation and
$S_\epsilon$ is the dimensionally regularized ${\cal N}=4$ SYM
action. For ``standard'' Wilson loops $C_n$ is a closed contour in
position space, parametrized by an affine parameter $\tau$:
$C_n=\{x^\mu=x^\mu(\tau),\,\tau\in[0,\,1] \}$.
In the present case, however, $C_n$ is a curve in a configuration
space defined such that the difference of the coordinates of two
points is a momentum; it is in fact the same Wilson loop described
at strong coupling. The ${\cal N}=4$ SYM is defined on the same
space.
In this section we will review this remarkable conjecture and the
evidence for it.

For ${\cal N}=4$ SYM theory one may identify the space with
coordinates $x$ with the position space the theory is originally
defined on. Expectation values of light-like Wilson loops defined on
this space are then translated to momentum space through the
identifications $2\pi k_i=x_i-x_{i+1}$ with $x_i$ labeling the cusps
of the Wilson loop. Such identifications are specific to conformal
field theories and are related to the fact that the position and
momentum space can be trivially mapped into each other. The strong
coupling arguments described in section
\ref{strong_coupling} imply however a more general relation -- that
 scattering amplitudes of a gauge theory with a string theory dual can
 be evaluated as expectation values of Wilson loops in the dual
 momentum space regardless of whether the relevant T-duality
 transformation leaves the boundary geometry invariant.
For this reason we will interpret in the following the Wilson loops
as defined directly in the configuration space with coordinates $x$
related to particle momenta by $2\pi k_i=x_i-x_{i+1}$. The
corresponding conformal group is that acting in momentum space,
i.e. the dual conformal group.

It was mentioned briefly in section \ref{weak_coupling} that
higher-loop and higher-multiplicity rescaled MHV amplitudes possess
a parity-odd component, proportional to the Levi-Civita tensor.
Such components cannot appear in a Wilson loop calculation. In all
available explicit calculation the parity-odd part of MHV
amplitudes exponentiates (following the BDS ansatz) and this drops
out of the left hand side of equation (\ref{Amp_WL_relation}).
This is consistent with the right hand side of that equation being
expressible in terms of the expectation value of a Wilson loop. It
thus follows that the equation (\ref{Amp_WL_relation}) suggests
that a odd remainder function does not exist to any loop order.

Before proceeding it is worth mentioning that
(\ref{WL_general_YM}) is the standard definition of a Wilson loop
in a generic YM theory. In ${\cal N}=4$ SYM this definition is
usually modified to include scalar fields; for a generic curve $C$
describing also some contour on $S^5$ this is \be
W_{C}=\frac{1}{N}\langle 0 | \Tr P \exp{\left(i g \int_{C} d\tau
\left[A_\mu(x(\tau))
\dot{x}^\mu(\tau)+\phi_i(x(\tau))\dot{y}^i(\tau)\right]\right)} |0\rangle
\label{general_WL} \ee
The condition that this Wilson loop is a protected operator is that
\be
{\dot x}^2={\dot y}^2~~.
\ee
As described in section \ref{strong_coupling}, $C_n$ has no component
on $S^5$, so $y=0$. Nevertheless, since it is constructed out of
light-like segments, ${\dot x}^2=0$ and thus this loop is (locally)
BPS away from the cusp. It however enjoys little or no protection from
supersymmetry; since the supersymmetry generators preserved by each
segment are different, the complete loop is not supersymmetric.

\subsection{The MHV amplitudes -- Wilson loop relation at
one-loop \label{1loop_Amp_WL}}

The perturbative evaluation of the expectation value of the Wilson
loops (\ref{WL_general_YM}) (and, generally, of any Wilson loop) is
quite straightforward: one expands the exponent while keeping track of
the path ordering and then one evaluates the expectation value in
equation (\ref{WL_general_YM}) by Wick-contracting the resulting
fields either among themselves or with fields from the ${\cal N}=4$
Lagrangian. The gluon two-point function on this configuration space
parametrized by $x$ is constructed in the standard way, by
transforming from the Fourier conjugate space. In Feynman gauge the
result reads\footnote{It is important to keep in mind that, similarly
to Feynman diagram calculations of scattering amplitudes, different
gauges lead to different forms for the two-point function and thus to
different expressions for each contribution to the expectation value
of the Wilson loop and some gauges may be more useful than others for
exposing specific properties of the final result.}
\begin{equation}
G_{\mu \nu}(x)=-\eta_{\mu \nu} \frac{\Gamma(1-\epsilon_{UV})}{4\pi^2}
              \frac{(\pi \tilde \mu^2)^{\epsilon_{UV}}}
                   {(-x^2+i \epsilon)^{1-\epsilon_{UV}}}~~.
\label{gluon_prop}
\end{equation}
The dimensional regulator $D=4-2\epsilon_{UV}$ with $\epsilon_{UV}>0$
appears due to the regularization of the action (and thus of the
Fourier transform) and is necessary because the cusps introduce
divergences. If this were a ``standard'' Wilson loop in Minkowski
space the divergences appearing due to the presence of cusps would be
due to short distance effects. Thus, the regulator is labeled as
``ultraviolet'' \cite{Drummond:2007aua} . In the present context, the
divergences are due to low energy and to low transverse momenta, so
the divergences should be labeled as ``infrared''. This
re-identification induces nontrivial transformations on $\epsilon$ as
well as on the unit mass $\mu$ of dimensional regularization
\cite{Drummond:2007aua}.

Using this two-point function, to lowest order in perturbation theory
the expectation value is given by
\begin{equation}
\langle W_{C_n} \rangle=1+\frac{1}{2}(i g)^2 C_F \int_{C_n} d\tau \int_{C_n}
d\tilde{\tau} \; \dot{x}^\mu(\tau)\dot{x}^\nu(\tilde{\tau})
G_{\mu \nu}(x(\tau)-x(\tilde{\tau}))+{\cal O}(g^4)~~.
\label{Wn_1loop}
\end{equation}
Here $C_F=(N^2-1)/(2N)$ is the quadratic Casimir of the fundamental
representation of $SU(N)$ and $G_{\mu \nu}$ is the free gluon
propagator in configuration space (\ref{gluon_prop}).\footnote{While
this is not very relevant for the expression above, it is important to
enforce the ordering of the gluons along the Wilson loop. For example,
the next term in the expansion (\ref{Wn_1loop}) contributing to the
two-loop expectation value of the Wilson loop, contains a contribution
with two gluons attached to one segment and a third attached to a
different one. To enforce the path ordering, the two gluons attached
to the same segment are integrated with the constraint
$\tau_1>\tau_2$. This will become clearer later.} We will now discuss
in some detail this expectation value and its comparison with the
one-loop MHV amplitudes discussed in section \ref{weak_coupling}.

\subsubsection{Four-sided polygon \label{four_poly}}

The simplest example of Wilson loop of the type described in
section \ref{strong_coupling} is a four-sided polygon. The contour
$C_4$ consists of four light-like segments $C_4={\cal C}_1 \cup
{\cal C}_2 \cup {\cal C}_3 \cup {\cal C}_4$; They are parametrized
in terms of affine parameters $\tau_i$ as
\be
{\cal C}_i=\{x^\mu(\tau_i)=x_i^\mu+\tau_i(x_{i+1}^\mu-x_i^\mu)
=x_i^\mu-\tau_i k_i^\mu\}~,~~ \tau_i ~\in ~[0,1]~~.
\ee
The calculation of the integral (\ref{Wn_1loop}) breaks up in several
contributions $I_{ij}$ depending on the segments connected by the
gluon propagator:\\ ~~ $(a)$ both ends of the gluon propagator are
attached to the same segment, denoted by $I_{i\,i}$\\[0pt] ~~ $(b)$
the ends are attached to adjacent segments, denoted by
$I_{i\,i+1}$\\[0pt] ~~ $(c)$ the ends are attached to different
non-adjacent segments, denoted by $I_{i\,i+2}$\\[0pt]

\begin{figure}[ht]
\centering
\includegraphics[scale=1.4]{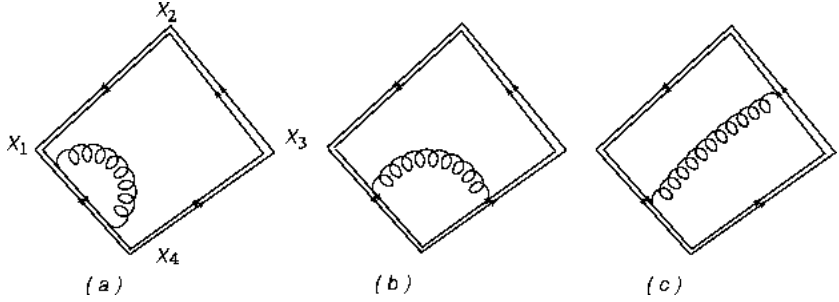}
\caption{Different type of diagrams contributing to the one loop
computation of the expectation value of the four cusps Wilson
loop.} \label{fig:fourpolydiag}
\end{figure}

\noindent The integrals just introduced are quite
straightforward to construct explicitly; one simply replaces in
(\ref{Wn_1loop}) the parametrization of the various segments
constructing $C_4$. The generic integral $I_{ij}$ is
\begin{equation}
I_{ij}=-\int_0^1 d\tau_i \int_0^1 d\tau_j \frac{(k_i \cdot k_j)
\Gamma(1-\epsilon_{UV}) (\pi \tilde \mu^2)^{\epsilon_{UV}}}
{\left(-(x_i-x_j-\tau_i k_i+\tau_j k_j)^2
                  +i \epsilon \right)^{1-\epsilon_{UV}}}~~.
\label{integral_Iij}
\end{equation}
These integrals can be conveniently represented as Feynman diagrams
(see fig.(\ref{fig:fourpolydiag})).  The one-loop correction to the
expectation value of $W_4$ may be isolated by simply taking the
logarithm of (\ref{Wn_1loop}) and, in terms of the integrals above it
reads
\begin{equation}
\ln W_4=-\frac{g^2 C_F}{4\pi^2}\sum_{1 \leq j \leq k \leq 4}I_{ij}~~.
\label{WL4_assembly}
\end{equation}

It is not hard to evaluate the integrals appearing in the three
cases above \cite{Drummond:2007aua}:

\begin{itemize}

\item Integrals of type $(a)$ vanish identically, being proportional to the
square of the corresponding momentum. Consider $I_{11}$:
\be
I_{11}\propto \int_{0}^1 d\tau_1\int _{\tau_1}^1 d\tau_2
\frac{(k_1^2)^{\epsilon_{UV}}}
{((\tau_1-\tau_2)^2+i\epsilon)^{1-\epsilon_{UV}}}=0
\label{Ia}
\ee
since the $\epsilon_{UV}>0$.

\item
Integrals of type $(b)$ are divergent, since they capture the
expectation value of a cusp Wilson loop (with sides of finite
extent). Consider for example $I_{12}$; it yields
\begin{equation}
I_{12}=-\int_0^1 d\tau_1 \int_0^1 d\tau_2 \frac{(k_1 \cdot
k_2)\Gamma(1-\epsilon_{UV})(\pi \tilde
\mu^2)^{\epsilon_{UV}}}{\left(-2(k_1 \cdot
k_2)(1-\tau_1)\tau_2\right)^{1-\epsilon_{UV}}}=(-s \pi \tilde
\mu^2
)^{\epsilon_{UV}}\frac{\Gamma(1-\epsilon_{UV})}{2\epsilon_{UV}^2}~~,
\label{Ib}
\end{equation}
where $s=(k_1+k_2)^2=(x_3-x_1)^2\equiv x_{13}^2$ is a two-particle
momentum invariant.  The double pole in $\epsilon_{UV}$ arises from
integration in the vicinity of the cusp at the point $x_2$. As
mentioned before, the formal similarity of this calculation with that
in position space suggests identifying $\epsilon$ with an ultraviolet
regulator.  This identification will be reinterpreted to account for
the fact that in dual ``configuration'' space short distances are
identified with small energies.

\item
Integrals of type $(c)$ are finite and thus can be evaluated directly
in four dimensions. For instance, $I_{13}$ yields:
\begin{eqnarray}
I_{13}&=&\int_0^1 d\tau_1 \int_0^1 d\tau_3 \frac{k_1 \cdot k_3}
{\left(k_1(1-\tau_1)+k_2+k_3 \tau_3\right)^2}\cr &=&-\frac{1}{2}
\int_0^1 d\tau_1 \int_0^1 d\tau_3 \frac{s+t}{s(1-\tau_1)+t
\tau_3+(s+t)(1-\tau_1)\tau_3}\cr
&=&-\frac{1}{4}\left(\ln^2(s/t)+\pi^2 \right)+{\cal
O}(\epsilon_{UV}) \label{Ic}
\end{eqnarray}
Similarly to the integrals of type $(b)$, $s=(k_1+k_2)^2=x_{13}^2$ and
$t=(k_2+k_3)^2=x_{24}^2$.

\end{itemize}

\noindent All other integrals can be obtained from $I_{12}$ and
$I_{13}$ by simple relabeling of the momenta. It is not hard to
see that the features of the three types of integrals described
here are quite general, regardless of the number of sides of the
polygon; integrals of type $(a)$ always vanish, integrals of type
$(b)$ contain only double-poles (and the associated momentum
dependence) and diagrams of type $(c)$ are completely finite.

The complete one-loop contribution to the expectation value of the
four-sided Wilson loop can now be pieced together. Inserting
(\ref{Ia}), (\ref{Ib}), (\ref{Ic}) and their relabeled versions into
(\ref{WL4_assembly}) leads to
\begin{equation}
\ln W_4=
\frac{g^2 N}{8\pi^2}\left(\Div_4+\frac{1}{2} \left(\ln\frac{s}{t}\right)^2+2\zeta_2
+{\cal O}(\epsilon_{UV})\right)+{\cal O}(g^4)
\label{WL_4pt_1loop}
\end{equation}
where the divergent part, denoted by $\Div$, arises entirely from
integrals of type $(b)$ and it reads
\be
\Div_4=-\frac{1}{\epsilon_{UV}^2} \left(
\left(\frac{\mu_{UV}^2}{-s} \right)^{-\epsilon_{UV}}
+
\left(\frac{\mu_{UV}^2}{-t} \right)^{-\epsilon_{UV}}\right)~~.
\label{div4}
\ee
An important point to note is the similarity of this divergence and
the expression of the logarithm of the Sudakov form factor
(\ref{soft_collinear}) with ${\cal G}_0^{(1)}=0$; this should not be
surprising, in light of the fact that part of the Sudakov form factor
may be computed in the eikonal approximation. The similarity may be
sharpened by the following identifications:
\be
\epsilon_{UV} = -\epsilon_{IR}\equiv-\epsilon
~~~~~~~~~
\mu_{UV}^2=({\tilde \mu}^2 \pi e^\gamma)^{-1}~~.
\label{identifications_UV_IR}
\ee
The first relation captures the fact that while from the standpoint of
the calculation described here the singularities have a short distance
nature, they are in fact due to gluons with low energy and/or low
transverse momentum when viewed from the position space perspective.
The second relation, inverting the unit scale of dimensional
regularization, is a reflection of the fact that the coordinates of
this configuration space have positive mass dimension.

A remarkable feature of equation (\ref{WL_4pt_1loop}) is that its
finite part, defined as the remainder after the subtraction of all
infrared divergences and of the associated terms depending on the unit
scale $\mu$ -- i.e. $\Div_4$ in equation (\ref{div4}), reproduces up
to an additive constant the finite part of the one-loop amplitude
(\ref{4pt_1loop_amp}).

\subsubsection{Higher polygons \label{higher_poly}}

\begin{figure}[ht]
\centering
\includegraphics[scale=1]{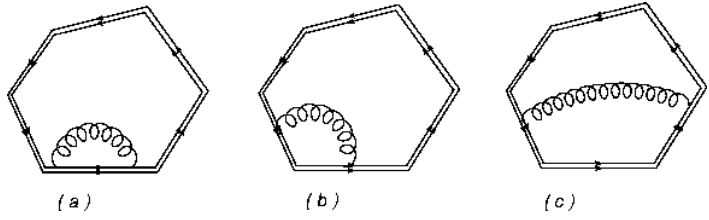}
\caption{The three generic diagrams contributing to the
expectation value of the $n$-sided polygonal Wilson loop at
one-loop order.} \label{fig:npolydiag}
\end{figure}

The calculation of the expectation value of Wilson loops constructed
on higher polygons bears a certain similarity with the expectation
value of the four-sided loop. The curve $C_n$ is now given by
$C_n=C_1\cup \dots \cup C_n$ where each segment $C_i$ ($i=1,\dots,n$)
is parametrized as before
$C_i=\{x^\mu(\tau_i)=x_i^\mu+\tau_i(x_{i+1}^\mu-x_i^\mu)
=x_i^\mu-\tau_i k_i^\mu,\,\tau_i\in[0,1]\}$. Similarly to the
four-sided loop calculation, the result for the expectation value at
one-loop is given in terms of three types of integrals whose generic
types are shown (for a six-sided loop) in figure
(\ref{fig:npolydiag}).
\be
\langle W_n\rangle = 1-\frac{g^2C_F}{4\pi^2}\sum_{1\le i\le j\le n}
I_{ij}+{\cal O}(g^4)
\label{assembly_n}
\ee
The main difference compared to the
four-sided loop involves the diagram in figure
\ref{fig:npolydiag}(c) and is that at least one of the two
momentum invariants that may be constructed from the coordinates
of the endpoints of the edges connected by the gluon exchange (and
not involving any of the momenta of the connected edges) is
nonvanishing. One may consider treating separately the case of a
vanishing invariant; it turns out however that this may be
obtained smoothly from the general situation of two nonvanishing
invariants.

The diagrams in figures \ref{fig:npolydiag}(a) and
\ref{fig:npolydiag}(b) are identical to those in figures
\ref{fig:fourpolydiag}(a) and \ref{fig:fourpolydiag}(b); the first one
vanishes identically ($I_{ii}=0$) because the edges of the Wilson loop
are light-like while second one yields\footnote{Recall that
$k_i=x_{i}-x_{i+1}$.}
\begin{equation}
\label{divfactor}
I_{i,i+1}=(-x_{i,i+2}^2 \pi \tilde \mu^2)^{\epsilon_{UV}}
\frac{\Gamma(1-\epsilon_{UV})}{2\epsilon_{UV}^2}~~.
 \end{equation}

Similarly to the integrals in figure \ref{fig:fourpolydiag}(c),
the integrals in figure \ref{fig:npolydiag}(c) are finite and thus may
be computed directly in four dimensions. Denoting by $i$ and $j$
the beginning vertex of the edges connected by the gluons and by
$k_i=x_i-x_{i+1}$ and $k_j=x_j-x_{j+1}$ the corresponding momenta,
the gluon propagator is a function of
\begin{equation}
\label{xprop}
(x(\tau_i)-x(\tau_j))^2
     =\left[\sum_{l=i}^{j-1}(x_l-x_{l+1})-\tau_i k_i+\tau_j k_j \right]^2
     =\left[k_i (1-\tau_i)+k_j\tau_j+\sum_{l=i+1}^{j-1}(x_l-x_{l+1})
\right]^2~~.
\end{equation}
The sum $\sum_{l=i+1}^{j-1}(x_l-x_{l+1})$ is nothing but the sum
of momenta on one side of the edges connected by the gluon.
Denoting it by $P\equiv k_{i+1,\dots,j-1}$ and also introducing
$Q$ as the sum of momenta on the other side of the edges connected
by the gluon, $Q\equiv k_{j+1,\dots,i-1}$ (such that
$P+Q+k_i+k_j=0$), $s=(k_i+P)^2$ and $t=(P+k_j)^2$, it is not hard
to find that
\begin{equation}
 (x(\tau_i)-x(\tau_j))^2=P^2+(s-P^2)(1-\tau_i)+(t-P^2)\tau_j+
(-s-t+P^2+Q^2)(1-\tau_i) \tau_j~~.
\end{equation}
The resulting double-integral representation of the diagram $I_{ij}$
shown in figure
\ref{fig:npolydiag}(c) is then
\be
I_{ij}
   &=&\frac{1}{8\pi^2}{\cal F}(s,t,P,Q)\\
   &=&\frac{1}{8\pi^2}\int_0^1 d\tau_i d \tau_j \frac{P^2+Q^2-s-t}
{-\left(P^2+(s-P^2)\tau_i+(t-P^2)\tau_j+(-s-t+P^2+Q^2)\tau_i
\tau_j \right)}~~. \nonumber
\ee
Upon integration \cite{BHT}, it surprisingly yields the finite (and
$\mu$-independent) part of the easy two-mass box function introduced
in section
\ref{weak_coupling}
\begin{eqnarray}
{\cal F}(s,t,P,Q)&=&-Li_2(1-{\hat a} s)-Li_2(1-{\hat a} t)
                                  +Li_2(1-{\hat a} P^2)+Li_2(1-{\hat a}Q^2)\\
{\hat a}&=&\frac{P^2+Q^2-s-t}{P^2 Q^2-s t}~~.
\end{eqnarray}
Note that the limits $P^2\rightarrow 0$ and $Q^2\rightarrow 0$ are
smooth; this justifies treating the configurations with $j=i+2$ as
limits of the generic integrals $I_{ij}$.

Collecting all contributions to the expectation value of the
$n$-sided Wilson loop following (\ref{assembly_n}) one finds
\cite{BHT} \be \langle W_n\rangle = \Div_n+\Fin_n+C_n \ee with
$C_n$ being an additive constant. Here $\Div_n$ is, up to the
identifications (\ref{identifications_UV_IR}), the appropriate sum
of the logarithms of Sudakov form factors\footnote{That is,
$\Div_n$ is the sum of one-loop form factors evaluated on all
adjacent two-particle invariants $s_{i,i+1}$. Its explicit
expression follows from (\ref{divergence_general}) by replacing
$f^{(-2)}$ and $g^{(-1)}$ with their one-loop values:
$f^{(-2)}(x)=4x$ and $g^{(-1)}(x)=0$.} and $\Fin_n$ reproduces the
finite part of the one-loop MHV amplitude
(\ref{OneLoopFiniteRemainder})-(\ref{DLodd}). \footnote{ The
relation between Wilson loops and scattering amplitudes may be
extended, at one-loop level, to an ``integral-by-integral''
identity. As discussed in \cite{BHNST_gravity}, one may pick a
gauge in the dimensionally regularized theory in which the
diagrams \ref{fig:npolydiag}(b) also vanishes identically. Then,
the complete contribution to the expectation value of the Wilson
loop comes from the integrals \ref{fig:npolydiag}(c) each of
which, in this gauge, equals one easy two-mass box function.}

It is interesting that in Feynman gauge, used in the preceeding
calculation, there is such a clean separation of the divergent and the
finite contributions to the expectation value of the Wilson loop. As
we will see in later sections, this separation does not survive at
higher loops. It would be interesting to identify a different gauge
which realizes such a separation to all orders in perturbation theory.

Before proceeding let us  remark that,
from the details described above following
\cite{Drummond:2007aua, BHT}, it is apparent that
the one-loop expectation value of the Wilson
loop under discussion is entirely independent on the matter content of
the gauge theory it is evaluated in. This suggests that, at least at
one-loop level, MHV scattering amplitudes in ${\cal N}=4$ SYM theory
capture a universal structure of all gauge theories. Since the BDS
ansatz is constructed -- up to the explicit expression of the cusp
anomaly -- from the one-loop amplitude, this also suggests that the
BDS ansatz captures a universal part of scattering amplitudes in all gauge
theories -- the part determined by infrared singularities and dual
conformal invariance.

\subsection{A conformal Ward identity}

Wilson loops are generically not invariant under coordinate
transformations, since the latter changes the contour defining
them; if the Lagrangian is invariant and if the Wilson loop is
well-defined (finite), then one finds instead that
\begin{equation}
\langle W(\tilde{C}) \rangle = \langle W(C)\rangle
\end{equation}
where $\tilde{C}$ is the image of curve $C$ under this coordinate
transformations.

By restricting the class of coordinate transformations to the
conformal group $SO(2,4)$ it is possible to find more interesting
information. While being operators of vanishing classical
dimension, generic Wilson loops do not enjoy definite conformal
properties, because the contour defining them changes under
general conformal transformations. Light-like Wilson loops are
however special due to the fact that conformal transformations
preserve the light-cone up to perhaps dilatations. In the case at
hand the contour is composed of light-like segments
$x^\mu=x^\mu(\tau)$. It is not hard to check that an inversion
transformation $\tilde{x}^{\mu}=\frac{x^\mu}{x^2}$ preserves this
property:
\begin{eqnarray}
&&
x^\mu(\tau_i)=\tau_i x_i^\mu+(1-\tau_i)x_{i+1}^\mu \rightarrow
\tilde{x}^\mu(\tilde{\tau}_i)= \tilde{\tau}_i
\tilde{x}_i^\mu+(1-\tilde{\tau}_i)\tilde{x}_{i+1}^\mu \\
&&~~
\tilde{\tau}_i=\frac{\tau_i\,x_i^2}{\tau_i\,x_i^2+(1-\tau_i)\,x_{i+1}^2},
\hspace{0.9in}(\tilde{x}_{i+1}-\tilde{x}_{i})^2=0
\end{eqnarray}
Thus, the light-like polygons used to construct the Wilson loops
related to scattering amplitudes are invariant under inversion.
They are of course not invariant under translations and Lorentz
transformations, but their expectation value is, since it depends
only on the norm of the difference of coordinates of the endpoints
of the edges. The same then holds true for special conformal
transformations.

Thus, since the ${\cal N}=4$ SYM theory is invariant under conformal
transformations, if the expectation value of light-like polygonal
Wilson loops were finite and well defined in four dimensions, they
would also be invariant under conformal transformations.
\begin{equation}
W(\tilde{C}_n)=W(C_n)~~,
\end{equation}
since the action compensates\footnote{I.e. a coordinate transformation
is necessary.} for the changes in the contour $C_n \mapsto
\tilde{C}_n$ under $SO(2,4)$ transformations.
\footnote{This conclusion holds despite the fact that the Wilson loops
are defined in a configuration space related to momentum space rather
than in the usual position space. One only needs to interpret the
conformal group acting on this configuration space -- in this case the
dual conformal group.}

As we have seen previously however, the light-like polygonal
Wilson loops require regularization both because of the presence
of cusps as well as because of the presence of light-like edges which
also lead to collinear singularities. Any regularization breaks
(dual) conformal invariance and thus potentially leads to anomalies.
They were studied in \cite{Drummond:2007au}; in the following we
will review the results obtained there.

\subsubsection{General Properties of Cusp Singularities}

A general feature of the polygonal Wilson loops conjecturally
related to MHV scattering amplitudes is the presence of cusps. As
we have seen in explicit calculations, their presence combined
with the fact that the edges are light-like leads to an
$\epsilon^{-2}$ short distance singularity at
one-loop.\footnote{We remind the reader that in the present
context the distance is ``short'' in the dual momentum space.} The
divergences arise from diagrams similar to those in figures
\ref{fig:fourpolydiag}(b) and \ref{fig:npolydiag}(b); each of
them contributes a factor (\ref{divfactor}). As a result, the
divergent piece of the Wilson loop expectation value at one-loop
has the form
\begin{equation}
\ln W_n\Big|_{\Div}=\frac{g^2}{4\pi^2}\frac{C_F}{2}\left(-\frac{1}{\epsilon^2}
\sum_{i=1}^n (-x_{i-1,i+1}^2 \mu^2)^\epsilon +{\cal
O}(1)\right)+{\cal O}(g^4)
\end{equation}
with $x_i$ the position of the cusps and the indices are cyclic --
$x_{i+n}=x_i$. As alluded to earlier, these singularities arise in two
regimes: $(a)$ when the gluon propagates a very short distance (which
can happen only if the points connected by the gluon are near the
cusp) and $(b)$ if the gluon propagates parallel to one of the edges
of the cusp. They are in direct correspondence with the usual soft and
collinear divergences which appear in one-loop scattering
amplitudes. Indeed, if the gluon propagates a short distance in
this configuration space its energy is very small; thus, this corresponds
to the soft divergence (\ref{soft}).  Analogously, if the gluon
propagates parallel to one of the light-like edges, then its transverse
momentum is very small and the corresponding singularity is analogous
to a collinear divergence (\ref{collinear}).

At higher loops the divergence structure is somewhat more
complicated to disentangle; since in the same diagram one finds
gluons attached to different segments and in different
configurations, various soft and collinear regimes can be realized
simultaneously, increasing the strength of the divergence.
Additional divergences can arise from subintegrals whose external
legs are not directly linked to the edges of the cusp.  It
nevertheless continues to be true that at higher loops all cusp
singularities are a combination of soft and collinear ones (in the
sense described above), similarly to the infrared singularities of
scattering amplitudes; to $L$-loop order, the singularities of
light-like cusped Wilson loops are poles starting at order $2L$
\be
\langle W_{\rm cusp}\rangle \approx \frac{(\WLa\,
\mu^{2\epsilon})^L}{\epsilon^{2L}}+{\cal O}(\epsilon^{2L-1})
\ee

Since from the standpoint of the evaluation of the Wilson loop
expectation value these divergences are due to short distance
effects, they may (and should) be renormalized.  The
renormalization properties of light-like Wilson loops have been
extensively studied. An important result obtained in
\cite{Korchemsky_Radyushkin} is that Wilson loop operators $W_n$
are multiplicatively renormalized. In other words, divergences may
be subtracted recursively and, after subtraction of all
subdivergences, the remaining overall divergence is local and can
also be subtracted by a counterterm, i.e.
\begin{equation}
\langle W_{\rm cusp}\rangle =Z_{\rm cusp} F_{\rm cusp}~~,
\label{ren_cusp}
\end{equation}
where $F_{\rm cusp}$ is finite as the regulator is removed while
$Z_{\rm cusp}$ contains all divergences. \footnote{ In general,
singularities in the expectation value of Wilson loops arise due to
the presence of cusps. For a multi-cusped light-like Wilson loop the
relation (\ref{ren_cusp}) generalizes in the obvious way $\langle
W_{n}\rangle =Z_{n} F_{n}$.}
The structure of the divergent factor $Z_n$ was analyzed in detail;
the result obtained in \cite{Korchemskaya:1992je, Bassetto:1993xd} is
that cusp singularities exponentiate and the divergent factor has the
special form
\begin{equation}
\label{alldiv}
\log Z_n = -\frac{1}{8}\sum_{l \geq 1}\WLa^{l}
\sum_{i=1}^n (-x_{i-1,i+1}^2 \mu^2)^{l \epsilon}
\left(\frac{\gamma_{K}^{(l)}}{l^2\epsilon^2}+
\frac{2\Gamma^{(l)}}{l \epsilon} \right)
\end{equation}
with $\gamma_{K}^{(l)}$ the expansion coefficients of the cusp
anomalous dimension (cf. equation (\ref{DefCusp})) and $\Gamma^{(l)}$
the so-called cusp collinear anomalous dimension, defined in the
adjoint representation of $SU(N)$. The first few coefficients of the
weak coupling expansion of $\gamma_{K}$ are shown in
(\ref{weak_coupling_cusp}) while the first term in the expansion of
the cusp collinear anomalous dimension is
%
\begin{eqnarray}
\Gamma(\WLa)&=&\sum_{\ell \geq 1}\WLa^\ell \Gamma^{(\ell)}=-7 \zeta_3
\WLa^2+{\cal O}(\WLa^3)
\end{eqnarray}
Integrability allows the construction of an integral equation
\cite{BES} determining the cusp anomaly or universal scaling function
$f(a)$ to all orders in a weak (and/or strong) coupling expansion. It
would be interesting to construct a similar equation for $\Gamma(a)$
and/or $G(a)$. \footnote{It has been argued in
\cite{Dixon_Magnea_Sterman} that $\Gamma(a)$ and $G(a)$ are related by
a scheme-independent multiple of the first subleading term in the
large spin expansion of the anomalous dimension of twist-two
operators.}

The structure of cusp divergences implies that the logarithm of the
expectation value of the light-like Wilson loops $W_n$ conjecturally
related to MHV amplitudes may be written as
\begin{equation}
\label{genericform}
\ln \langle W_n\rangle =
-\frac{1}{8}\sum_{l \geq 1}\WLa^{l}\sum_{i=1}^n
(-x_{i-1,i+1}^2 \mu^2)^{l \epsilon}
\left(\frac{\gamma_{K}^{(l)}}{l^2 \epsilon^2}+
\frac{2\Gamma^{(l)}}{l \epsilon} \right)
+F_n
\end{equation}
where $F_n$ is a finite contribution independent on $\mu$. This structure
matches the gauge theory expectations for the logarithm of the
rescaled MHV amplitudes ${\cal M}_n$. In fact, the universality of
infrared divergences, emphasized by the soft and collinear factorization
and exponentiation theorem and by the discussion above, suggests that
if non-MHV amplitudes are captured by some types of Wilson loops, then
these loops necessarily must exhibit $n$ cusps.

\subsubsection{Conformal properties of light-like Wilson loops}

On general grounds, when a symmetry is broken by a regulator it
may develop anomalies at the quantum level. They typically appear
in the form of the product between a factor vanishing as the
regulator is removed and a factor that would diverge in the same
limit.  Wilson loops in dimensional regularization could (and, as
we will see following \cite{Drummond:2007au}, actually do) exhibit
such anomalies in (dual) dilatation and (dual) special conformal
transformations.

Dimensional regularization preserves the dimension of fields due to
the introduction of the unit scale $\mu$:
\be
S_{\epsilon}=\frac{1}{g^2 \mu^{2\epsilon}}\int d^D x {\cal L}(x),
\hspace{0.3in}
{\cal L}=-\frac{1}{2}\Tr(F_{\mu \nu}^2)+...
\ee
where $D=4-2\epsilon$ and we suppressed terms in the Lagrangian which
have other fields besides gluons.  The fact that the dimension of
fields is unchanged implies that in the regularized theory the Wilson
loop operator (\ref{WL_general_YM}) continues to be conformally
invariant up to a coordinate transformation.
Both the Lagrangian and the integration measure relating it to the
action transform homogeneously under this coordinate transformation;
in four dimensions the action is invariant. A
$D=4-2\epsilon$-dimensional integration measure breaks this invariance
and induces an ${\cal O}(\epsilon)$ violation of dual conformal
invariance (more precisely, it is no longer invariant under dilatations
and conformal boosts). In the path integral evaluation of the Wilson
loop expectation value this leads to an anomaly under these
transformations.

The existence of anomalies may be captured systematically in
several different ways. In the current context the most efficient
one is through a Ward identity whose form -- if the regulator
preserved dilatations and conformal boosts -- would be \be \langle
\mathbb{D}W_n \rangle=0, ~~~~~~~~~~~~~~ \langle \mathbb{K}^\mu
W_n\rangle=0 ~~; \label{perfect_world_Ward_id} \ee the anomalies
will appear as inhomogeneous terms on the right hand side of these
equations.
The starting point in the derivation of the relevant anomalies would
be to perform the relevant transformations on the action in the
presence of the regulator. It is possible however to bypass this
step\footnote{\label{conformal_transf} The following arguments may, of
course, be checked using the conformal transformations of ${\cal N}=4$
SYM fields, denoted collectively by $\phi_I$:
\begin{eqnarray}
\mathbb{D}\phi_I&=&x^\mu \partial_\mu \phi_I+d_\phi \phi_I\cr
\mathbb{K}^mu \phi_I&=&(2 x^\mu x^\nu \partial_\nu-x^2 \partial^\mu) \phi_I+2
x^\mu d_\phi \phi_I+2 x^\nu(M^{\mu}{}_{\nu})_I^{~J}\phi_J~~;
\nonumber
\end{eqnarray}
here $M^{\mu}{}_{\nu}$ are the Lorentz generators:
$M^{\mu}{}_{\nu}=x^\mu \partial_{x^\nu} - x^\nu \partial_{x^\mu}$. }
by recalling that the action is invariant if $\epsilon=0$ and
that the transformation rules for fundamental fields does not
depend explicitly on the dimension. It thus follows that
the Lagrangian transforms homogeneously with weight four:
\be
\mathbb{D}:{\cal L}(x)\mapsto \Lambda^4 {\cal L}(\Lambda x)
\label{Ltranformation}
\ee
%
%
Thus, under infinitesimal dilatations, the regularized action
transforms as\footnote{In the presence of the regulator the measure
transforms as $\mathbb{D}:d^Dx\mapsto \Lambda^{-4+2\epsilon}d^Dx$. Using
$\Lambda=1+\delta\Lambda$ and assuming that for an infinitesimal
transformation $\delta\Lambda$ is small, it is easy to find
(\ref{infinitesimal_dil}) upon expansion in $\delta\Lambda$.}
\begin{equation}
\delta_{\mathbb{D}}S_\epsilon=\frac{2\epsilon}{g^2
\mu^{2\epsilon}}\int d^D x {\cal L}(x)~~.
\label{infinitesimal_dil}
\end{equation}
From here one immediately concludes that
\be
\label{dileq}
\mathbb{D} \langle W_n \rangle=\sum_{i=1}^n( x_i \cdot \partial_i)
\langle W(C_n)\rangle
=\langle \delta_{\mathbb{D}}S_\epsilon W_n \rangle =\frac{2
i \epsilon }{g^2 \mu^{2\epsilon}}\int d^Dx \langle {\cal L}(x) W_n
\rangle
\ee
Because of the presence of divergences in the
correlation function on the right hand side, it is not possible to
set it to zero, despite its manifest proportionality to the
dimensional regulator. In light of the discussion of singularities
of cusped Wilson loops, one may in fact wonder whether a single
power of $\epsilon$ is sufficient to render the right hand side
finite. As we shall see, this is indeed the case.

A similar analysis yields an expression for the special conformal
transformations of the expectation value of the Wilson loop. The
derivative part of special conformal transformations is a genuine
invariance of the Lagrangian; this may be seen from the fact that the
transformation of the measure is trivial. Special conformal
transformations also exhibit a dilatation component, whose presence is
necessary for the closure of the algebra. Since the conformal boost
generators carry a vector index, their dilatation component must be
multiplied by $x^\mu$. Thus, up to this factor of the coordinate, the
action of these generators on the Wilson loop is the same as that of
the dilatation generator:
\begin{equation}
\label{Keq_WL}
\mathbb{K}^\mu \langle W_n \rangle=(2 x^\mu x^\nu \partial_\nu-x^2
\partial^\mu) \langle W_n \rangle=\frac{4 i
\epsilon }{g^2 \mu^{2\epsilon}}\int d^Dx \;x^\mu \langle {\cal L}(x)
W_n \rangle
\end{equation}
The precise normalization may be derived by making use of the
commutation relation $[P_\mu,\,K_\nu]=-2i[\eta_{\mu\nu}D+M_{\mu\nu}]$
or from the transformation rules in footnote
\ref{conformal_transf}. Similarly to (\ref{dileq}), the presence
of divergences prevents at this stage setting $\epsilon\rightarrow
0$ on the right hand side of the equation above; these terms are
the origin of the conformal boost anomaly \cite{Drummond:2007au}.

The renormalization properties of Wilson loops (\ref{ren_cusp})
suggest that it would be convenient to organize the equations
(\ref{dileq}) and (\ref{Keq_WL}) such that the divergent and
finite parts are separated. Thus, a convenient rewriting should
involve the logarithm of the expectation value of the Wilson loop
by simply dividing by $\langle W_n\rangle$:
\begin{eqnarray}
\label{Deq_WL_d}
\mathbb{D}\ln \langle W_n \rangle
=\frac{2i\epsilon}{g^2 \mu^{2\epsilon}}\int d^Dx \frac{\langle
{\cal L}(x) W_n
\rangle}{\langle W_n \rangle}\\
\label{Keq_WL_d}
\mathbb{K}^\mu\ln \langle W_n \rangle =\frac{4i\epsilon}{g^2
\mu^{2\epsilon}}\int d^Dx x^\mu \frac{\langle {\cal L}(x) W_n
\rangle}{\langle W_n \rangle}~~,
\end{eqnarray}
where the left hand side is written in terms of logarithms because
$\mathbb{D}$ and $\mathbb{K}$ are linear differential operators. The
explicit multiplication by $\epsilon$ on the right hand side implies
that only the divergent terms need to be evaluated since only they can
contribute to this product in the $\epsilon\rightarrow 0$ limit.

The right hand side of the equations (\ref{Deq_WL_d}) and
(\ref{Keq_WL_d}) have been explicitly evaluated at one-loop order
and the result extended to all loops in \cite{Drummond:2007au}.
For the most part, the details are common between the two
equations: one first evaluates ${\langle {\cal L}(x) W_n
\rangle}/{\langle W_n \rangle}$ as a function of the insertion
point $x$ and then reconstructs the two inhomogeneous terms above
by direct integration.

\begin{figure}[ht]
\centering
\includegraphics[scale=1]{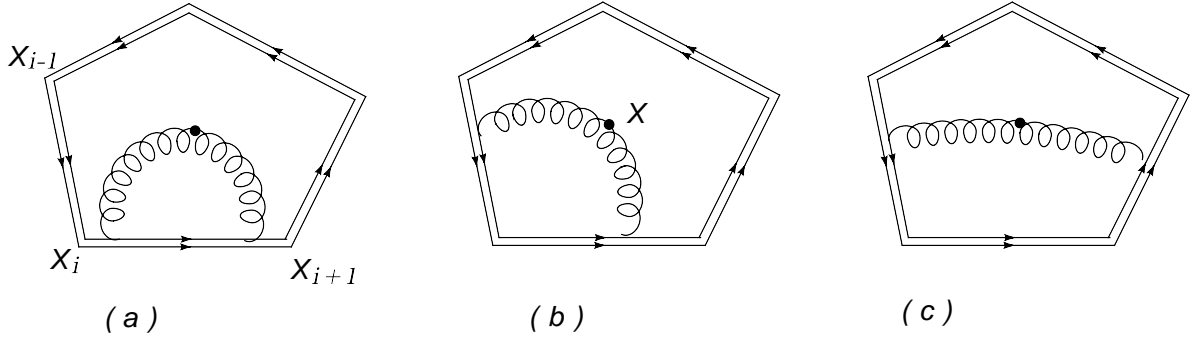}
\caption{Feynman diagrams contributing to  $\langle {\cal L}(x)
W_n \rangle$ at lowest order in perturbation theory. The wiggly
line represents the gluon propagator and the blob the insertion
point.} \label{fig:ward}
\end{figure}

To lowest nontrivial order, ${\langle {\cal L}(x) W_n \rangle}/{\langle W_n
\rangle}$ is simply given by the same matrix element as the
leading contribution to the
expectation value of the Wilson loop with an additional insertion of
the Lagrangian, the only contributing part of which is the gluon
kinetic term
\begin{equation}
\frac{\langle {\cal L}(x) W_n \rangle}{\langle W_n
\rangle}=-\frac{1}{8N}\langle \Tr\left[(\partial_\mu
A_\nu(x)-\partial_\nu A_\mu(x))^2 \right] P
\left(\oint_{C_n}dy\cdot A(y) \right)^2 \rangle+{\cal O}(g^6)~~.
\end{equation}
%
%
This insertion slightly modifies the diagrams in figure
\ref{fig:npolydiag} to those in figure \ref{fig:ward}.  As we have
seen, in the absence of additional insertions, only the diagram
\ref{fig:npolydiag}(b) develops singularities in Feynman gauge. The
same continues to hold here and only the diagram
\ref{fig:ward}(b) develops singularities in Feynman gauge.

The final result for this expectation value,
obtained in \cite{Drummond:2007au}, is
\begin{equation}
\frac{2i}{g^2 \mu^{2\epsilon}}\frac{\langle {\cal L}(x) W_n
\rangle}{\langle W_n \rangle}=- \WLa \sum_{i=1}^n
(-x^2_{i-1,i+1}\mu^2)^\epsilon
\left(\frac{1}{\epsilon^2}\delta^{(D)}(x-x_i)
     +\frac{1}{\epsilon}\Upsilon^{(1)}(x|x_{i-1},x_i,x_{i+1})+{\rm finite}
\right)
\label{1loop_CWI}
\end{equation}
where the function $\Upsilon^{(1)}$
\begin{equation}
\Upsilon^{(1)}(x|x_{i-1},x_i,x_{i+1})=\int_0^1
\frac{ds}{s}\left(\delta^{(D)}(x-x_i-s x_{i-1,i})
                 +\delta^{(D)}(x-x_i+s x_{i,i+1})
                -2\delta^{(D)}(x-x_i)
\right)
\end{equation}
is first term in the weak coupling expansion of a function $\Upsilon(a)$.
Not entirely unexpectedly, the double-pole in (\ref{1loop_CWI}) is
localized at the cusps; the simple poles in (\ref{1loop_CWI}) receive
contributions both from cusps as well as from the edges. There is thus
a similarity between the origin of various pole terms at weak and
strong coupling (cf. section \ref{CWI_strong_coupling}, up to the
change of regularization from cut-off to dimensional regularization).

Upon integration with and without the $x^\mu$ weight in equations
(\ref{Deq_WL_d}) and (\ref{Keq_WL_d}) the contribution of
$\Upsilon^{(1)}$ simplifies to
\begin{eqnarray}
\int d^Dx \Upsilon^{(1)}(x|x_{i-1},x_i,x_{i+1})&=&0\\
\int d^Dx x^\mu
\Upsilon^{(1)}(x|x_{i-1},x_i,x_{i+1})&=&x_{i-1}^\mu+x_{i+1}^\mu-2x_i^\mu~~;
\end{eqnarray}
upon summation over the labels of the cusps, the contribution of
$\Upsilon^{(1)}$ to the right hand side of the Ward identity for
(dual) conformal boosts vanishes as well.  Thus, collecting
everything, it is easy to find that
\begin{eqnarray}
\label{KDeq_WL_fin}
\mathbb{D}\ln \langle W_n \rangle &=&-\WLa
\frac{1}{\epsilon}\sum_{i=1}^n
(-x^2_{i-1,i+1}\mu^2)^\epsilon +{\cal O}(\WLa^2)\\
\mathbb{K}^\mu\ln \langle W_n \rangle &=&-\WLa
\frac{1}{\epsilon}\sum_{i=1}^n x_i^\mu
(-x^2_{i-1,i+1}\mu^2)^\epsilon +{\cal O}(\WLa^2)
\end{eqnarray}
These equations may be turned into an anomalous Ward identities for
the (logarithm of the) finite factor of the Wilson loop expectation
value.  Substituting the expression of $\ln \langle W_n\rangle$ in
terms of $Z_n$ and $F_n$ -- $\ln \langle W_n\rangle=\ln Z_n+\ln F_n$
-- and working out the action of $\mathbb{D}$ and $\mathbb{K}$ on
$\ln Z_n$, one quickly finds that
\be
\mathbb{D}\ln F_n &=&0 +{\cal O}(\WLa^2)\\
\mathbb{K}^\mu\ln F_n &=&\WLa \sum_{i=1}^n x_i^\mu
\ln \frac{x_{i,i+2}^2}{x^2_{i-1,i+1}} +{\cal O}(\WLa^2)
\ee
The complete cancellation of singularities also identifies the
numerical coefficient of the leading singularity on the right hand
side of (\ref{KDeq_WL_fin}) with the leading term in the weak coupling
expansion of the cusp anomalous dimension. This observation was used
in \cite{Drummond:2007au} to gain information on the all-loop
structure of the anomaly.

\subsubsection{An all-loop generalization of the conformal Ward identity}

Unlike axial anomalies, the anomalies of the dual conformal
symmetry are not one-loop exact. It is possible, though not
completely straightforward, to generalize to higher loops the
calculation described in the previous section. A key point which
makes the calculations tractable is that, as repeatedly mentioned
above, the complete contribution to the anomaly is governed by the
singular terms in the expectation value ${\langle {\cal L}(x) W_n
\rangle}/{\langle W_n \rangle}$.

Information on the structure of anomaly can be gained from the
observations made before -- that at one loop the coefficient of the
leading pole is the one-loop cusp anomaly -- and from the fact that
the Ward identity for the finite part of the Wilson loop should relate
finite quantities. From here it seems reasonable to conclude that the
coefficient of the leading singularity of ${\langle {\cal L}(x) W_n
\rangle}/{\langle W_n \rangle}$ should in fact be proportional to
first logarithmic integral of the cusp anomaly (with the
proportionality coefficient determined by the one-loop calculation)
and that the coefficient of the local term $\delta(x-x_i)$ should also
contain $\Gamma(a)$:
\begin{eqnarray}
\label{allops}
&& \frac{2i\epsilon}{g^2 \mu^{2\epsilon}}
\frac{\langle {\cal L}(x) W_n \rangle}{\langle W_n \rangle}=\\
&=& -\sum_{l=1}\WLa^l \sum_{i=1}^n (-x_{i-1,i+1} \mu^2)^{\ell
\epsilon}
\left\{\frac{1}{4} \left(\frac{\gamma_{K}^{(l)}}{l \epsilon}
                        +2\Gamma^{(l)} \right)\delta^{(D)}(x-x_i)
       +\Upsilon^{(l)}(x|x_{i-1},x_i,x_{i+1})\right\}~~. \nonumber
\end{eqnarray}
The structure of the coefficient of $\delta^{(D)}(x-x_i)$ guarantees
that, as desired, all singularities cancel in the action of
$\mathbb{D}$ and $\mathbb{K}^\mu$ on $\ln F_n$.

Ref.\cite{Drummond:2007au} argued that, order by order in the weak
coupling expansion, the contribution of
$\Upsilon^{(l)}(x|x_{i-1},x_i,x_{i+1})$ to the Ward identity vanishes
identically upon integration over the insertion point $x$ and
summation over cusp labels.  The argument is based on dimensional
analysis, the scheme independence of $\Upsilon$
and the fact that the form of $\Upsilon$ is, up to its arguments,
independent of the cusp it originates from.\footnote{Independently of
these arguments, it is possible to show explicitly that
\begin{eqnarray}
\int d^Dx \Upsilon^{(l)}(x|x_{i-1},x_i,x_{i+1})=0~~,\nonumber
\end{eqnarray}
a calculation which strengthens the arguments of
\cite{Drummond:2007au}. }

Inserting (\ref{allops}) into the first equation of
(\ref{KDeq_WL_fin}) and making use of the vanishing arguments for the
contributions of $\Upsilon$ one finds the all-order Ward identities
for dilatation and conformal boosts
\begin{eqnarray}
\label{almostD}
\mathbb{D}\ln \langle W_n \rangle&=&
-\frac{1}{4}\sum_{l=1} \WLa^l \sum_{i=1}^n (-x_{i-1,i+1} \mu^2)^{l
\epsilon} \left(\frac{\gamma_{k}^{(l)}}{l
\epsilon}+2\Gamma^{(l)} \right)
\\
\label{almostK}
\mathbb{K}^\mu\ln \langle W_n \rangle&=&
-\frac{1}{2}\sum_{l=1} \WLa^l \sum_{i=1}^n \;x_i^\mu\,(-x_{i-1,i+1} \mu^2)^{l
\epsilon} \left(\frac{\gamma_{K}^{(l)}}{l
\epsilon}+2\Gamma^{(l)} \right)
\end{eqnarray}
Similarly to the one-loop calculation in the previous section, it is
useful to recast these equations as constraints on the finite factor
$F_n$ in (\ref{ren_cusp}). As expected and desired, all singularities
cancel and the regulator can be removed ($\epsilon\rightarrow
0$). Thus, (\ref{almostD}) and (\ref{almostK}) become
\begin{eqnarray}
\label{finiteD}
\mathbb{D}\ln F_n&=&\sum_{i=1}^n (x_i\cdot\partial_i)F_n=0 \\
\label{finiteK}
\mathbb{K}^\mu\ln F_n
&=&\sum_{i=1}^n(2 x_i^\mu x_i \cdot \partial_i-x_i^2 \partial_i^\mu)\ln F_n
=\frac{1}{4}\,f(\WLa)\,\sum_{i=1}^n \,x_{i,i+1}^\mu
\log{\frac{x_{i,i+2}^2}{x_{i-1,i+1}^2}}
\end{eqnarray}
The first equation implies that the finite part of the expectation
value of cusped Wilson loops discussed here are functions of
dimensionless, scale invariant ratios of products of cusp
coordinates $x_i$. The same conclusion may be reached on
dimensional grounds, making use of the fact that, in the presence
of the regulator, $\langle W_n \rangle$ depends on the unit scale
of dimensional regularization. Then, \be \sum_i\left(x_i\cdot
\partial_i-\mu\partial_\mu \right)\ln \langle W_n\rangle=0~~, \ee
which becomes (\ref{finiteD}) upon use of (\ref{ren_cusp}).  The
consequences of the conformal boost Ward identity will be
discussed in the next section; not surprisingly, the results
reproduce the structure of constraints on scattering amplitudes
following from dual conformal invariance.

\subsubsection{Constraints on expectation values of Wilson loops
               \label{constraints_WL_vev}}

As we have seen previously, the equation (\ref{finiteD}) implies a
relatively simple constraint on the finite part $F_n$ of the
Wilson loop. The Ward identity for conformal boosts requires
further analysis.

The notation may be slightly simplified by making use of the maximal
nonabelian exponentiation theorem \cite{max_nonabelian_exp1,max_nonabelian_exp2,max_nonabelian_exp3} which
states that the expectation value of Wilson loops have a natural
exponential structure and the exponent itself has an diagrammatic
interpretation in that it receives contributions only from the Feynman
diagrams which are two-particle irreducible. From this standpoint it is perhaps more natural to
replace $\ln F_n$ by its logarithm
\be
\ln F_n(\WLa)=\frac{1}{4}\,f(\WLa) \, {\rm F}_n(\WLa)~~.
\ee
The reason for introducing the factor of the cusp anomaly $f(\WLa)$ is
that the right hand side of (\ref{finiteK}) also exhibits a similar
overall factor and thus the resulting equation does not
contain any further explicit coupling constant dependence:
\begin{equation}
\label{dualward}
\sum_{i=1}^n(2 x_i^\mu x_i\cdot\partial_i-x_i^2 \partial_i^\mu)
{\rm F}_n=\sum_{i=1}^n x_{i,i+1}^\mu\log{\frac{x_{i,i+2}^2}{x_{i-1,i+1}^2}}~~.
\end{equation}

As usual, the solution to this type of equation is a sum between a
solution of the inhomogeneous equation and a general solution of the
homogeneous one. Because of its lack of coupling constant dependence,
the equation (\ref{dualward}) implies that the solution to the
inhomogeneous equation is determined by a one-loop calculation. It is
thus clear that, up to the coupling constant dependence, ${\rm F}_n$
is just the one-loop expectation value of the $n$-sided Wilson
loop.\footnote{It is quite remarkable that, interpreting the cusp
anomalous dimension as the ``physical'' coupling constant makes the
anomaly of dual conformal conformal boosts into an one-loop-exact
quantity. This is reminiscent of suggestions in reference
\cite{physical_coupling1,physical_coupling2}.}

The calculations in sections \ref{four_poly} and \ref{higher_poly}
imply then that
\be
{\rm F}_4=\frac{1}{2}F_4^{(1)}(0)+C_4
~~~~~~~~~~~~~~~~~
{\rm F}_n=\frac{1}{2}F_n^{(1)}(0)+C_n
\label{sol_CWI}
\ee
with $F_4^{(1)}(0)$ and $F_n^{(1)}(0)$ given by equations
\ref{4pt_1loop_amp} and \ref{OneLoopFiniteRemainder}-\ref{DLodd},
respectively. It is indeed possible to check \cite{Drummond:2007au}
that these expressions solve the equation (\ref{dualward}).

To this solution one should add a solution to the homogeneous version
of equation (\ref{dualward}) (i.e. the equation with the right hand
side removed). Instead of finding the general solution by directly
solving this differential equation, it is more convenient to make use
of the fact that conformal boosts are just a combination of inversion
and translation -- $K=IPI$ -- and construct inversion and translation
invariants known as conformal cross-ratios which also appeared in our
discussion of conformal properties of scattering amplitudes in section
\ref{BDS_and_departures}. This
strategy is extensively applied in two-dimensional conformal field
theories; in four dimensional theories it was initially applied, in a
related context, in \cite{broadhurst}.

As we briefly mentioned in section \ref{conf_ints}, inversion
transformations act as
\be
I:x_i^\mu\mapsto \frac{x_i^\mu}{x_i^2}
~~~~~~{\rm and}~~~~~~
I:x_{ij}^2\mapsto \frac{x_{ij}^2}{x_i^2x_j^2}~~.
\ee
Thus, using the fact that $x_{i,i+1}^2=0$, inversion and translation
invariants are constructed from the coordinates of any four cusps
whose labels differ ($\mod$ the number of sides of the polygon) by at
least two units. All relevant invariants may be identified by choosing
four unordered labels $(i,j,k,l)$ and constructing
\be
u_{ijkl}=\frac{x_{ij}^2x_{kl}^2}{x_{ik}^2x_{jl}^2}~~;
\label{cr}
\ee
reordering of the labels $(i,j,k,l)$ may lead to different invariants
corresponding to the same choice of cusps.
The solution to the
homogeneous form of the equation (\ref{dualward}) is then an arbitrary
function of all possible such conformal cross-ratios and of the coupling
constant $\WLa$.

Clearly, the cross-ratios (\ref{cr}) are even under parity
transformations. It is also possible to construct dual conformal
invariants which are parity-odd. Products of an even numbers of such
quantities are expressible in terms of the cross-ratios (\ref{cr}). We
are assuming throughout that the solution to the anomalous Ward
identity does not contain terms linear in such parity-odd invariants.

Not surprisingly, the number of invariants depends strongly on the
number of cusps. Simple counting implies that for $n=4$ and $n=5$ no
such conformal ratios may be constructed; thus, the only solution of
the homogeneous equation (\ref{dualward}) for four- and five-sided
polygons is just a constant. Thus,
\be
{\rm F}_{4,5}=\frac{1}{2}F^{(1)}_{4,5}(0)+C_{4,5}~~.
\ee

Conformal cross-ratios exist for polygons of at least six sides. The
minimum number of invariants occurs for $n=6$; it is trivial to check
that
\begin{eqnarray}
u_1=\frac{x_{13}^2x_{46}^2}{x_{14}^2x_{36}^2},\hspace{0.3in}
u_2=\frac{x_{24}^2x_{15}^2}{x_{25}^2x_{14}^2},\hspace{0.3in}
u_3=\frac{x_{35}^2x_{26}^2}{x_{36}^2x_{25}^2}~~.
\label{6pt_crossratios}
\end{eqnarray}
are indeed solutions of the homogeneous equation (\ref{dualward}) and,
because of the linearity of $\mathbb{K}$, so is any function of
them. Thus,
\be
{\rm F}_6=\frac{1}{2}F^{(1)}_{6}(0)+\Remainder_{W6}(u_1,u_2,u_3;\WLa)~~,
\label{sixpt_Remainder_WL}
\ee
where we denote by $\Remainder_{W6}(u_1,u_2,u_3;a)$ the solution of
the non-anomalous Ward identity which is necessary to relate
(\ref{sol_CWI}) to the expectation value of the six-sided Wilson loop.
A similar function may be defined for Wilson loops with an arbitrary
number of edges
\be
{\rm F}_n=\frac{1}{2}F^{(1)}_{n}(0)+\Remainder_{Wn}({\bf u};\WLa)
\label{general_Remainder_WL}
\ee
and represents the Wilson loop analog of the remainder function
$\Remainder_{An}$ capturing the departure of scattering amplitudes
from the BDS ansatz (cf. section
\ref{BDS_and_departures}).\footnote{The first argument of
$\Remainder_{Wn}({\bf u};\WLa)$
denotes the set of all conformal cross-ratios that can be constructed
from the coordinates of the $n$ cusps.}

\subsection{Higher-loop tests of the amplitude/Wilson loop relation}

The discussion in previous sections exposes a relation between two
apparently different quantities: MHV gluon scattering amplitudes in
${\cal N}=4$ SYM theory and the expectation value of Wilson loops
constructed in a special way from light-like segments. Manipulations
using $T$-duality transformations for strings in $AdS_5\times S^5$
suggest that this is indeed the case at strong coupling (cf. section
\ref{strong_coupling}).  At weak coupling we have seen evidence for
this relation at one-loop level and we will see further two-loop
evidence in the next section. Before proceeding with describing these
calculations \cite{DHKS1_2loop,Drummond:2007au,DHKS3_2loop,DHKS4_2loop},
let us pause and discuss the general structure of both amplitudes and
Wilson loop expectation values which have emerged from our symmetry
considerations.

The structure of MHV scattering amplitudes follows from the BDS
ansatz, explicit calculations and the assumption that the dual
conformal symmetry observed at low orders in perturbation theory holds
to all orders in perturbation theory: \footnote{As mentioned previously
in section \ref{weak_coupling}, in the absence of an explicit proof,
this assumption must to be tested by explicit calculations.}
\begin{equation}
\label{Amp_Gen}
\ln {\cal M}_n=\Div_n+\frac{f(\WLa)}{4}{\cal M}_{n}^{(1)}(k_1,...,k_n)+
\Remainder_{An}({\bf u}, \lambda)+C(\WLa)+n k(\WLa)
\end{equation}
where ${\cal M}_{n}^{(1)}$ is the rescaled one-loop amplitude,
$R_A({\bf u}, \WLa)$ is a function of the 't~Hooft couplings and
all conformal ratios consistent with the kinematics of the process and
$C(\WLa)$ and $k(\WLa)$ are functions that are independent
of the kinematics and the number of gluons. The first two and last two
terms above are part of the BDS ansatz while the remainder function
$\Remainder_A$ captures the potential departures from it. It is
important to stress that $R_{A4,5}({\bf u}, \WLa)=0$ and thus the
BDS ansatz holds true in these cases.

The structure of the expectation value of light-like cusped Wilson
loops follows, as seen in the previous section, from explicit
calculations and dual conformal invariance:
\begin{equation}
\label{WL_Gen}
\ln \langle W_n \rangle=\widetilde{\Div}{}_n
+w_{n}^{(1)}(k_1,...,k_n,\WLa)+\Remainder_{Wn}({\bf u}, \WLa)
+c(\WLa)+n d(\WLa)
\end{equation}
where $w_{n}^{(1)}(k_1,...,k_n)$ is, up to a factor of the cusp
anomaly, the one-loop expectation value of the Wilson loop, $R_W({\bf
u}, \lambda)$ is a function of the 't~Hooft couplings and all
conformal ratios consistent with the light-like nature of the loop and
$c(\lambda)$ and $d(\lambda)$ are functions that are independent of
the kinematics or the number of edges. Explicit computations show that
$w_n^{(1)}=\frac{f(\lambda)}{4}{\cal M}_{n}^{(1)}$.  It is also important to
mention that this separation of the divergent part of the logarithm of
the Wilson loop expectation value from its finite part is consistent
with the strong coupling analysis described in section
\ref{strong_coupling}.

If both the amplitude/Wilson loop relation as well as the BDS ansatz
were indeed to hold, then together they would imply that the logarithm
of the expectation value of the light-like cusped Wilson loops,
naturally given by the maximal nonabelian exponentiation theorem in
terms of Feynman diagrams, would equal up to a factor of the cusp
anomaly the one-loop expectation values of the same light-like cusped
Wilson loops. This nontrivial statement holds for four- and five-sided
loops, as a consequence of dual conformal invariance which implies
that the remainder functions $\Remainder_{W4,5}$ are constants. As we
will review in section \ref{sixsides_2loops}, starting with six-sided
loops this no longer holds, implying the existence of nontrivial
remainder functions.

On the explicit expressions (\ref{Amp_Gen})-(\ref{WL_Gen}) one may
take the strong coupling limit and contrast the result with the
expectations stemming from the analysis in section
\ref{strong_coupling}. If the relation between Wilson loops and MHV
amplitudes holds, then
\be
\lim_{\lambda\rightarrow\infty} (w_{n}^{(1)}(k_1,...,k_n,\WLa)
+\Remainder_{Wn}({\bf u}, \WLa))
=
\lim_{\lambda\rightarrow\infty}
\left(\frac{f(\WLa)}{4}{\cal M}_{n}^{(1)}(k_1,...,k_n)+
\Remainder_{An}({\bf u}, \WLa)\right)\;.~~
\label{comparison}
\ee
One may further make use of the fact that
$w_{n}^{(1)}(k_1,...,k_n,\WLa)=
\frac{f(\WLa)}{4}{\cal M}_{n}^{(1)}(k_1,...,k_n)$ to conclude that the two remainder
functions $\Remainder_{Wn}$ and $\Remainder_{An}$ must be equal.  It
is however more instructive to temporarily leave equation
(\ref{comparison}) unchanged.

\subsubsection{Rectangular configuration with a large number of gluons}

To test the relation between Wilson loops and MHV amplitude at strong
coupling in the form of the equation (\ref{comparison}) it appears
necessary, at least at first sight, to find the minimal surface
corresponding to some $n$-sided polygonal boundary conditions. As seen
in section \ref{strong_coupling}, such a construction is challenging
for any number of sides larger than four. As we have discussed,
agreement is guaranteed in this instance by dual conformal invariance.

\begin{figure}[ht]
\centering
\includegraphics[scale=0.6]{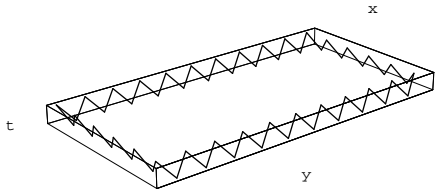}
\caption{Zigzag configuration approaching the space-like
rectangular Wilson loop.} \label{Rlines}
\end{figure}

A technically simpler construction corresponds to a relatively singular
kinematics corresponding to infinitely many gluons with alternating
positive and negative energies. In the dual configuration space this
is represented by a sequence of light-like segments spanning a zigzag
following a light-like rectangular contour of width $L$ and height $T$,
as shown in figure \ref{Rlines}.

In the limit of a large number of edges and for very large $T$ and $L$ such
that $T \gg L$ the contribution to the scale-invariant part of the result
may be identified as being given by the same minimal surface that yields the
quark-antiquark potential \cite{Rey:1998ik,{Maldacena:1998im}}. Indeed, in
this limit and for the purpose of the evaluation of the scale-invariant
part of the area, the minimal surface corresponding to the zigzag boundary
condition may be approximated by the minimal surface ending of the rectangle.
It follows then \cite{Rey:1998ik,{Maldacena:1998im}} that
\be
\ln \langle W_{rect}\rangle =\frac{\sqrt{\lambda}}{4}\,\frac{16\pi^2}
{\Gamma(\frac{1}{4})^4}\,\frac{T}{L}~~.
\label{inf_pt}
\ee
This is the strong coupling limit of the scale-invariant part of the
left hand side of the equation (\ref{comparison}). The first term on
the right hand side may be trivially computed, as it is given by the
strong coupling limit of the cusp anomaly and the particular
kinematic limit of the expectation value of the $n$-sided Wilson loop
at one-loop -- i.e. by the scale-invariant part of the one-loop
correction to the quark-antiquark potential:
\be
\frac{1}{4}f(\WLa){\cal M}_{n}^{(1)}(k_1,...,k_n)
{\longrightarrow}\frac{\sqrt{\lambda}}{4} \frac{T}{L}~~,
\ee
where we used the fact that, as stated in (\ref{Amp_WL_relation}),
$\WLa=\lambda/8\pi^2$.  Since this first term on the right hand side
of the equation (\ref{comparison}) does not reproduce the expression
in equation (\ref{inf_pt}), it follows that the remainder functions
$\Remainder_A$ and/or $\Remainder_W$ are nontrivial functions of
kinematic invariants\footnote{It is worth mentioning that this
reasoning was the first indication that the BDS ansatz departs from
the true expression of scattering amplitudes. Additional evidence in
the same direction comes from the analysis of the BFKL equation
\cite{BLSA}.}. This is consistent with the observation in section
\ref{weak_coupling} that $\Remainder_{A6}$ is nontrivial.  To
strengthen this conclusion we will summarize in the next section the
two-loop correction to the expectation value of light-like cusped
Wilson loops.

\subsection{Two loops and beyond}

The arguments in the previous sections, leading to the conjectured
relation between scattering amplitudes and light-like cusped Wilson
loops, are very compelling. Dual conformal symmetry  $SO(2,4)$ is,
however, yet to be proven to be a symmetry of the (MHV) amplitudes. Thus,
its presence and consequences need to be tested.

%
%
%

One way to explore further the scattering amplitudes/ Wilson loops
equivalence is by comparing the results of explicit calculations
on both sides. This includes both four- and five-point amplitudes
and the corresponding Wilson loops. The rationale behind low-point
calculations is to test the consequences of dual conformal invariance.
In the following we review two-loop computations for the expectation
value of light-like Wilson loops for four and six cusps\footnote{The
two-loop correction to the five-sided cusped Wilson loop was evaluated
in \cite{Drummond:2007au}.}. The loop with $n=6$ is
particularly important since, as we have already seen, it is the
lowest number of edges for which the result is not fixed by conformal
invariance and thus a remainder function can appear.

Both for $n=4$ and for $n=6$ the goal is to evaluate the expectation value
(\ref{WL_general_YM}) up to order $g^4$. One may organize the
contributions to the
expectation value of the loop following the order in the expansion of
the path-ordered exponential which participates in the calculation. To
order $g^4$ there are are terms with two, three and four gauge fields:
\be
W_n\ &=&1 \\
 &+&\frac{1}{2}(i g)^2 \Tr P\int_{C_n} d\tau_1 d{\tau_2}\;
\dot{\hat x}^\mu_1\dot{\hat x}^\nu_2A_\mu({\hat x}_1)A_\nu({\hat x}_2)\cr
 &+&\frac{1}{3!}(i g)^3 \Tr P\int_{C_n} d\tau_1 d{\tau_2} d{\tau_3}
\dot{\hat x}^\mu_1\dot{\hat x}^\nu_2\dot{\hat x}^\rho_3
A_\mu({\hat x}_1)A_\nu({\hat x}_2)A_\rho({\hat x}_3)\cr
 &+&\frac{1}{4!}(i g)^4 \Tr P\int_{C_n} d\tau_1 d{\tau_2} d{\tau_3}
d{\tau_4}
\dot{\hat x}^\mu_1\dot{\hat x}^\nu_2\dot{\hat x}^\rho_3 \dot{\hat x}^\sigma_4
A_\mu({\hat x}_1)A_\nu({\hat x}_2)A_\rho({\hat x}_3)A_\sigma({\hat x}_4)~~,
 \nonumber
\ee
where we used the notation ${\hat x}_i\equiv x(\tau_i)$.
The two-gluon term contributed to the one-loop expectation
value as well. To order $g^4$ some of the interaction terms in the
Lagrangian become relevant. Indeed, to this order, the two-gluon
term above combines with two three-field terms or one four-field
term from the Lagrangian, while the three-gluon term above
combines with a three-field term in ${\cal L}$.

\begin{figure}[ht]
\centering
\includegraphics[scale=0.7]{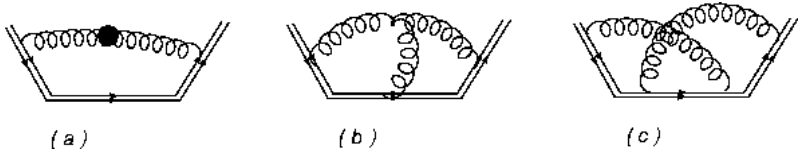}
\caption{Generic diagrams appearing at two loops when computing
the expectation value of cusped Wilson loops.} \label{diftwoloops}
\end{figure}

As in the one-loop computation, all terms may be conveniently
represented in terms of Feynman diagrams. Depending on the number
of edges of the loop they may be further classified following the
number of adjacent edges the gluons are attached to. Some of the
diagrams which appear in all calculations are shown in figure
\ref{diftwoloops}. The diagrams \ref{diftwoloops}(a) have the same
topology as the one loop diagrams except that the gluon propagator
is dressed with a self-energy insertion.  The diagrams
\ref{diftwoloops}(b) contain three gluon propagators joined by a
three-gluon vertex. Last, diagrams \ref{diftwoloops}(c) do not
contain any interaction terms from the Lagrangian.

It is interesting to note that these diagrams are almost
independent of the matter content of the theory. Indeed, all
diagrams except the self-energy insertion receive contributions
only from gauge fields. Moreover, the one-loop gluon self-energy
insertions are very similar in all four-dimensional conformal
field theories as fundamental role of the matter content at this
order is to guarantee the vanishing of the beta function. A
possible interpretation of this observation is that, similarly to
the BDS ansatz, the expectation value of cusped Wilson loops
captures a universal part of the physics of four-dimensional
conformal field theories in general and of their scattering
amplitudes in particular (a part which is nonetheless different
from that captured by the BDS ansatz).

An important tool in higher-loop computations is the so called
nonabelian exponentiation
theorem
\cite{max_nonabelian_exp1,max_nonabelian_exp2,max_nonabelian_exp3}.
As mentioned previously, it states that the expectation value of
Wilson loops has a natural exponential structure; the logarithm of
$\langle W \rangle$ is itself given in terms of Feynman diagrams which
are a subset of the complete set of diagrams\footnote{Up to a few
planar diagrams with two gluon propagators} contributing to $\langle W
\rangle$. This subset is identified by the color factors:
\begin{equation}
\ln{\langle W \rangle}=\log{\left(1+\sum_{l=1}^\infty
\left(\frac{g^2}{4\pi^2}\right)^l W^{(l)}
\right)=\sum_{l=1}^\infty \left(\frac{g^2}{4\pi^2}\right)^l
c^{(l)}w^{(l)}}
\end{equation}
where $W^{(l)}$ denotes the $l$ loops contribution to the
expectation value of the Wilson loop and $w^{(l)}$ denotes the
contribution to $W^{(l)}$ with the "maximally nonabelian"
color factor $c^{(l)}$. Roughly, only the subset of $l$
loops diagrams with maximally interconnected gluon propagators
contribute to $w^{(l)}$. The examples in the following section will
hopefully clarify this notion.
To the first few orders in the loop expansion, $l=1,2,3$, the maximally
nonabelian factor is $c^{(l)}=C_F N^{l -1}$, but starting from
four loops it is not expressible in terms of the Casimirs $C_F$
and $C_A$ \cite{max_nonabelian_exp2,max_nonabelian_exp3}.

The diagrams with non-maximal color factor factorize in products of
lower-loop contributions and, in the calculation of $\langle
W\rangle$, may be identified with terms in the expansion of
$\exp\left[{\sum_{l=1}^\infty \left(\frac{g^2}{4\pi^2}\right)^l
c^{(l)}w^{(l)}}\right]$. From this perspective one may intuitively
relate the appearance of maximal nonabelian color factors with
two-particle irreducibility. To conclude these preparations, by using
the nonabelian exponentiation theorem, $w^{(2)}$ is completely
determined by the contribution to $W^{(2)}$ proportional to $C_F N$.
\footnote{Indeed, by using the nonabelian exponentiation theorem
we find that $W^{(1)}=C_F w^{(1)}$ and $W^{(2)}=C_F N
w^{(2)}+\frac{1}{2}C_F^2(w^{(1)})^2$, so the piece of $W^{(2)}$
proportional to $C_F^2$ is determined by the one loop correction.}

\subsubsection{Polygon with four cusps \label{4cusps_2loops}}

\begin{figure}[ht]
\centering
\includegraphics[scale=0.8]{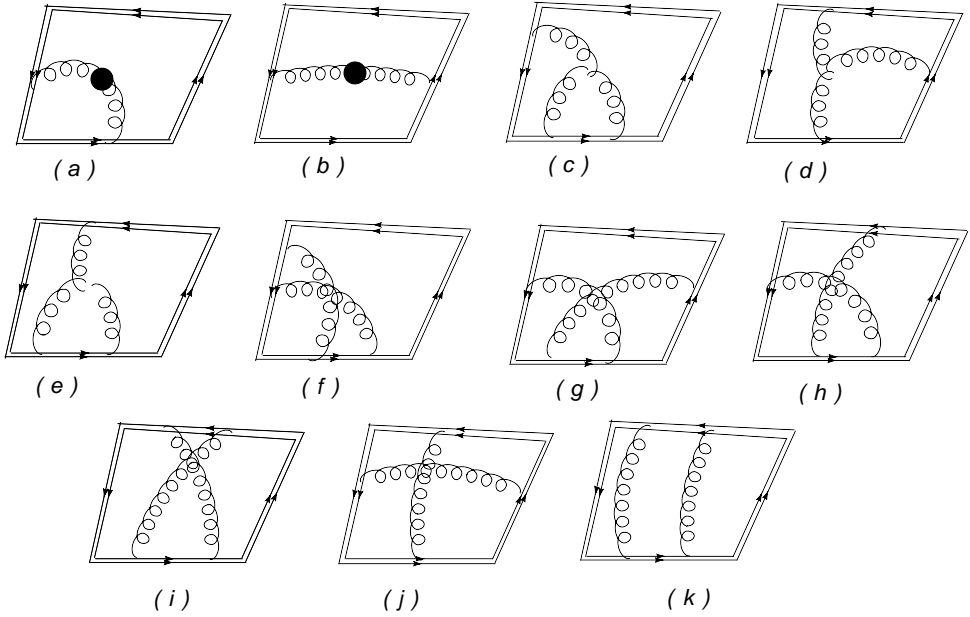}
\caption{Diagrams contributing to the expectation value of the
four-sided Wilson loop at two-loop level; all diagrams
contributing to the logarithm of the Wilson loop expectation value
(i.e. those with maximal nonabelian color factor) are shown. We
also show an example (diagram $k$) of diagram with non-maximal
color factor; these diagrams don't need to be evaluated, as their
contribution disappears in the logarithm of the expectation value
of the Wilson loop.} \label{fourtwoloops}
\end{figure}

The basic ingredients of two loops computations already appear when
studying two loops corrections to the four cusps Wilson loop, so we
begin by reviewing this calculation in some detail \cite{DHKS1_2loop}.
The complete set of diagrams, not making use of the nonabelian
exponentiation theorem, is shown in figure \ref{fourtwoloops}.

It is not hard to argue that, similarly to their one-loop
counterparts, two-loop diagrams with both ends of a gluon propagator
on the same light-like edge vanish identically (essentially because
they depends on a single momentum and no nonvanishing invariant may be
constructed from it).  They have not been included in figure
\ref{fourtwoloops}. The properties of nonvanishing diagrams are
summarized below \cite{DHKS1_2loop}:

\begin{itemize}

\item Diagrams with a single (dressed) gluon propagator are
shown in fig. \ref{fourtwoloops}, (a) and (b). Their color factor is
$C_F N$, and thus they contribute to $w^{(2)}$. Both contain contain
divergences due to the presence of the cusp, from the light-like edges
and also intrinsic to the self-energy insertion.  Their pole expansion
takes the form $A^{(a)}\simeq \frac{1}{\epsilon^3}+...$ and
$A^{(b)}\simeq \frac{1}{\epsilon}+...$.

\item The color factor of diagrams with three gluon propagators
joined by a interaction vertex, shown in figures \ref{fourtwoloops},
(c), (d) and (e), is also $C_F N$; they therefore also contribute to
$w^{(2)}$. All of them are divergent and their pole expansion is
$A^{(c)}\simeq \frac{1}{\epsilon^4}+...$, $A^{(d)}\simeq
\frac{1}{\epsilon^2}+...$ and $A^{(e)}\simeq
\frac{1}{\epsilon}+...$. The strength of the leading singularity may
be understood in terms of the number gluon propagators which may
collapse at cusp singularities.

\item Non-planar diagrams with two gluon propagators, shown
in figures \ref{fourtwoloops}(f)--(j), have a color factor equal to
$C_F(C_F-N/2)$. Hence, the term proportional to $C_F N$, obtained by
replacing $C_F(C_F-N/2) \rightarrow -C_F N/2$, contributes to
$w^{(2)}$ \footnote{Notice that the contribution of these diagrams to
$W^{(2)}$ is suppressed in the large $N$ limit, however, their
contribution to $w^{(2)}$ is not}. Diagrams of type (g), (i) and (j)
are finite, while diagrams $(f)$ and $(h)$ diverge as $A^{(f)}\simeq
\frac{1}{\epsilon^4}+...$ and $A^{(h)}\simeq
\frac{1}{\epsilon}+...$. Similarly to the previous item, the strength
of these singularities may be understood in terms of the number gluon
propagators which may collapse at cusp singularities.

\item The planar diagram with iterated gluon propagators shown in
figure \ref{fourtwoloops}(k) has $C_F^2$ as color factor. Therefore,
this diagram does not contribute to $w^{(2)}$. In fact, it combines
with the terms proportional to $C_F^2$ ignored in the previous item to
yield unrestricted integrals over the end-points of the two gluon
propagators; these terms are nothing but the finite part of
$(w^{(1)})^2$ which is necessary to reconstruct $W^{(2)}$.

\end{itemize}

It is possible to find analytic expressions for all the integrals
shown in figure \ref{fourtwoloops}. Combining them in the appropriate
permutations leads to an expression for $w^{(2)}$:
\begin{equation}
w^{(2)}=\left((-x_{13}^2\mu^2)^{2\epsilon}+
(-x_{24}^2\mu^2)^{2\epsilon}\right)
\left(\frac{1}{\epsilon^2}\frac{\pi^2}{48}+\frac{1}{\epsilon}\frac{7}{8}\zeta_3
\right)-\frac{\pi^2}{24}\log^2\left(\frac{x_{13}^2}{x_{24}^2}\right)
-\frac{37}{720}\pi^4+{\cal
O}(\epsilon)
\end{equation}
Combining this and the expression for $w^{(1)}$ constructed in section
\ref{four_poly} leads to an expression for the logarithm of the
expectation value of the Wilson loop as a sum of a divergent term and
a finite term, which matches the structure of equation
(\ref{genericform})
\begin{equation}
\ln \langle W_4\rangle=\Div(-x_{13}\mu^2)+\Div(-x_{24}\mu^2)
+{F}_4\left(\frac{x_{13}^2}{x_{24}^2}\right)~~.
\end{equation}
The divergent $\Div(-x_{13}\mu^2)$ and finite ${F}_4$ parts are
\begin{eqnarray}
\Div(-x_{13}\mu^2)=-\frac{\WLa}{\epsilon^2}(-x_{13}\mu^2)^\epsilon
                  +\WLa^2 (-x_{13}\mu^2)^{2\epsilon}
\left(\frac{1}{(2\epsilon)^2}\frac{\pi^2}{6}+\frac{1}{2\epsilon}\frac{7}{2}\zeta_3
\right)+{\cal O}(\WLa^3)\\
F_4\left(\frac{x_{13}^2}{x_{24}^2}\right)
=\frac{1}{4}\;\left(2\WLa-\frac{\pi^2}{3}\WLa^2+{\cal O}(\WLa^3)
\right)\;
\left(\ln\frac{x_{13}^2}{x_{24}^2}\right)^2
+\left(\frac{\pi^2}{3}\WLa-\frac{37}{360}\pi^4 \WLa^2
\right)+{\cal O}(\WLa^3)
\end{eqnarray}
The identifications (\ref{identifications_UV_IR}) imply that the
leading pole of $\Div(-x_{13}\mu^2)$ agrees with the general form of
the infrared poles of scattering amplitudes
(\ref{divergence_general}). The subleading poles in the same equation,
evaluated for the universal scaling function and $G$-function of
${\cal N}=4$ SYM, are also reproduced after a further
$\lambda$-dependent redefinition of $\mu$.
The double-logarithm in the finite term $F_4$ reproduces the kinematic
dependence of the finite part of the two loops MHV amplitude for four
gluons (\ref{OneLoopFiniteRemainder}). Its coefficient may be
recognized as the two-loop cusp anomaly (\ref{weak_coupling_cusp}).
In light of the constraints imposed by dual conformal invariance
this agreement is not surprising; this calculation however shows
the validity of higher-loop arguments based on it. \footnote{In
\cite{Korchemskaya} it was argued that a similar duality relation
between the four point amplitude and the four edges Wilson loop
holds also for QCD in the Regge limit.}

\subsubsection{Polygon with six cusps \label{sixsides_2loops}}

As discussed at length in section \ref{constraints_WL_vev}, dual
conformal symmetry is not sufficiently powerful to completely fix the
expectation value of light-like Wilson loops with at least six
cusps. The expectation value is instead fixed up to the addition of
some function of conformal ratios and the coupling constant
(cf. equation (\ref{general_Remainder_WL}).
Thus, the finite part of the logarithm of the six-sided loop (defined by
$\ln{W_n}=\ln Z_n+\ln F_n$) is
\begin{equation}
\ln F_6(\WLa)=\frac{1}{4} f(\WLa)\,F_6^{(1)}(0)
                              +\Remainder_{W6}(u_1,u_2,u_3; \WLa)
\label{ln_finite_6pt}
\end{equation}
where $F_6^{(1)}(0)$ is the one-loop expectation value of the
six-sided loop and $\Remainder_{W6}(u_1,u_2,u_3; a)$ is an
arbitrary function of the coupling constant and the three
nontrivial conformal cross-ratios which may be constructed from
the coordinates of the six cusps (see equation
(\ref{6pt_crossratios})). Its loop expansion reads
\be
\Remainder_{W6}(u_1,u_2,u_3; \WLa)=\sum_{l=1}^\infty\;\WLa^l\,
\Remainder_{W6}^{(l)}(u_1,u_2,u_3)~~.
\label{Remainder_WL_expansion}
\ee
By construction, the one-loop remainder function
$\Remainder_{W6}^{(1)}(u_1,u_2,u_3)$
is at most a constant independent of the kinematics.  The properties
of the higher loop terms in (\ref{Remainder_WL_expansion}) may be
found (at the time of this writing) only by explicit calculations of
the higher loop expectation value of the Wilson loop.

As for the four-sided loop, the calculation makes use of the
nonabelian exponentiation theorem, which identifies, at the level
of Feynman diagrams, the contributions to $\ln \langle
W_6\rangle$. It is however unclear whether the contributions to
$\Remainder_{W6}^{(2)}(u_1,u_2,u_3)$ have by themselves a Feynman
diagram interpretation.\footnote{\label{differences_4_vs_6} One
may attempt to identify them with the diagrams which do not exists
for loops with four and five cusps. However, even the diagrams
that do exists in those cases are structurally different since
some of the differences of cusp coordinates are no longer
light-like. It is therefore not clear whether such an
identification is appropriate.} Thus, the only available method of
identifying $\Remainder_{W6}^{(2)}(u_1,u_2,u_3)$ is to first
evaluate the complete two-loop contribution to $\ln \langle
W_6\rangle$ and then subtract the divergences as well as the
two-loop contribution of the first term in equation
(\ref{ln_finite_6pt}).

The calculation was carried out in \cite{DHKS3_2loop, DHKS4_2loop};
some of the diagram topologies contributing to $w^{(2)}_6$ are
analogous to those discussed in section \ref{4cusps_2loops} (see
however the differences pointed out in footnote
\ref{differences_4_vs_6}). Others are completely different, due to the
additional freedom provided by the existence of additional edges.

\begin{figure}[ht]
\centering
\includegraphics[scale=1]{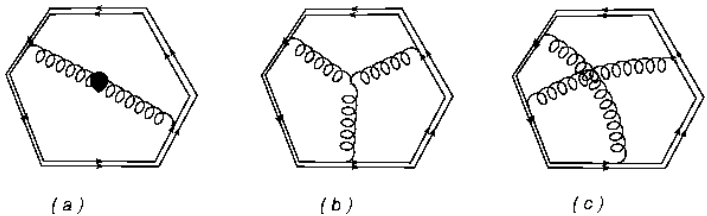}
\caption{Diagram topologies contributing to the expectation value of the
light-like hexagon Wilson loop which have no counterpart for Wilson
loops based on lower polygons.}
\label{sixtwoloops}
\end{figure}

The calculation is technically more involved that that of the
four- and five-sided polygon; the final result may currently be
found only numerically. An immediate test of the result is that
indeed the Wilson loop remainder function depends only on the
three conformal cross-ratios, as the initial construction implies.
Besides this consequence of dual conformal invariance, it is also
possible to identify other properties of the
$\Remainder_{W6}^{(2)}(u_1,u_2,u_3)$ and compare them with
expectations based on the conjectured relation with MHV scattering
amplitudes, as captured by the equation (\ref{comparison}).
Perhaps the main consequence of that equation is that in the
Wilson loop analog of a collinear limit,  consistency with the
results for the expectation value of the five-sided loop requires
that the Wilson loop remainder function
$\Remainder_{W6}^{(2)}(u_1,u_2,u_3)$ should become a
constant.\footnote{This is analogous to the requirement that the
amplitude remainder function $\Remainder_{A6}^{(2)}(u_1,u_2,u_3)$
vanishes in this limit.}

\begin{figure}[ht]
\centering
\includegraphics[scale=0.5]{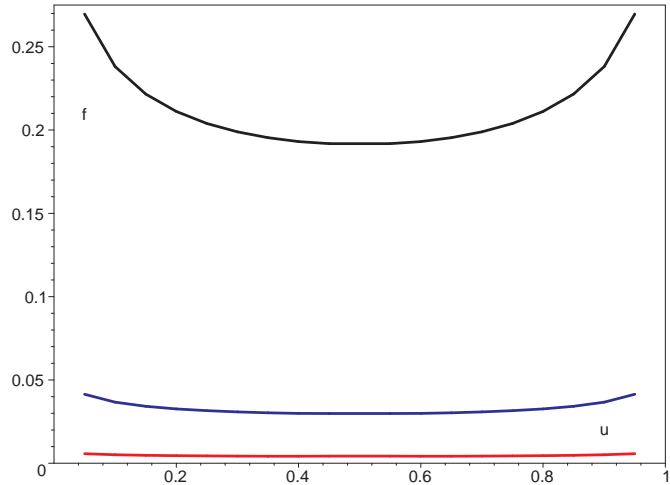}
\caption{The $u$ dependence of the difference
$(c-\Remainder_{W6}^{(2)}(\gamma,u,1-u))$ for different values of
$\gamma=0.001$ (lower curve), $\gamma=0.01$ (middle curve) and
$\gamma=0.1$ (upper curve).} \label{f_vs_u}
\end{figure}

The precise definition of collinear limit was described in
equation (\ref{collinear_limit}). By inspecting the expressions of
the cross-rations $u_{1,2,3}$ in equations
(\ref{SixPtConformalCrossRatios}) and (\ref{6pt_crossratios}) it
is easy to see that any collinear limit corresponds to exactly one
vanishing cross-ratio. Without loss of generality one may choose
it to be $u_1$. It is then not hard to see that, in this limit,
the remaining cross-ratios are related by \be u_2+u_3=1~~. \ee
With the notation $u\equiv u_2$, the Wilson loop remainder
function should therefore behave as \be
\Remainder_{W6}^{(2)}(0,u,1-u)=c ~~, \label{collin_lim_WL} \ee
independently of $u$. Thorough numerical tests of this relation
were carried out in \cite{DHKS3_2loop} with the result that
equation (\ref{collin_lim_WL}) does indeed hold.  Figure
\ref{f_vs_u} illustrates the numerical results \footnote{We thank
the authors of \cite{Drummond:2007aua} for providing us with this
figure.}. The constant value of $\Remainder_{W6}^{(2)}(0,u,1-u)$
was found in \cite{DHKS4_2loop} to be \be
\Remainder_{W6}^{(2)}(0,u,1-u)=12.1756~~.
\label{constant_Reminder} \ee Due to the singular nature of
collinear limits, the numerical error associated to this value
($\sim 10^{-3}$) is larger than for generic $u_1\equiv \gamma\ne
0$ kinematic  configurations.

Figure \ref{f_vs_u} also implies that
$\Remainder_{W6}^{(2)}(\gamma,u,1-u)$ is not a constant away from the
$u_1\equiv \gamma=0$ surface.  As discussed in section \ref{6gamp},
the same holds for the (MHV) amplitude remainder function
$\Remainder_{A6}^{(2)}$.  The comparison of the values of the two
functions at selected kinematic points was carried out in
\cite{BDKRSVV, DHKS4_2loop}. To avoid any mismatch due to unaccounted
constant shifts as well as loss of numerical precision it is
convenient to compare the difference of remainder functions from a
chosen reference point. A convenient one is the symmetric point
$K^{(0)}$ in table
\ref{RemainderTable} for which the conformal cross-ratios are
\be
(u_1,u_2,u_3)=\left(\frac{1}{4},\,\frac{1}{4},\,\frac{1}{4}\right)~~.
\ee
The differences between the amplitude and Wilson loop remainder
functions at the points $K^{(i)}$ and at $K^{(0)}$ (denoted by
$\Remainder_{A6}^{0}$ and $\Remainder_{W6}^{0}$, respectively) are
shown in table \ref{ComparisonTable}; the third column contains the
difference of remainders for the amplitude, while the fourth column
has the corresponding difference for the Wilson loop.

\begin{table}[ht]
\begin{tabular}{||c|c||c|c||}
\hline
\hline
kinematic point & $(u_1, u_2, u_3)$ & $\Remainder_{A6} - \Remainder_{A6}^{0}$ &
    $\Remainder_{W6} - \Remainder_{W6}^{0}$ \\
\hline
\hline
$K^{(1)}\vphantom{A^{B^{C^D}}}$
                  &$(1/4, 1/4, 1/4)$ & $-0.018 \pm  0.023 $ &  $ <10^{-5}$  \\
\hline
$K^{(2)}\vphantom{A^{B^{C^D}}}$
                & $\,(0.547253,\, 0.203822,\, 0.881270)\,$  & $-2.753 \pm 0.015$
 & $ -2.7553$ \\
\hline
$K^{(3)}\vphantom{A^{B^{C^D}}}$
                & $(28/17, 16/5, 112/85)$ & $\, -4.7445 \pm  0.0075\, $ & $ -4.7446$ \\
\hline
$K^{(4)}\vphantom{A^{B^{C^D}}}$
               & $(1/9, 1/9, 1/9)$  & $ 4.12  \pm  0.10$  & $   4.0914$  \\
\hline
$K^{(5)}\vphantom{A^{B^{C^D}}}$
               & $(4/81, 4/81, 4/81)$  & $ 10.00 \pm  0.50$  & $  9.7255$  \\
\hline
\end{tabular}

\caption{\label{ComparisonTable} The comparison between the remainder
functions $R_{A6}$ and $R_{W6}$ for the six-point MHV amplitude and
the six-sided Wilson loop.
The numerical agreement between the third column and fourth
columns provides strong evidence that the remainder function for the
Wilson loop is identical to that for the MHV amplitude.
}

\end{table}

The agreement between the remainder functions for the six gluon
amplitude and that of the six-sided Wilson loop shown by table
\ref{ComparisonTable}  suggests that MHV amplitudes and Wilson loops
continue to be related even when dual conformal invariance is not
sufficiently restrictive to uniquely fix them.

By construction, the remainder function for the six-gluon MHV
amplitude vanishes in all collinear limits. Using the constant
value (\ref{constant_Reminder}) one may define a Wilson loop
reminder with similar vanishing properties.  Such a remainder
function enforces the fact that all singularities as well as the
finite terms related to them are accounted for by the solution of
the anomalous Ward identity. It is worth mentioning that this
remainder function reproduces the amplitude one without additional
subtractions.

\section{Outlook \label{outlook}}

Scattering amplitudes remain one of the basic ingredients in our
understanding of quantum field theories. They are usually evaluated
order by order in a weakly-coupled perturbation theory and it is
rarely the case that the resulting series can be constructed and
resummed to all orders in perturbation theory, even only in the planar
limit. The ${\cal N}=4$ SYM theory is perhaps special in this respect,
being sufficiently simple to make higher order calculations feasible
yet being sufficiently nontrivial for the resulting scattering matrix
to (apparently) contain nontrivial dynamical information.

We have described techniques which allow, in ${\cal N}=4$ SYM,
efficient higher-loop and higher multiplicity calculations. Though not
reviewed here, some of these techniques have been extended to
phenomenologically-relevant theories, such as QCD (for a review see
e.g. \cite{Bern:2007dw}). General properties of infrared singularities
together with explicit higher loop calculations led to the formulation
of the BDS ansatz, expressing all higher-loop MHV amplitudes in terms
of their one-loop expressions.

In the strong coupling regime, the AdS/CFT duality suggests, through
the use of T-duality, that partial amplitudes may be evaluated as the
regularized area of the minimal surface ending on a special light-like
polygon whose specific properties depend on the particles being scattered.
Quite remarkably, explicit calculations show that this relation holds
at weak coupling as well. While explicit higher-loop high-multiplicity
calculations turn out to depart from existing conjectures for the
resummation of the perturbative series for planar MHV amplitudes, they
reproduce the results of Wilson loop calculations.

An important concept behind both weak- and strong-coupling amplitude
calculations in ${\cal N}=4$ SYM is that of dual conformal
invariance. Initially observed in explicit results for scattering
amplitudes, its origin remains unclear and its presence not proven to
all orders in perturbation theory at the level of scattering
amplitudes. It is however an almost manifest symmetry of Wilson loop
expectation values. This symmetry is sufficiently powerful to uniquely
fix the expressions of four- and five-point amplitudes to all orders
in perturbation theory if its presence is assumed to all orders in
perturbation theory. The resulting expressions reproduce the BDS
ansatz for four and five particle processes.

The full consequences and implications of the developments reviewed
here are yet to emerge and many questions, which will undoubtedly
contribute in this direction, remain to be addressed. Some of them are
included here:

\begin{itemize}

\item Despite partial progress outlined in section
\ref{generalizations}, the construction of minimal surfaces describing
the scattering of more than four gluons is still lacking. Such
solutions, or at least an expression for their regularized area, may
provide valuable input for understanding the iteration relations at
strong coupling.
Up to a choice of boundary conditions, the dynamics of strings in
$AdS_5\times S^5$ is described by an integrable two-dimensional field
theory \cite{BPR, Wadia}. It would be potentially profitable to
understand the consequences of integrability for the construction of
minimal surfaces in this space. While it is not clear whether the
integrability of the worldsheet theory survives in the presence of the
dimensional regulator, integrability at $\epsilon=0$ may suffice to
find the regularized area of the minimal surface without explicitly
constructing the solution.

\item
Besides scattering processes, the strategy described in section
\ref{strong_coupling} may be used to argue that the overlap
between some composite operator and some multi-gluon state also
has a minimal surface interpretation. Physically, this overlap has
the interpretation of the decay amplitude of a colorless scalar
(with particular couplings) into gluons. Minimal surfaces with
this interpretation are not known even in the simplest cases.
Besides their obvious interpretation in terms of decay amplitudes,
understanding in detail such processes may also lead to
understanding the calculations of anomalous dimensions of short
operators on the string theory side of the AdS/CFT correspondence.

\item
Besides dependence on momentum invariants, scattering amplitudes are
also sensitive to the polarization (helicity) of the scattering
states. While present if the calculation of scattering amplitudes is
organized in terms of open-string vertex operators,this information
is lost in the map to light-like Wilson loops. It would be very
interesting to understand whether polarization information may be
encoded in Wilson loop language. Attempts in this direction have
appeared in
\cite{McGreevy_Sever_pol}, based on earlier considerations of
ref.~\cite{polyakov_amplitudes}.
In a similar vein, the Wilson loop/amplitude relation should map the
loop equation obeyed by Wilson loops into apparently nontrivial
constraints on scattering amplitudes. It would be interesting to
understand their relevance.

\item
As reviewed in section \ref{strong_coupling}, the evaluation of
$1/\sqrt{\lambda}$ corrections to the expectation value of the
four-sided Wilson loop at strong coupling remains elusive.
Calculations carried out in \cite{Kruczenski:2007cy} suggest that the
definition of the strong coupling version of dimensional
regularization is subtle beyond classical level. While dimensional
regularization remains the ideal choice for comparison with weak
coupling calculations, intuition on the structure of the answer may be
gained by considering alternative regularization schemes, such as that
through a radial cut-off.

\item
A second set of corrections to the leading order strong coupling
calculation of scattering amplitudes amounts to relaxing the
requirement that the saddle-point surface (in the presentation of the
amplitude in terms of vertex operators) has disk topology.
Such higher genus corrections  translates into nonplanar
corrections to the planar scattering amplitudes.
It is an interesting question whether such calculations have a Wilson
loop counterpart.
It would be very interesting to understand whether dual conformal
symmetry is present in any one of these calculations and, if not,
what is the origin of its breaking.

\item
In section \ref{generalizations} we have briefly summarized attempts
of extending the strong coupling relation between Wilson loops and
scattering amplitudes to less supersymmetric and/or non-conformally
invariant situations. It seems important to complete this program, as
it may provide valuable clues for extending the strong coupling
calculations of scattering amplitudes to other, perhaps more
phenomenologically-relevant theories.

\item
While the existence of a dual conformal invariance for scattering
amplitudes is beyond doubt at low orders in perturbation theory, it
remains to be proven whether ${\cal N}=4$ SYM exhibits this symmetry
to all orders in the coupling constant expansion.
Establishing it may constitute a step towards understanding the
complete relation between amplitudes and Wilson loop expectation
values. It would be equally interesting to extent this equivalence
(initially formulated for MHV amplitudes) to non-MHV amplitudes.  From
a field theory standpoint little is known about such amplitudes beyond
leading order, where they are given in terms of pseudo-conformal
integrals with spinor-dependent coefficients. The dual conformal
properties of next-to-MHV amplitudes have been discussed in
\cite{DHKS_super, GK_ES_Paris}. A supersymmetric generalization of
dual conformal transformations played an important role in this
discussion. At strong coupling, similar supersymmetric generalizations
of dual conformal symmetries have been discussed in
\cite{NB_Paris,BRTW,NBJM}. The analysis described there also exposes
a close relation between the generators of the dual (super)conformal
group and the hidden (non-local) integrals of motion of the world
sheet theory in the original AdS$_5\times$S$_5$ (i.e. the worldsheet
theory prior to the T-duality transformations relating scattering
amplitudes and Wilson loops). The full implications of this relation
remain to be uncovered.

\item
Scattering amplitudes and Wilson loops are logically disconnected
quantities. The existence of a close relation between MHV amplitudes
and light-like cusped Wilson loops hints to the existence of new
symmetries which connect (and perhaps uniquely determine) both
quantities.

\item
While numerical comparison shows that the remainder functions are the
same in the amplitude and Wilson loop calculations, it should prove
interesting to construct an analytic expression for these
remainders. (It is worth mentioning that the Wilson loop approach
yields apparently simpler integrals.) Such expressions may hold a key
towards understanding the general structures of MHV amplitudes and to
their generalization to all values of the coupling constant.

\end{itemize}

It is clear that additional structure, waiting to be
uncovered, is present in ${\cal N}=4$ SYM and that it may be sufficiently
powerful to completely determine, at least in some sectors, the
kinematic dependence of the scattering matrix of the theory.

\

\section*{Acknowledgments}
L.F.A. thanks J.~Maldacena and R.R. thanks Z.~Bern, L.~Dixon,
D.~Kosower, M.~Kruczensky, M.~Spradlin, A.~Tirziu, A.Tseytlin,
C.~Vergu and A.~Volovich, for collaboration on the topics reviewed
here.
We would like to thank Z.~Bern,  D.~Kosover,
J.~Maldacena, G.~Korchemsky, A.~Tseytlin and especially L.~Dixon
for valuable discussions, suggestions  and comments on the draft. We
would also like to thank
A.~Sever, N.~Berkovits, J.~de Boer, P.~Caputa, J.~J.~Carrasco,
J.~Henn, P.~Heslop, V.~Khoze, A.~Murugan, E.~Sokatchev and
G.~Travaglini
for discussions.
RR thanks the ETH in Zurich for hospitality during the programme ``QCD
and Strings''.
The work of LFA is supported by VENI grant 680-47-113 while
the work of R.R. is supported in part by the US National Science
Foundation under grant PHY-0608114 and the A.~P.~Sloan Foundation.

\end{document}